# Problems of antimatter after Big Bang, dark energy and dark matter. Solutions in the frame of non-local physics.

By Boris V. Alexeev
*Moscow Academy of Fine Chemical Technology (MITHT)*
*Prospekt Vernadskogo, 86, Moscow 119570, Russia*

Quantum solitons are discovered with the help of generalized quantum hydrodynamics. The solitons have the character of the stable quantum objects in the self consistent electric field. The delivered theory demonstrates the great possibilities of the generalized quantum hydrodynamics in investigation of the quantum solitons. The theory leads to solitons as typical formations in the generalized quantum hydrodynamics. The principle of universal antigravitation is considered from positions of the Newtonian theory of gravitation and non-local kinetic theory. It is found that explanation of Hubble effect in the Universe and peculiar features of the rotational speeds of galaxies need not in introduction of new essence like dark matter and dark energy. Problems of antimatter after Big Bang are considered from positions of non-local physics. The origin of difficulties consists in total Oversimplification following from principles of local physics and reflects the general shortenings of the local kinetic transport theory.

**Keywords:** Foundations of the theory of transport processes; generalized Boltzmann physical kinetics; plasma – gravitational analogy; antigravitation; dark energy; dark matter; the theory of solitons; antimatter after Big Bang.

PACS: 67.55.Fa, 67.55.Hc

## 1. Introduction.

### (a) Preliminary remarks.

Several extremely significant problems challenge modern fundamental physics – missing antimatter after the Big Bang, and so called dark energy and dark matter. The last two problems lead to affirmation that only about four percents of matter leaved for us for direct investigation because the other matter is out of our diagnostic methods. No reason to follow tremendous scientific literature devoted to investigations of these problems. But nevertheless some remarks should be done. The basic ideas of 'big bang' theory were formulated by originators—George Lemaitre, (1927) and also by George Gamow, R.A. Alpher, and R. Herman who devised the basic Big Bang model in 1948. According to the 'big bang' theory for the origin of the universe, equal amounts of matter and antimatter should have formed. Antimatter is the same as matter except that each particle has the opposite charge, magnetic moment, etc. For instance, the antiparticle for the negatively charged electron is the positively charged positron. Antimatter is supposed to be an exact counterpart to matter, down to the same mass. This has been verified in many experiments and it was shown experimentally that a proton and an antiproton have the same mass to within one part in 10 billion. At the beginning, equal amounts of matter and antimatter were created in the "big bang". The problem is that, so far, no antimatter (AM) domains have been detected in space, at least in the nearby universe. This result creates a long-standing mystery: why the big bang displays such blatant favoritism towards matter. Now there seems to be only matter or regular matter (RM) in the following theory. There have been theoretical speculations about the disappearance of antimatter, but no experimental support.

More than ten years ago, the accelerated cosmological expansion was discovered in direct astronomical observations at distances of a few billion light years, almost at the edge of the observable Universe. This acceleration should be explained because mutual attraction of cosmic bodies is only capable of decelerating their scattering. It means that we reach the revolutionary situation not only in physics but in the natural philosophy on the whole. Practically we are in front of the new challenge since Newton's *Mathematical Principles of Natural Philosophy* was published. As result, new idea was introduced in physics about existing of a force with the opposite sign which is called universal antigravitation. Its physical source is called as dark energy that manifests itself only because of postulated property of providing antigravitation.

It was postulated that the source of antigravitation is "dark matter" which inferred to exist from gravitational effects on visible matter. But from the other side dark matter is undetectable by emitted or scattered electromagnetic radiation. It means that new essences – dark matter, dark energy – were introduced in physics only with the aim to account for discrepancies between measurements of the mass of galaxies, clusters of galaxies and the entire universe made through dynamical and general relativistic means, measurements based on the mass of the visible "luminous" matter. It could be reasonable if we are speaking about small corrections to the system of knowledge achieved by mankind to the time we are living. But mentioned above discrepancies lead to affirmation, that dark matter constitutes the most part of the matter in the universe. There is a variety in the corresponding estimations, but the situation is defined by maybe emotional, but the true exclamation which can be found between thousands Internet cues – "It is humbling, perhaps even humiliating, that we know almost nothing about 96% of what is "out there"!!

Dark matter was postulated by Swiss astrophysicist Fritz Zwicky of the California Institute of Technology in 1933. He applied the virial theorem to the Coma cluster of galaxies and obtained evidence of unseen mass. Zwicky estimated the cluster's total mass based on the motions of galaxies near its edge and compared that estimate to one based on the number of galaxies and total brightness of the cluster. He found that there was about 400 times more estimated mass than was visually observable. The gravity of the visible galaxies in the cluster would be far too small for such fast orbits, so something extra was required. This is known as the "missing mass problem". Based on these conclusions, Zwicky inferred that there must be some non-visible form of matter which would provide enough of the mass and gravity to hold the cluster together.

I do not intend to review the different speculations based on the principles of local physics. I see another problem. It is the problem of Oversimplification – but not "trivial" simplification of the important problem. The situation is much more serious – total Oversimplification based on principles of local physics, and obvious crisis, we see in astrophysics, simply reflects the general shortenings of the local kinetic transport theory.

The formulated above problems are solved further in the frame of non-local statistical physics and the Newtonian law of gravitation.

### (b) About the basic principles of the generalized Boltzmann physical kinetics, non-local statistical physics and generalized quantum hydrodynamics.

I begin with the short reminding of basic principles of the generalized Boltzmann physical kinetics, non-local statistical physics (Alekseev 1982, 2000, 2003, Alexeev 1994, 1995, 2004) and generalized quantum hydrodynamics (GQH) created in particular in (Alexeev 2008ab). As it is shown (see, for example Alexeev 2008ab) the theory of transport processes (including quantum mechanics) can be considered in the frame of unified theory based on the non-local physical description. In particular the generalized hydrodynamic equations represent an effective tool for solving problems in the very vast area of physical problems including turbulence. For simplicity in introduction, we will consider fundamental methodic aspects from the qualitative standpoint of view avoiding excessively cumbersome formulas. A rigorous description is found, for example, in the monograph (Alexeev 2004)

Transport processes in open dissipative systems are considered in physical kinetics. Therefore, the kinetic description is inevitably related to the system diagnostics. Such an element of diagnostics in the case of theoretical description in physical kinetics is the concept of the physically infinitely small volume $(\mathbf{PhSV})$. The correlation between theoretical description and system diagnostics is well-known in physics. Suffice it to recall the part played by test charge in electrostatics or by test circuit in the physics of magnetic phenomena. The traditional definition of **PhSV** contains the statement to the effect that the **PhSV** contains a sufficient number of particles for introducing a statistical description; however, at the same time, the **PhSV** is much smaller than the volume $V$ of the physical system under consideration; in a first approximation, this leads to local approach in investigating the transport processes. It is assumed in classical hydrodynamics that local thermodynamic equilibrium is first established within the **PhSV**, and only after that the transition occurs to global thermodynamic equilibrium if it is at all possible for the system under study. Let us consider the hydrodynamic description in more detail from this point of view. Assume that we have two neighboring physically infinitely small volumes $\mathbf{PhSV_1}$ and $\mathbf{PhSV_2}$ in a nonequilibrium system. The one-particle distribution function (DF) $f_{sm,1}(\mathbf{r}_1, \mathbf{v}, t)$ corresponds to the volume $\mathbf{PhSV_1}$, and the function $f_{sm,2}(\mathbf{r}_2, \mathbf{v}, t)$ — to the volume $\mathbf{PhSV_2}$. It is assumed in a first approximation that $f_{sm,1}(\mathbf{r}_1, \mathbf{v}, t)$ does not vary within $\mathbf{PhSV_1}$, same as $f_{sm,2}(\mathbf{r}_2, \mathbf{v}, t)$ does not vary within the neighboring volume $\mathbf{PhSV_2}$. It is this assumption of locality that is implicitly contained in the Boltzmann equation (BE). However, the assumption is too crude. Indeed, a particle on the boundary between two volumes, which experienced the last collision in $\mathbf{PhSV_1}$ and moves toward $\mathbf{PhSV_2}$, introduces information about the $f_{sm,1}(\mathbf{r}_1, \mathbf{v}, t)$ into the neighboring volume $\mathbf{PhSV_2}$. Similarly, a particle on the boundary between two volumes, which experienced the last collision in $\mathbf{PhSV_2}$ and moves toward $\mathbf{PhSV_1}$, introduces information about the DF $f_{sm,2}(\mathbf{r}_2, \mathbf{v}, t)$ into the neighboring volume $\mathbf{PhSV_1}$. The relaxation over translational degrees of freedom of particles of like masses occurs during several collisions. As a result, "Knudsen layers" are formed on the boundary between neighboring physically infinitely small volumes, the characteristic dimension of which is of the order of path length. Therefore, a correction must be introduced into the DF in the **PhSV**, which is proportional to the mean time between collisions and to the substantive derivative of the DF being measured, rigorous derivation is given in (Alexeev 1994, 2004) using BBGKY (Bogolyubov 1946, Born & Green1946, Green 1952, Kirkwood 1947) chain of kinetic equations.

Let a particle of finite radius be characterized as before by the position $\mathbf{r}$ at the instant of time $t$ of its center of mass moving at velocity $\mathbf{v}$. Then, the situation is possible where, at some instant of time $t$, the particle is located on the interface between two volumes. In so doing, the lead effect is possible (say, for $\mathbf{PhSV_2}$), when the center of mass of particle moving to the neighboring volume $\mathbf{PhSV_2}$ is still in $\mathbf{PhSV_1}$. However, the delay effect takes place as well, when the center of mass of particle moving to the neighboring volume (say, $\mathbf{PhSV_2}$) is already located in $\mathbf{PhSV_2}$ but a part of the particle still belongs to $\mathbf{PhSV_1}$. This entire complex of effects defines non-local effects in space and time.

The physically infinitely small volume (**PhSV**) is an *open* thermodynamic system *for any division of macroscopic system by a set of PhSVs.* However, the BE (Boltzmann 1872, 1912)

$$Df/Dt = J^B, \tag{1.1}$$

where $J^B$ is the Boltzmann collision integral and $D/Dt$ is a substantive derivative, fully ignores non-local effects and contains only the local collision integral $J^B$. The foregoing nonlocal effects are insignificant only in equilibrium systems, where the kinetic approach changes to methods of statistical mechanics.

This is what the difficulties of classical Boltzmann physical kinetics arise from. Also a weak point of the classical Boltzmann kinetic theory is the treatment of the dynamic properties of interacting particles. On the one hand, as follows from the so-called "physical" derivation of the BE, Boltzmann particles are regarded as material points; on the other hand, the collision integral in the BE leads to the emergence of collision cross sections.

The rigorous approach to derivation of kinetic equation relative to one-particle DF $f$ ($KE_f$) is based on employing the hierarchy of Bogolyubov equations. Generally speaking, the structure of $KE_f$ is as follows:

$$\frac{Df}{Dt} = J^B + J^{nl}, \tag{1.2}$$

where $J^{nl}$ is the non-local integral term. An approximation for the second collision integral is suggested by me in *generalized* Boltzmann physical kinetics,

$$J^{nl} = \frac{D}{Dt}\left(\tau \frac{Df}{Dt}\right). \tag{1.3}$$

Here, $\tau$ is the mean time *between* collisions of particles, which is related in a hydrodynamic approximation with dynamical viscosity $\mu$ and pressure $p$,

$$\tau\ p = \Pi \mu, \tag{1.4}$$

where the factor $\Pi$ is defined by the model of collision of particles: for neutral hard-sphere gas, $\Pi$=0.8 (Chapman & Cowling, 1952; Hirshfelder *et al.* 1954). All of the known methods of deriving kinetic equation relative to one-particle DF lead to approximation (1.3), including the method of many scales, the method of correlation functions, and the iteration method.

Fluctuation effects occur in any open thermodynamic system bounded by a control surface transparent to particles. Generalized BE (1.2) leads to generalized hydrodynamic equations (GHE) (Alexeev 1994, 2004) as the local approximation of non local effects, for example, to the continuity equation

$$\frac{\partial \rho^a}{\partial t} + \frac{\partial}{\partial \mathbf{r}} \cdot (\rho \mathbf{v}_0)^a = 0, \tag{1.5}$$

where $\rho^a$, ${\mathbf{v}_0}^a$, $(\rho \mathbf{v}_0)^a$ are calculated in view of non-locality effect in terms of gas density $\rho$, hydrodynamic velocity of flow $\mathbf{v}_0$, and density of momentum flux $\rho \mathbf{v}_0$; for locally Maxwellian distribution, $\rho^a$, $(\rho \mathbf{v}_0)^a$ are defined by the relations

$$\left(\rho - \rho^a\right)/\tau = \frac{\partial \rho}{\partial t} + \frac{\partial}{\partial \mathbf{r}} \cdot (\rho \mathbf{v}_0), \quad \left(\rho \mathbf{v}_0 - (\rho \mathbf{v}_0)^a\right)/\tau = \frac{\partial}{\partial t}(\rho \mathbf{v}_0) + \frac{\partial}{\partial \mathbf{r}} \cdot \rho \mathbf{v}_0 \mathbf{v}_0 + \overleftrightarrow{I} \cdot \frac{\partial p}{\partial \mathbf{r}} - \rho \mathbf{a}, \tag{1.6}$$

where $\overleftrightarrow{I}$ is a unit tensor, and $\mathbf{a}$ is the acceleration due to the effect of mass forces.

In the general case, the parameter $\tau$ is the non-locality parameter; in quantum hydrodynamics, its magnitude is defined by the "time-energy" uncertainty relation. The violation of Bell's inequalities (Bell 1964) is found for local statistical theories, and the transition to non-local description is inevitable.

The following conclusion of principal significance can be done from the previous consideration (Alexeev 2008ab):

1. Madelung's quantum hydrodynamics is equivalent to the Schrödinger equation (SE) and furnishes the description of the quantum particle evolution in the form of Euler equation and continuity equation. SE is consequence of the Liouville equation as result of the local approximation of non-local equations. Madelung's interpretation of SE (connected with wave function $\psi = \alpha \exp(i\beta)$ leads the probability density $\rho = \alpha^2$ and velocity $\mathbf{v} = \frac{\partial}{\partial \mathbf{r}}(\beta\hbar / m)$. Madelung quantum hydrodynamics does not lead to the energy equation on principal, the corresponding dependent variable $p$ in the energy equation of the generalized quantum hydrodynamics developed by me can be titled as the rest quantum pressure or simply quantum pressure.

2. Generalized Boltzmann physical kinetics brings the strict approximation of non-local effects in space and time and after transmission to the local approximation leads to parameter $\tau$, which on the quantum level corresponds to the uncertainty principle "time-energy".
3. GHE produce SE as a deep particular case of the generalized Boltzmann physical kinetics and therefore of non-local hydrodynamics.

On principal GHE (and therefore GQH) needn't in using of the "time-energy" uncertainty relation for estimation of the value of the non-locality parameter $\tau$. Moreover the "time-energy" uncertainty relation does not produce the exact relations and from position of non-local physics is only the simplest estimation of the non-local effects. Really, let us consider two neighboring physically infinitely small volumes $\mathbf{PhSV_1}$ and $\mathbf{PhSV_2}$ in a nonequilibrium system. Obviously the time $\tau$ should tends to diminish with increasing of the velocities $u$ of particles invading in the nearest neighboring physically infinitely small volume ($\mathbf{PhSV_1}$ or $\mathbf{PhSV_2}$):

$$\tau = H/u^n \, . \tag{1.7}$$

But the value $\tau$ cannot depend on the velocity direction and naturally to tie $\tau$ with the particle kinetic energy, then

$$\tau = H/mu^2 \, , \tag{1.8}$$

where $H$ is a coefficient of proportionality, which reflects the state of physical system. In the simplest case $H$ is equal to Plank constant $\hbar$ and relation (1.8) became compatible with the Heisenberg relation.

Strict consideration leads to the following system of the generalized hydrodynamic equations (GHE) (Alexeev 1994, 2004) written in the generalized Euler form:

Continuity equation for species $\alpha$:

$$\frac{\partial}{\partial t}\left\{\rho_\alpha - \tau_\alpha^{(0)}\left[\frac{\partial \rho_\alpha}{\partial t} + \frac{\partial}{\partial \mathbf{r}}\cdot(\rho_\alpha \mathbf{v}_0)\right]\right\} + \frac{\partial}{\partial \mathbf{r}}\cdot\left\{\rho_\alpha \mathbf{v}_0 - \tau_\alpha^{(0)}\left[\frac{\partial}{\partial t}(\rho_\alpha \mathbf{v}_0) + \right.\right.$$
$$\left.\left. + \frac{\partial}{\partial \mathbf{r}}\cdot(\rho_\alpha \mathbf{v}_0 \mathbf{v}_0) + \vec{I}\cdot\frac{\partial p_\alpha}{\partial \mathbf{r}} - \rho_\alpha \mathbf{F}_\alpha^{(1)} - \frac{q_\alpha}{m_\alpha}\rho_\alpha \mathbf{v}_0 \times \mathbf{B}\right]\right\} = R_\alpha \, , \tag{1.9}$$

Continuity equation for mixture:

$$\frac{\partial}{\partial t}\left\{\rho - \sum_\alpha \tau_\alpha^{(0)}\left[\frac{\partial \rho_\alpha}{\partial t} + \frac{\partial}{\partial \mathbf{r}}\cdot(\rho_\alpha \mathbf{v}_0)\right]\right\} +$$
$$+ \frac{\partial}{\partial \mathbf{r}}\cdot\left\{\rho \mathbf{v}_0 - \sum_\alpha \tau_\alpha^{(0)}\left[\frac{\partial}{\partial t}(\rho_\alpha \mathbf{v}_0) + \frac{\partial}{\partial \mathbf{r}}\cdot(\rho_\alpha \mathbf{v}_0 \mathbf{v}_0) + \vec{I}\cdot\frac{\partial p_\alpha}{\partial \mathbf{r}} - \right.\right. \tag{1.10}$$
$$\left.\left. - \rho_\alpha \mathbf{F}_\alpha^{(1)} - \frac{q_\alpha}{m_\alpha}\rho_\alpha \mathbf{v}_0 \times \mathbf{B}\right]\right\} = 0,$$

Momentum equation

$$\frac{\partial}{\partial t}\left\{\rho_\alpha \mathbf{v}_0 - \tau_\alpha^{(0)}\left[\frac{\partial}{\partial t}(\rho_\alpha \mathbf{v}_0) + \frac{\partial}{\partial \mathbf{r}}\cdot \rho_\alpha \mathbf{v}_0\mathbf{v}_0 + \frac{\partial p_\alpha}{\partial \mathbf{r}} - \rho_\alpha \mathbf{F}_\alpha^{(1)} - \right.\right.$$

$$\left.\left. - \left(\frac{q_\alpha}{m_\alpha}\right)\rho_\alpha \mathbf{v}_0 \times \mathbf{B}\right]\right\} - \mathbf{F}_\alpha^{(1)}\left[\rho_\alpha - \tau_\alpha^{(0)}\left(\frac{\partial \rho_\alpha}{\partial t} + \frac{\partial}{\partial \mathbf{r}}\cdot(\rho_\alpha \mathbf{v}_0)\right)\right] -$$

$$- \frac{q_\alpha}{m_\alpha}\left\{\rho_\alpha \mathbf{v}_0 - \tau_\alpha^{(0)}\left[\frac{\partial}{\partial t}(\rho_\alpha \mathbf{v}_0) + \frac{\partial}{\partial \mathbf{r}}\cdot \rho_\alpha \mathbf{v}_0\mathbf{v}_0 + \frac{\partial p_\alpha}{\partial \mathbf{r}} - \rho_\alpha \mathbf{F}_\alpha^{(1)} - \right.\right.$$

$$\left.\left. - \frac{q_\alpha}{m_\alpha}\rho_\alpha \mathbf{v}_0 \times \mathbf{B}\right]\right\} \times \mathbf{B} + \frac{\partial}{\partial \mathbf{r}}\cdot\left\{\rho_\alpha \mathbf{v}_0\mathbf{v}_0 + p_\alpha \overset{\leftrightarrow}{I} - \tau_\alpha^{(0)}\left[\frac{\partial}{\partial t}(\rho_\alpha \mathbf{v}_0\mathbf{v}_0 + \right.\right.$$

$$+ p_\alpha \overset{\leftrightarrow}{I}) + \frac{\partial}{\partial \mathbf{r}}\cdot\left(\rho_\alpha(\mathbf{v}_0\mathbf{v}_0)\mathbf{v}_0 + \rho_\alpha(\mathbf{v}_0\overline{\mathbf{V}_\alpha)\mathbf{V}_\alpha} + \rho_\alpha\overline{(\mathbf{V}_\alpha\mathbf{v}_0)\mathbf{V}_\alpha} + \right.$$

$$\left. + \rho_\alpha\overline{(\mathbf{V}_\alpha\mathbf{V}_\alpha)}\mathbf{v}_0\right) - \mathbf{F}_\alpha^{(1)}\rho_\alpha \mathbf{v}_0 - \rho_\alpha \mathbf{v}_0\mathbf{F}_\alpha^{(1)} -$$

$$- \frac{q_\alpha}{m_\alpha}\rho_\alpha[\mathbf{v}_0 \times \mathbf{B}]\mathbf{v}_0 - \frac{q_\alpha}{m_\alpha}\rho_\alpha\overline{[\mathbf{V}_\alpha \times \mathbf{B}]\mathbf{V}_\alpha} -$$

$$\left.\left. - \frac{q_\alpha}{m_\alpha}\rho_\alpha \mathbf{v}_0[\mathbf{v}_0 \times \mathbf{B}] - \frac{q_\alpha}{m_\alpha}\rho_\alpha\overline{\mathbf{V}_\alpha[\mathbf{V}_\alpha \times \mathbf{B}]}\right]\right\} =$$

$$= \int m_\alpha \mathbf{v}_\alpha J_\alpha^{st,el} d\mathbf{v}_\alpha + \int m_\alpha \mathbf{v}_\alpha J_\alpha^{st,inel} d\mathbf{v}_\alpha. \qquad (1.11)$$

Energy equation for mixture:

$$\frac{\partial}{\partial t}\left\{\frac{\rho v_0^2}{2} + \frac{3}{2}p + \sum_\alpha \varepsilon_\alpha n_\alpha - \sum_\alpha \tau_\alpha^{(0)}\left[\frac{\partial}{\partial t}\left(\frac{\rho_\alpha v_0^2}{2} + \frac{3}{2}p_\alpha + \varepsilon_\alpha n_\alpha\right) + \right.\right.$$

$$\left.\left. + \frac{\partial}{\partial \mathbf{r}}\cdot\left(\frac{1}{2}\rho_\alpha v_0^2 \mathbf{v}_0 + \frac{5}{2}p_\alpha \mathbf{v}_0 + \varepsilon_\alpha n_\alpha \mathbf{v}_0\right) - \mathbf{F}_\alpha^{(1)}\cdot \rho_\alpha \mathbf{v}_0\right]\right\} +$$

$$+ \frac{\partial}{\partial \mathbf{r}}\cdot\left\{\frac{1}{2}\rho v_0^2 \mathbf{v}_0 + \frac{5}{2}p \mathbf{v}_0 + \mathbf{v}_0\sum_\alpha \varepsilon_\alpha n_\alpha - \sum_\alpha \tau_\alpha^{(0)}\left[\frac{\partial}{\partial t}\left(\frac{1}{2}\rho_\alpha v_0^2 \mathbf{v}_0 + \right.\right.\right.$$

$$\left. + \frac{5}{2}p_\alpha \mathbf{v}_0 + \varepsilon_\alpha n_\alpha \mathbf{v}_0\right) + \frac{\partial}{\partial \mathbf{r}}\cdot\left(\frac{1}{2}\rho_\alpha v_0^2 \mathbf{v}_0\mathbf{v}_0 + \frac{7}{2}p_\alpha \mathbf{v}_0\mathbf{v}_0 + \frac{1}{2}p_\alpha v_0^2 \overset{\leftrightarrow}{I} + \right.$$

$$\left. + \frac{5}{2}\frac{p_\alpha^2}{\rho_\alpha}\overset{\leftrightarrow}{I} + \varepsilon_\alpha n_\alpha \mathbf{v}_0\mathbf{v}_0 + \varepsilon_\alpha \frac{p_\alpha}{m_\alpha}\overset{\leftrightarrow}{I}\right) - \rho_\alpha \mathbf{F}_\alpha^{(1)}\cdot \mathbf{v}_0\mathbf{v}_0 - p_\alpha \mathbf{F}_\alpha^{(1)}\cdot \overset{\leftrightarrow}{I} -$$

$$- \frac{1}{2}\rho_\alpha v_0^2 \mathbf{F}_\alpha^{(1)} - \frac{3}{2}\mathbf{F}_\alpha^{(1)}p_\alpha - \frac{\rho_\alpha v_0^2}{2}\frac{q_\alpha}{m_\alpha}[\mathbf{v}_0 \times \mathbf{B}] - \frac{5}{2}p_\alpha \frac{q_\alpha}{m_\alpha}[\mathbf{v}_0 \times \mathbf{B}] -$$

$$\left.\left. - \varepsilon_\alpha n_\alpha \frac{q_\alpha}{m_\alpha}[\mathbf{v}_0 \times \mathbf{B}] - \varepsilon_\alpha n_\alpha \mathbf{F}_\alpha^{(1)}\right]\right\} - \left\{\mathbf{v}_0\cdot\sum_\alpha \rho_\alpha \mathbf{F}_\alpha^{(1)} - \sum_\alpha \tau_\alpha^{(0)}\left[\mathbf{F}_\alpha^{(1)}\cdot\right.\right.$$

$$\left.\left.\cdot\left(\frac{\partial}{\partial t}(\rho_\alpha \mathbf{v}_0) + \frac{\partial}{\partial \mathbf{r}}\cdot \rho_\alpha \mathbf{v}_0\mathbf{v}_0 + \frac{\partial}{\partial \mathbf{r}}\cdot p_\alpha \overset{\leftrightarrow}{I} - \rho_\alpha \mathbf{F}_\alpha^{(1)} - q_\alpha n_\alpha[\mathbf{v}_0 \times \mathbf{B}]\right)\right]\right\} = 0. \qquad (1.12)$$

Here $\mathbf{F}_\alpha^{(1)}$ are the forces of the non-magnetic origin, $\mathbf{B}$ - magnetic induction, $\overset{\leftrightarrow}{I}$ - unit tensor, $q_\alpha$ - charge of the $\alpha$-component particle, $p_\alpha$ - static pressure for $\alpha$-component, $\mathbf{V}_\alpha$ - thermal velocity, $\varepsilon_\alpha$ - internal energy for the particles of $\alpha$-component, $\mathbf{v}_0$ - hydrodynamic velocity for mixture.

## 2. Nonstationary 1D generalized hydrodynamic equations in the self consistent electrical field. Quantization in the generalized quantum hydrodynamics.

In the following we intend to obtain the soliton's type of solution of the generalized hydrodynamic equations for plasma in the self-consistent electrical field. All elements of possible formation like quantum soliton should move with the same translational velocity. Then the system of GHE consist from the generalized Poisson equation reflecting the effects of the charge and the charge flux perturbations, two continuity equations for positive and negative species (in particular, for ion and electron components), one motion equation and two energy equations for ion and electron components. This system of six equations for non-stationary 1D case can be written in the form (Alexeev 2008ab, 2009ab):

(Poisson equation)

$$\frac{\partial^2 \varphi}{\partial x^2} = -4\pi e\left\{\left[n_i - \tau_i\left(\frac{\partial n_i}{\partial t} + \frac{\partial}{\partial x}(n_i u)\right)\right] - \left[n_e - \tau_e\left(\frac{\partial n_e}{\partial t} + \frac{\partial}{\partial x}(n_e u)\right)\right]\right\}, \tag{2.1}$$

(continuity equation for positive ion component)

$$\frac{\partial}{\partial t}\left\{\rho_i - \tau_i\left[\frac{\partial \rho_i}{\partial t} + \frac{\partial}{\partial x}(\rho_i u)\right]\right\} + \frac{\partial}{\partial x}\left\{\rho_i u - \tau_i\left[\frac{\partial}{\partial t}(\rho_i u) + \frac{\partial}{\partial x}(\rho_i u^2) + \frac{\partial p_i}{\partial x} - \rho_i F_i\right]\right\}, \tag{2.2}$$

(continuity equation for electron component)

$$\frac{\partial}{\partial t}\left\{\rho_e - \tau_e\left[\frac{\partial \rho_e}{\partial t} + \frac{\partial}{\partial x}(\rho_e u)\right]\right\} + \frac{\partial}{\partial x}\left\{\rho_e u - \tau_e\left[\frac{\partial}{\partial t}(\rho_e u) + \frac{\partial}{\partial x}(\rho_e u^2) + \frac{\partial p_e}{\partial x} - \rho_e F_e\right]\right\}, \tag{2.3}$$

(momentum equation)

$$\begin{aligned}
&\frac{\partial}{\partial t}\left\{\rho u - \tau_i\left[\frac{\partial}{\partial t}(\rho_i u) + \frac{\partial}{\partial x}(p_i + \rho_i u^2) - \rho_i F_i\right] - \tau_e\left[\frac{\partial}{\partial t}(\rho_e u) + \frac{\partial}{\partial x}(p_e + \rho_e u^2) - \rho_e F_e\right]\right\} - \\
&- \rho_i F_i - \rho_e F_e + F_i \tau_i\left(\frac{\partial \rho_i}{\partial t} + \frac{\partial}{\partial x}(\rho_i u)\right) + F_e \tau_e\left(\frac{\partial \rho_e}{\partial t} + \frac{\partial}{\partial x}(\rho_e u)\right) + \\
&+ \frac{\partial}{\partial x}\left\{\begin{matrix}\rho u^2 + p - \tau_i\left[\frac{\partial}{\partial t}(\rho_i u^2 + p_i) + \frac{\partial}{\partial x}(\rho_i u^3 + 3p_i u) - 2\rho_i u F_i\right] - \\ - \tau_e\left[\frac{\partial}{\partial t}(\rho_e u^2 + p_e) + \frac{\partial}{\partial x}(\rho_e u^3 + 3p_e u)\right] - 2\rho_e u F_e\end{matrix}\right\} = 0
\end{aligned} \tag{2.4}$$

(energy equation for positive ion component)

$$\begin{aligned}
&\frac{\partial}{\partial t}\left\{\rho_i u^2 + 3p_i - \tau_i\left[\frac{\partial}{\partial t}(\rho_i u^2 + 3p_i) + \frac{\partial}{\partial x}(\rho_i u^3 + 5p_i u) - 2\rho_i F_i u\right]\right\} + \\
&+ \frac{\partial}{\partial x}\left\{\rho_i u^3 + 5p_i u - \tau_i\left[\frac{\partial}{\partial t}(\rho_i u^3 + 5p_i u) + \frac{\partial}{\partial x}\left(\rho_i u^4 + 8p_i u^2 + 5\frac{p_i^2}{\rho_i}\right) - F_i(3\rho_i u^2 + 5p_i)\right]\right\}, \\
&- 2u\rho_i F_i + 2\tau_i F_i\left[\frac{\partial}{\partial t}(\rho_i u) + \frac{\partial}{\partial x}(\rho_i u^2 + p_i) - \rho_i F_i\right] = -\frac{p_i - p_e}{\tau_{ei}}
\end{aligned} \tag{2.5}$$

(energy equation for electron component)

$$\frac{\partial}{\partial t}\left\{\rho_e u^2 + 3p_e - \tau_e\left[\frac{\partial}{\partial t}\left(\rho_e u^2 + 3p_e\right) + \frac{\partial}{\partial x}\left(\rho_e u^3 + 5p_e u\right) - 2\rho_e F_e u\right]\right\} +$$

$$+\frac{\partial}{\partial x}\left\{\rho_e u^3 + 5p_e u - \tau_e\left[\frac{\partial}{\partial t}\left(\rho_e u^3 + 5p_e u\right) + \frac{\partial}{\partial x}\left(\rho_e u^4 + 8p_e u^2 + 5\frac{p_e^2}{\rho_e}\right) - F_e\left(3\rho_e u^2 + 5p_e\right)\right]\right\}, \quad (2.6)$$

$$-2u\rho_e F_e + 2\tau_e F_e\left[\frac{\partial}{\partial t}\left(\rho_e u\right) + \frac{\partial}{\partial x}\left(\rho_e u^2 + p_e\right) - \rho_e F_e\right] = -\frac{p_e - p_i}{\tau_{ei}}$$

where $u$ is translational velocity of the quantum object, $\varphi$ - scalar potential, $n_i$ and $n_e$ are the number density of the charged species, $F_i$ and $F_e$ are the forces acting on the unit mass of positive and negative particles like ions and electrons.

Approximations for non-local parameters $\tau_i$, $\tau_e$ and $\tau_{ei}$ need the special consideration. In the following for the $\tau_i$ and $\tau_i$ approximation the relation (2.6) is used in the forms

$$\tau_i = H / m_i u^2, \quad \tau_e = H / m_e u^2. \qquad (2.7)$$

For non-local parameter of electron-ion interaction $\tau_{ei}$ is applicable the relation

$$\frac{1}{\tau_{ei}} = \frac{1}{\tau_e} + \frac{1}{\tau_i}. \qquad (2.8)$$

In this case parameter $\tau_{ei}$ serves as relaxation time in the process of the particle interaction of different kinds. Transformation (2.8) for the case $H = \hbar$ leads to the obvious compatibility with the Heisenberg principle

$$\frac{1}{\tau_{ei}} = \frac{\tau_e + \tau_i}{\tau_e \tau_i} = \frac{\dfrac{\hbar}{m_e u^2} + \dfrac{\hbar}{m_i u^2}}{\dfrac{\hbar^2}{u^4}\dfrac{1}{m_e m_i}} = \frac{u^2}{\hbar}\left(m_e + m_i\right). \qquad (2.9)$$

Then

$$u^2\left(m_e + m_i\right)\tau_{ei} = \hbar. \qquad (2.10)$$

Equality (2.10) is consequence of “time-energy” uncertainty relation for combined particle with mass $m_i + m_e$.

In principal the time values $\tau_i$ and $\tau_e$ should be considered as sums of mean times between collisions ($\tau_i^{tr}, \tau_e^{tr}$) and discussed above non-local quantum values ($\tau_i^{qu}$, $\tau_e^{qu}$), namely for example

$$\tau_i = \tau_i^{tr} + \tau_i^{qu}. \qquad (2.11)$$

For molecular hydrogen in standard conditions mean time between collisions is equal to $6.6 \cdot 10^{-11}$ s. For quantum objects moving with velocities typical for plasmoids or lightning balls $\tau^{qu}$ is much more than $\tau^{tr}$ and the usual static pressure $p$ transforms in the pressure which can be named as the rest non-local pressure. In the definite sense this kind of pressure can be considered as analogue of the Bose condensate pressure.

The following formulae are valid for acting forces

$$F_i = -\frac{e}{m_i}\frac{\partial \varphi}{\partial x},\ F_e = -\frac{e}{m_e}\frac{\partial \varphi}{\partial x}. \tag{2.12}$$

Let us consider now the introduction of quantization in quantum hydrodynamics. With this aim write down the expression for the total energy $E$ of a particle moving along the positive direction of $x$-axis with velocity $u$ in the attractive field of Coulomb forces

$$E = \frac{mu^2}{2} - \frac{Ze^2}{x}, \tag{2.13}$$

where $Z$ is the charge number and $x$ is the distance from the center of forces. If this movement obeys to the condition of non-locality $mu^2 = H/\tau$ and $x = u\tau$, then

$$E = \frac{H^2}{2mx^2} - \frac{Ze^2}{x}. \tag{2.14}$$

Minimal total energy corresponds to the condition $\left(\frac{\partial E}{\partial x}\right)_{x=x_B} = 0$ and

$$\frac{H^2}{mx_B{}^3} = \frac{Ze^2}{x_B{}^2}. \tag{2.15}$$

From (2.15) follows

$$x_B = \frac{H^2}{Zme^2}. \tag{2.16}$$

and from (2.14), (2.16)

$$E = -\frac{Z^2me^4}{2H^2}. \tag{2.17}$$

For atom with single electron in the shell moving on the Bohr's orbit of radius $r_B$, Eq. (2.15) with taking into account the relation

$$\frac{H^2}{mr_B{}^3} = \frac{mu^2}{r_B} \tag{2.18}$$

leads to equality of Coulomb and inertial forces for the orbit electron

$$\frac{m_eu^2}{r_B} = \frac{Ze^2}{r_B{}^2} \tag{2.19}$$

and to the same expression for energy

$$E = -\frac{Z^2m_ee^4}{2H^2}. \tag{2.20}$$

The comparison of Eq. (2.20) with the Balmer's relation leads to condition

$$H = n\hbar \tag{2.21}$$

with integer $n = 1,2,...$ known as principal quantum number and to well known relation

$$E = -\frac{Z^2m_ee^4}{2\hbar^2}\frac{1}{n^2}. \tag{2.22}$$

Eqs. (2.16), (2.18) lead also to the character velocity for this obviously model problem

$$u = \frac{Ze^2}{H} \tag{2.23}$$

with the velocity $2.187\cdot 10^{8}$ cm/s.

As we see the mentioned simple considerations allow on principal to introduce quantization in the quantum hydrodynamics without direct application of Schrödinger equation. Important to notice that conditions of quantization are not the intrinsic feature of Schrödinger equation, for example the appearance of quantization in Schrödinger's theory is connected with the truncation of infinite series and transformation in polynomials with the finite quantity of terms.

### 3. Quantum solitons in self consistent electric field.

Let us introduce the coordinate system moving along the positive direction of $x$ - axis in ID space with velocity $C = u_0$ equal to phase velocity of considering quantum object

$$\xi = x - Ct. \tag{3.1}$$

Taking into account the De Broglie relation we should wait that the group velocity $u_g$ is equal $2u_0$. In moving coordinate system all dependent hydrodynamic values are function of $(\xi, t)$.We investigate the possibility of the quantum object formation of the soliton type. For this solution there is no explicit dependence on time for coordinate system moving with the phase velocity $u_0$. Write down the system of equations (2.1) - (2.6) for the two component mixture of charged particles without taking into account the component's internal energy in the dimensionless form, where dimensionless symbols are marked by tildes. We begin with introduction the scales for velocity

$$[u] = u_0 \tag{3.2}$$

and for coordinate $x$

$$\frac{\hbar}{m_e u_0} = x_0. \tag{3.3}$$

Generalized Poisson equation (2.1)

$$\frac{\partial^2 \varphi}{\partial x^2} = -4\pi e\left\{\left[n_i - \frac{\hbar}{m_i u^2} u_0\left(-\frac{\partial n_i}{\partial x} + \frac{\partial}{\partial x}(n_i \tilde{u})\right)\right] - \left[n_e - \frac{\hbar}{m_e u^2} u_0\left(-\frac{\partial n_e}{\partial x} + \frac{\partial}{\partial x}(n_e \tilde{u})\right)\right]\right\} \tag{3.4}$$

now is written as

$$\frac{\partial^2 \tilde{\varphi}}{\partial \tilde{\xi}^2} = -\left\{\frac{m_e}{m_i}\left[\tilde{\rho}_i - \frac{1}{\tilde{u}^2}\frac{m_e}{m_i}\left(-\frac{\partial \tilde{\rho}_i}{\partial \tilde{\xi}} + \frac{\partial}{\partial \tilde{\xi}}(\tilde{\rho}_i \tilde{u})\right)\right] - \left[\tilde{\rho}_e - \frac{1}{\tilde{u}^2}\left(-\frac{\partial \tilde{\rho}_e}{\partial \tilde{\xi}} + \frac{\partial}{\partial \tilde{\xi}}(\tilde{\rho}_e \tilde{u})\right)\right]\right\}, \tag{3.5}$$

if the potential scale $\varphi_0$ and the density scale $\rho_0$ are chosen as

$$\varphi_0 = \frac{m_e}{e} u_0^2, \tag{3.6}$$

$$\rho_0 = \frac{m_e^4}{4\pi\hbar^2 e^2} u_0^4. \tag{3.7}$$

Scaled forces will be described by ($e$ - absolute electron charge) relations

$$\rho_i F_i = -\frac{u_0^2}{x_0}\rho_0 \frac{m_e}{m_i}\frac{\partial \tilde{\varphi}}{\partial \tilde{\xi}}\tilde{\rho}_i, \tag{3.8}$$

$$\rho_e F_e = \frac{u_0^2}{x_0}\rho_0 \frac{\partial \tilde{\varphi}}{\partial \tilde{\xi}}\tilde{\rho}_e. \tag{3.9}$$

Analogical transformations should be applied to the other equations of the system (2.1) - (2.6). We have the following system of six ordinary non-linear equations:

$$\frac{\partial^2\tilde{\varphi}}{\partial\tilde{\xi}^2}=-\left\{\frac{m_e}{m_i}\left[\tilde{\rho}_i-\frac{1}{\tilde{u}^2}\frac{m_e}{m_i}\left(-\frac{\partial\tilde{\rho}_i}{\partial\tilde{\xi}}+\frac{\partial}{\partial\tilde{\xi}}\left(\tilde{\rho}_i\tilde{u}\right)\right)\right]-\left[\tilde{\rho}_e-\frac{1}{\tilde{u}^2}\left(-\frac{\partial\tilde{\rho}_e}{\partial\tilde{\xi}}+\frac{\partial}{\partial\tilde{\xi}}\left(\tilde{\rho}_e\tilde{u}\right)\right)\right]\right\}, \tag{3.10}$$

$$\frac{\partial\tilde{\rho}_i}{\partial\tilde{\xi}}-\frac{\partial\tilde{\rho}_i\tilde{u}}{\partial\tilde{\xi}}+\frac{m_e}{m_i}\frac{\partial}{\partial\tilde{\xi}}\left\{\frac{1}{\tilde{u}^2}\left[\frac{\partial}{\partial\tilde{\xi}}\left(\tilde{p}_i+\tilde{\rho}_i+\tilde{\rho}_i\tilde{u}^2-2\tilde{\rho}_i\tilde{u}_i\right)+\frac{m_e}{m_i}\tilde{\rho}_i\frac{\partial\tilde{\varphi}}{\partial\tilde{\xi}}\right]\right\}=0, \tag{3.11}$$

$$\frac{\partial\tilde{\rho}_e}{\partial\tilde{\xi}}-\frac{\partial\tilde{\rho}_e\tilde{u}}{\partial\tilde{\xi}}+\frac{\partial}{\partial\tilde{\xi}}\left\{\frac{1}{\tilde{u}^2}\left[\frac{\partial}{\partial\tilde{\xi}}\left(\tilde{p}_e+\tilde{\rho}_e+\tilde{\rho}_e\tilde{u}^2-2\tilde{\rho}_e\tilde{u}_e\right)-\tilde{\rho}_e\frac{\partial\tilde{\varphi}}{\partial\tilde{\xi}}\right]\right\}=0, \tag{3.12}$$

$$\begin{aligned}
&\frac{\partial}{\partial\tilde{\xi}}\left\{\left(\tilde{\rho}_i+\tilde{\rho}_e\right)\tilde{u}^2+\left(\tilde{p}_i+\tilde{p}_e\right)-\left(\tilde{\rho}_i+\tilde{\rho}_e\right)\tilde{u}\right\}+\\
&+\frac{\partial}{\partial\tilde{\xi}}\left\{\begin{aligned}&\frac{1}{\tilde{u}^2}\frac{m_e}{m_i}\left[\frac{\partial}{\partial\tilde{\xi}}\left(2\tilde{p}_i+2\tilde{\rho}_i\tilde{u}^2-\tilde{\rho}_i\tilde{u}-\tilde{\rho}_i\tilde{u}^3-3\tilde{p}_i\tilde{u}\right)+\tilde{\rho}_i\frac{m_e}{m_i}\frac{\partial\tilde{\varphi}}{\partial\tilde{\xi}}\right]+\\&+\frac{1}{\tilde{u}^2}\left[\frac{\partial}{\partial\tilde{\xi}}\left(2\tilde{p}_e+2\tilde{\rho}_e\tilde{u}^2-\tilde{\rho}_e\tilde{u}-\tilde{\rho}_e\tilde{u}^3-3\tilde{p}_e\tilde{u}\right)-\tilde{\rho}_e\frac{\partial\tilde{\varphi}}{\partial\tilde{\xi}}\right]\end{aligned}\right\}+\\
&+\tilde{\rho}_i\frac{m_e}{m_i}\frac{\partial\tilde{\varphi}}{\partial\tilde{\xi}}-\tilde{\rho}_e\frac{\partial\tilde{\varphi}}{\partial\tilde{\xi}}-\frac{\partial\tilde{\varphi}}{\partial\tilde{\xi}}\frac{1}{\tilde{u}^2}\left(\frac{m_e}{m_i}\right)^2\left(-\frac{\partial\tilde{\rho}_i}{\partial\tilde{\xi}}+\frac{\partial}{\partial\tilde{\xi}}\left(\tilde{\rho}_i\tilde{u}\right)\right)+\\
&+\frac{\partial\tilde{\varphi}}{\partial\tilde{\xi}}\frac{1}{\tilde{u}^2}\left(-\frac{\partial\tilde{\rho}_e}{\partial\tilde{\xi}}+\frac{\partial}{\partial\tilde{\xi}}\left(\tilde{\rho}_e\tilde{u}\right)\right)-2\frac{\partial}{\partial\tilde{\xi}}\left\{\frac{1}{\tilde{u}}\frac{\partial\tilde{\varphi}}{\partial\tilde{\xi}}\left[\left(\frac{m_e}{m_i}\right)^2\tilde{\rho}_i-\tilde{\rho}_e\right]\right\}=0,
\end{aligned} \tag{3.13}$$

$$\begin{aligned}
&\frac{\partial}{\partial\tilde{\xi}}\left\{\tilde{\rho}_i\tilde{u}^3+5\tilde{p}_i\tilde{u}-\tilde{\rho}_i\tilde{u}^2-3\tilde{p}_i\right\}+\\
&+\frac{\partial}{\partial\tilde{\xi}}\left\{\frac{1}{\tilde{u}^2}\frac{m_e}{m_i}\left[\begin{aligned}&\frac{\partial}{\partial\tilde{\xi}}\left(2\tilde{\rho}_i\tilde{u}^3+10\tilde{p}_i\tilde{u}-\tilde{\rho}_i\tilde{u}^4-8\tilde{p}_i\tilde{u}^2-5\frac{\tilde{p}_i^2}{\tilde{\rho}_i}-\tilde{\rho}_i\tilde{u}^2-3\tilde{p}_i\right)+\\&+\frac{m_e}{m_i}\frac{\partial\tilde{\varphi}}{\partial\tilde{\xi}}\left(2\tilde{\rho}_i\tilde{u}-3\tilde{\rho}_i\tilde{u}^2-5\tilde{p}_i\right)\end{aligned}\right]\right\}\\
&+2\frac{m_e}{m_i}\tilde{\rho}_i\tilde{u}\frac{\partial\tilde{\varphi}}{\partial\tilde{\xi}}-\\
&-2\frac{\partial\tilde{\varphi}}{\partial\tilde{\xi}}\frac{1}{\tilde{u}^2}\left(\frac{m_e}{m_i}\right)^2\left[\frac{\partial}{\partial\tilde{\xi}}\left(\tilde{\rho}_i\tilde{u}^2+\tilde{p}_i-\tilde{\rho}_i\tilde{u}\right)+\tilde{\rho}_i\frac{m_e}{m_i}\frac{\partial\tilde{\varphi}}{\partial\tilde{\xi}}\right]=-\left(\tilde{p}_i-\tilde{p}_e\right)\tilde{u}^2\left(1+\frac{m_i}{m_e}\right).
\end{aligned} \tag{3.14}$$

$$\frac{\partial}{\partial \tilde{\xi}}\left\{\tilde{\rho}_e\tilde{u}^3 + 5\tilde{p}_e\tilde{u} - \tilde{\rho}_e\tilde{u}^2 - 3\tilde{p}_e\right\} +$$

$$+\frac{\partial}{\partial \tilde{\xi}}\left\{\frac{1}{\tilde{u}^2}\left[\begin{array}{l}\frac{\partial}{\partial \tilde{\xi}}\left(2\tilde{\rho}_e\tilde{u}^3 + 10\tilde{p}_e\tilde{u} - \tilde{\rho}_e\tilde{u}^4 - 8\tilde{p}_e\tilde{u}^2 - 5\frac{\tilde{p}_e^2}{\tilde{\rho}_e} - \tilde{\rho}_e\tilde{u}^2 - 3\tilde{p}_e\right) + \\ +\frac{\partial \tilde{\varphi}}{\partial \tilde{\xi}}\left(3\tilde{\rho}_e\tilde{u}^2 + 5\tilde{p}_e - 2\tilde{\rho}_e\tilde{u}\right)\end{array}\right]\right\} -$$

$$-2\tilde{\rho}_e\tilde{u}\frac{\partial \tilde{\varphi}}{\partial \tilde{\xi}} +$$

$$+2\frac{\partial \tilde{\varphi}}{\partial \tilde{\xi}}\frac{1}{\tilde{u}^2}\left[\frac{\partial}{\partial \tilde{\xi}}\left(\tilde{\rho}_e\tilde{u}^2 + \tilde{p}_e - \tilde{\rho}_e\tilde{u}\right) - \tilde{\rho}_e\frac{\partial \tilde{\varphi}}{\partial \tilde{\xi}}\right] = -\left(\tilde{p}_e - \tilde{p}_i\right)\left(1 + \frac{m_i}{m_e}\right)\tilde{u}^2 .$$

(3.15)

Some comments to equations (3.10) – (3.15):

1. Every equation from the system is of the second order and needs two conditions. The problem belongs to the class of Cauchy problems.
2. In comparison with the Schrödinger theory connected with behavior of the wave function, no special conditions are applied for dependent variables including the domain of the solution existing. This domain is defined automatically in the process of the numerical solution of the concrete variant of calculations.
3. From the introduced scales

$$u_0,\ x_0 = \frac{\hbar}{m_e}\frac{1}{u_0},\ \varphi_0 = \frac{m_e}{e}u_0^2,\ \rho_0 = \frac{m_e^4}{4\pi\hbar^2 e^2}u_0^4,\ p_0 = \rho_0 u_0^2 = \frac{m_e^4}{4\pi\hbar^2 e^2}u_0^6$$

only two parameters are independent – the phase velocity $u_0$ of the quantum object, and external parameter $H$, which is proportional to Plank constant $\hbar$ and in general case should be inserted in the scale relation as $x_0 = \frac{H}{m_e u_0} = \frac{n\hbar}{m_e u_0}$. It leads to exchange in all scales $\hbar \leftrightarrow H$. But the value $\nu^{qu} = \hbar / m_e$ has the dimension $[cm^2/s]$ and can be titled as quantum viscosity, $\nu^{qu} = 1.1577\ cm^2/s$. Of course on principal the electron mass can be replaced in scales by mass of other particles with the negative charge. From this point of view the obtained solutions which will be discussed below have the universal character defined only by Cauchy conditions.

## 4. Results of mathematical modeling in the theory of quantum solitons.

The system of generalized quantum hydrodynamic equations (3.10) – (3.15) have the great possibilities of mathematical modeling as result of changing of twelve Cauchy conditions describing the character features of initial perturbations which lead to the soliton formation.

On this step of investigation I intend to demonstrate the influence of difference conditions on the soliton formation. The following figures reflect some results of calculations realized according to the system of equations (3.10) - (3.15) with the help of Maple 9. The following notations on figures are used: r- density $\tilde{\rho}_i$ (solid black line), s- density $\tilde{\rho}_e$ (solid line), u- velocity $\tilde{u}$ (dashed line), p - pressure $\tilde{p}_i$ (black dash dotted line), $q$ - pressure $\tilde{p}_e$ (dash dotted line) and v - self consistent potential $\tilde{\varphi}$. Explanations placed under all following figures, Maple program contains Maple's notations – for example the expression $D(u)(0) = 0$ means in usual notations $\frac{\partial \tilde{u}}{\partial \tilde{\xi}}(0) = 0$, independent variable $t$ responds to $\tilde{\xi}$.

We begin with investigation of the problem of principle significance – is it possible after a perturbation (defined by Cauchy conditions) to obtain the quantum object of the soliton's kind as result of the self-organization of ionized matter? In the case of the positive answer, what is the origin of existence of this stable object? By the way the mentioned questions belong to the typical problem in the theory of the ball lightning. With this aim let us consider the initial perturbations (4.1)

```
v(0)=1,r(0)=1,s(0)=1,u(0)=1,p(0)=1,q(0)=.95,
D(v)(0)=0,D(r)(0)=0,D(s)(0)=0,D(u)(0)=0,D(p)(0)=0,D(q)(0)=0
```
(4.1)

in the mixture of positive and negative ions of equal masses if the pressure $\tilde{p}_i(0)$ of positive particles is larger than $\tilde{p}_e(0)$ of the negative ones (for the variant under consideration `p(0)=1,q(0)=.95`). The following Figures 1 – 3 reflect the result of solution of equations (3.10) – (3.15).

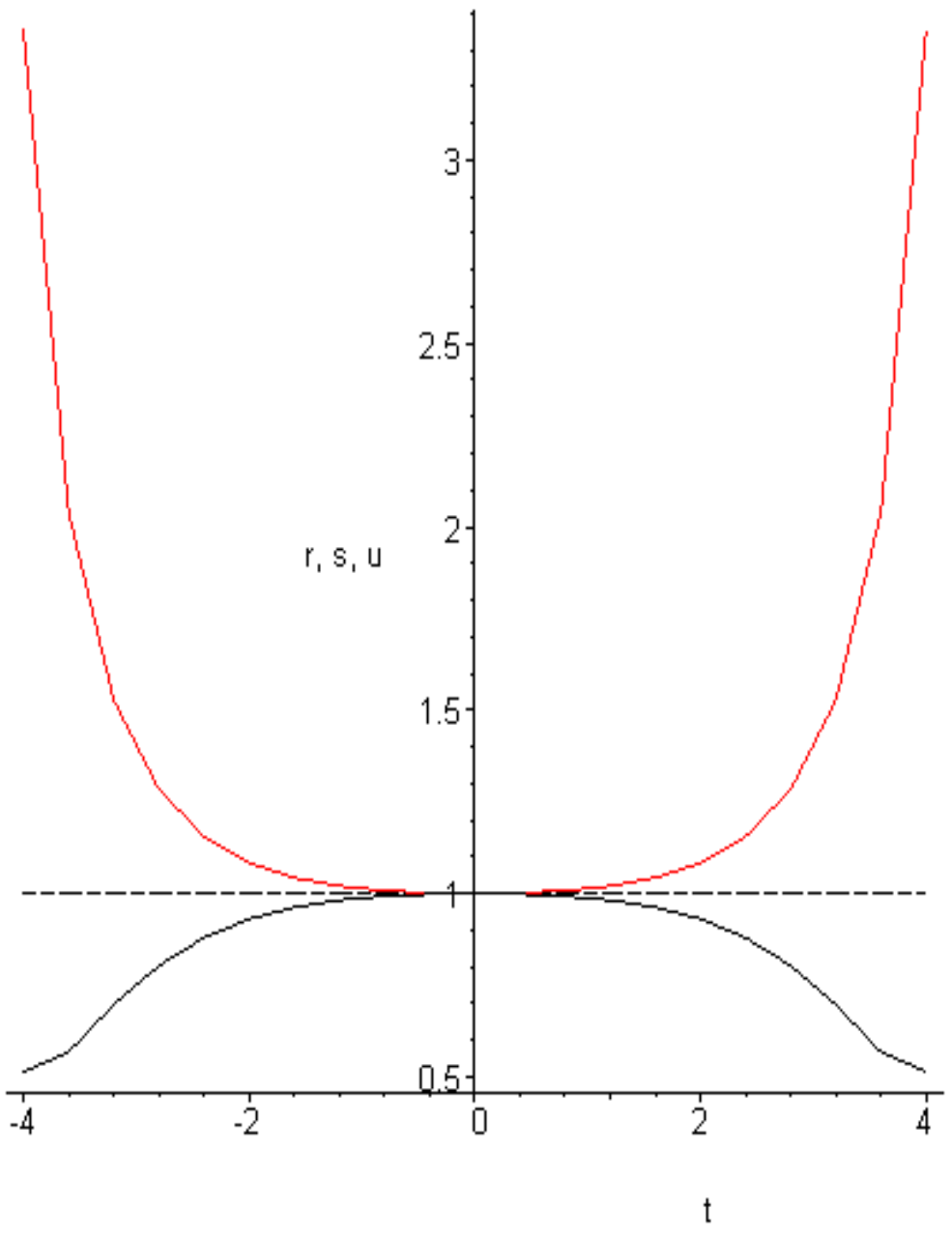

Figure1. r- density $\tilde{\rho}_i$, u- velocity $\tilde{u}$, s- density $\tilde{\rho}_e$ in quantum soliton.

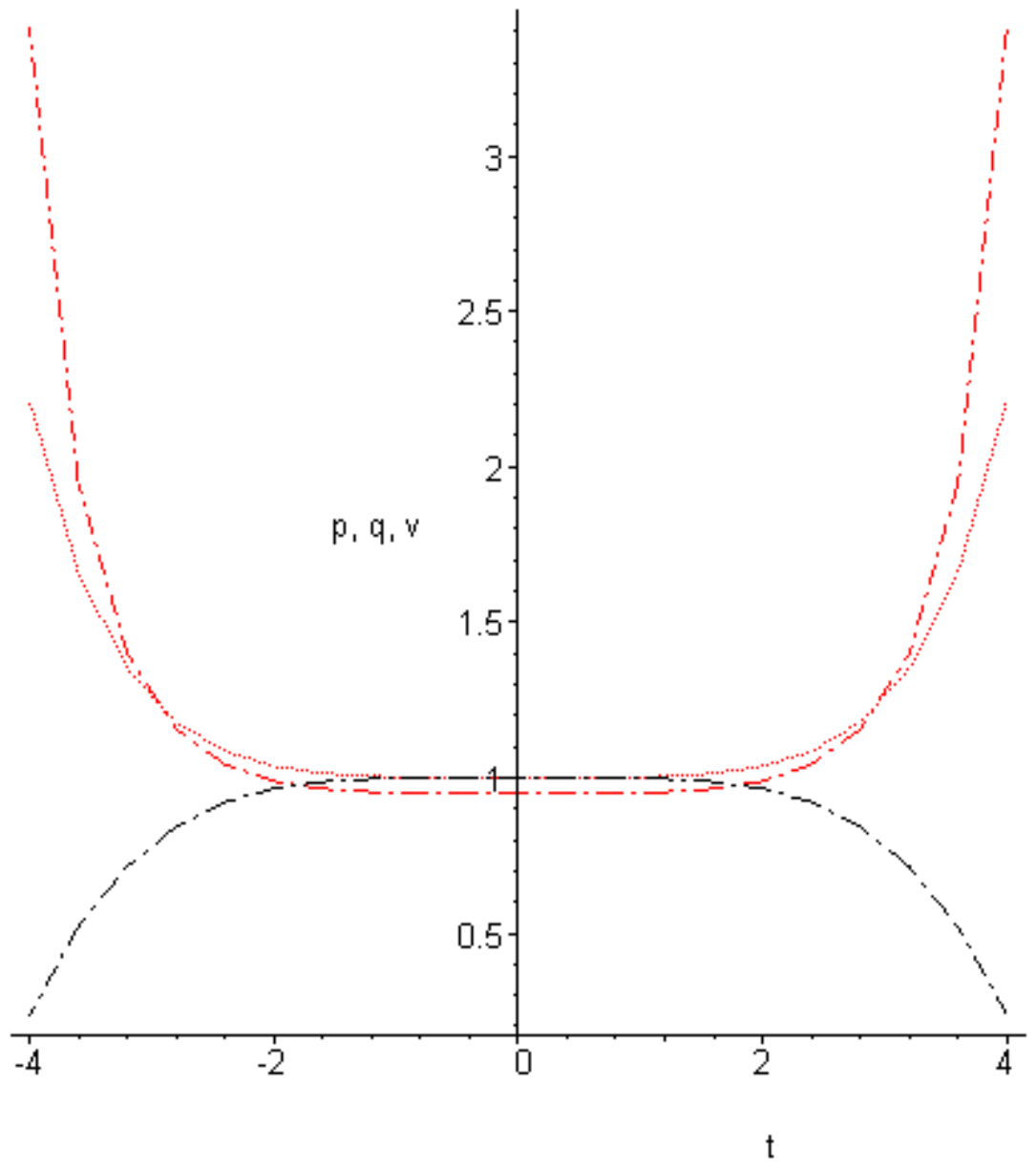

Figure 2. p - pressure $\tilde{p}_i$, $q$ - pressure $\tilde{p}_e$, v - self-consistent potential $\tilde{\varphi}$ in quantum soliton.

Figure 1 displays the quantum object placed in bounded region of 1D space; all parts of this object are moving with the same velocity. Important to underline that no special boundary conditions were used for this and all following cases. Then this soliton is product of the self-organization of ionized matter. Fig. 2, 3 contain the answer for formulated above questions about stability of the object. Really the object is restricted by negative shell. The derivative $\partial\tilde{\varphi}/\partial\tilde{\xi}$ is proportional to the self-consistent forces acting on the positive and negative parts of the soliton. Consider for example the right side of soliton. The self consistent force of the electrical origin compresses the positive part of this soliton and provokes the movement of the negative part along the positive direction of the $\tilde{\xi}$ - axis ($t$ – axis in nomination of Figure 1). But the increasing of quantum pressure prevent to destruction of soliton. Therefore the stability of the quantum object is result of the self-consistent influence of electric potential and quantum pressures.

Interesting to notice that stability can be also achieved if soliton has *positive* shell and *negative* kernel but $\tilde{p}_i(0)<\tilde{p}_e(0)$, see Fig. 4 – 6 obtained as result of mathematical modeling for the case (4.2)

```
v(0)=1,r(0)=1,s(0)=1,u(0)=1,p(0)=1,q(0)=1.05,
D(v)(0)=0,D(r)(0)=0,D(s)(0)=0,D(u)(0)=0,D(p)(0)=0,D(q)(0)=0.
```
(4.2)

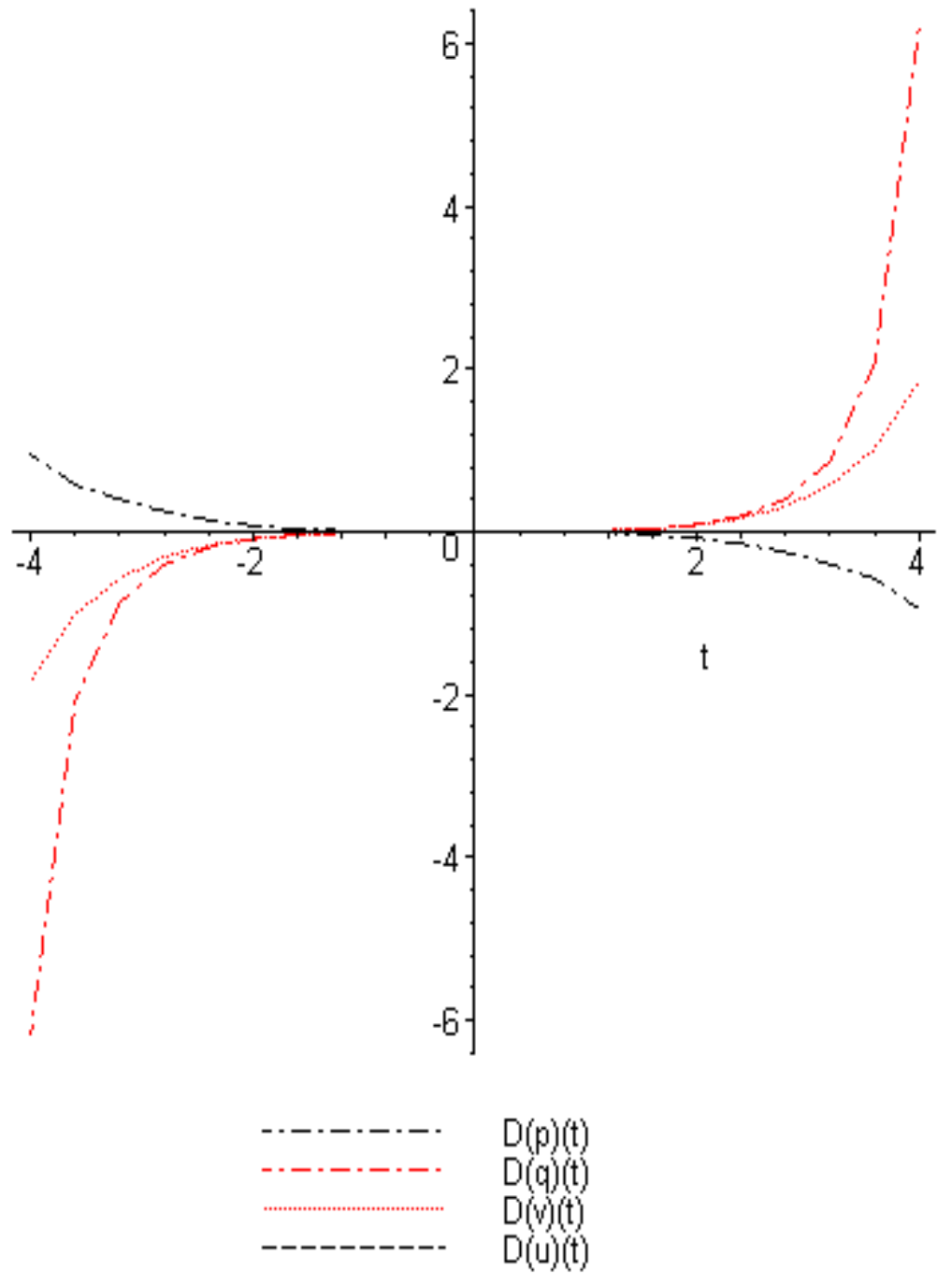

Figure 3. The derivative of pressure of the positive component $\partial\widetilde{p}_i/\partial\widetilde{\xi}$, the derivative of pressure of negative component $\partial\widetilde{p}_e/\partial\widetilde{\xi}$, the derivative of the self-consistent potential $\partial\widetilde{\varphi}/\partial\widetilde{\xi}$ in quantum soliton.

The explanation for this case has practically the same character as in the previous case but positive and negative species change their roles.

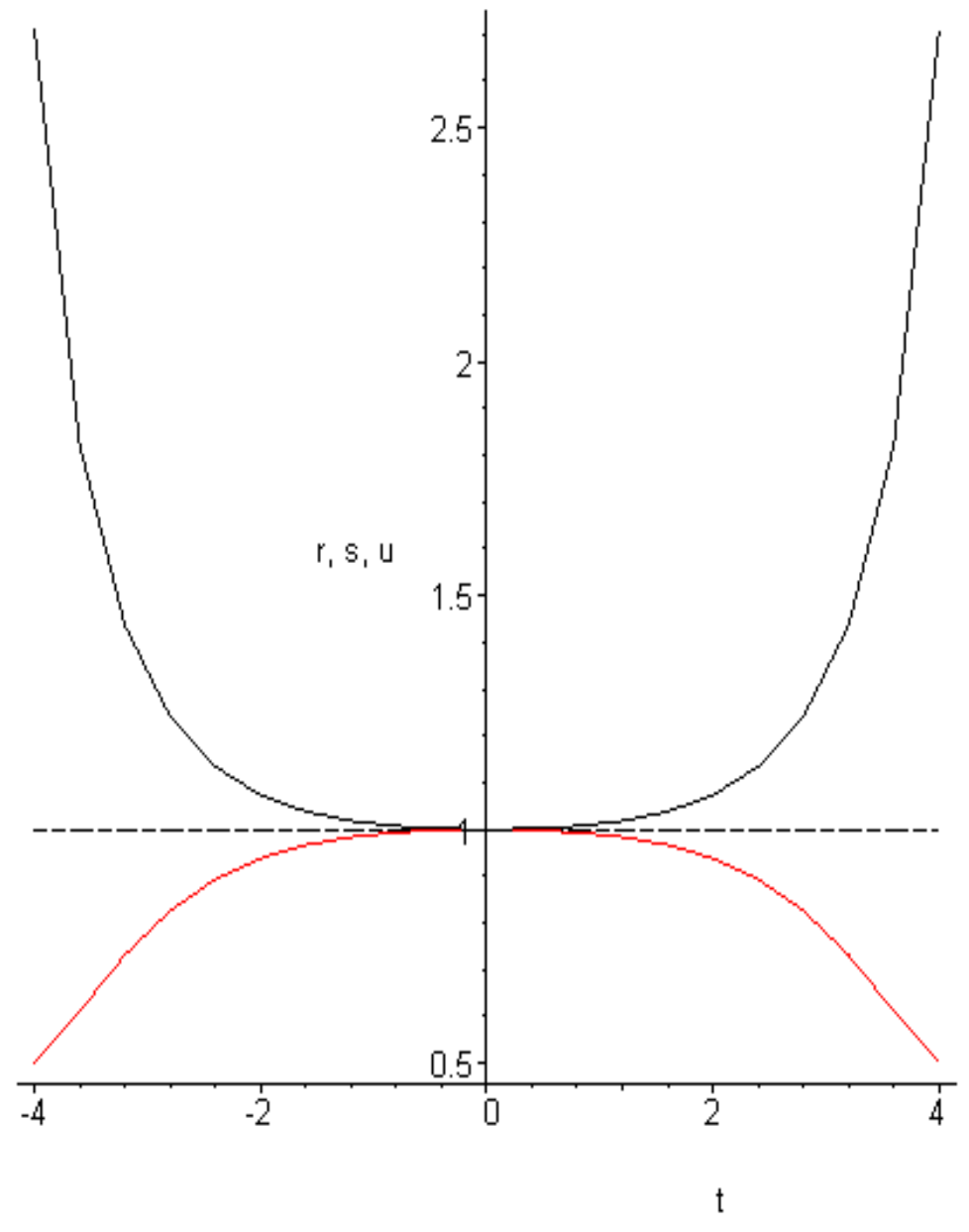

Figure 4. r- density $\widetilde{\rho}_i$, u- velocity $\widetilde{u}$, s- density $\tilde{\rho}_e$- in quantum soliton.

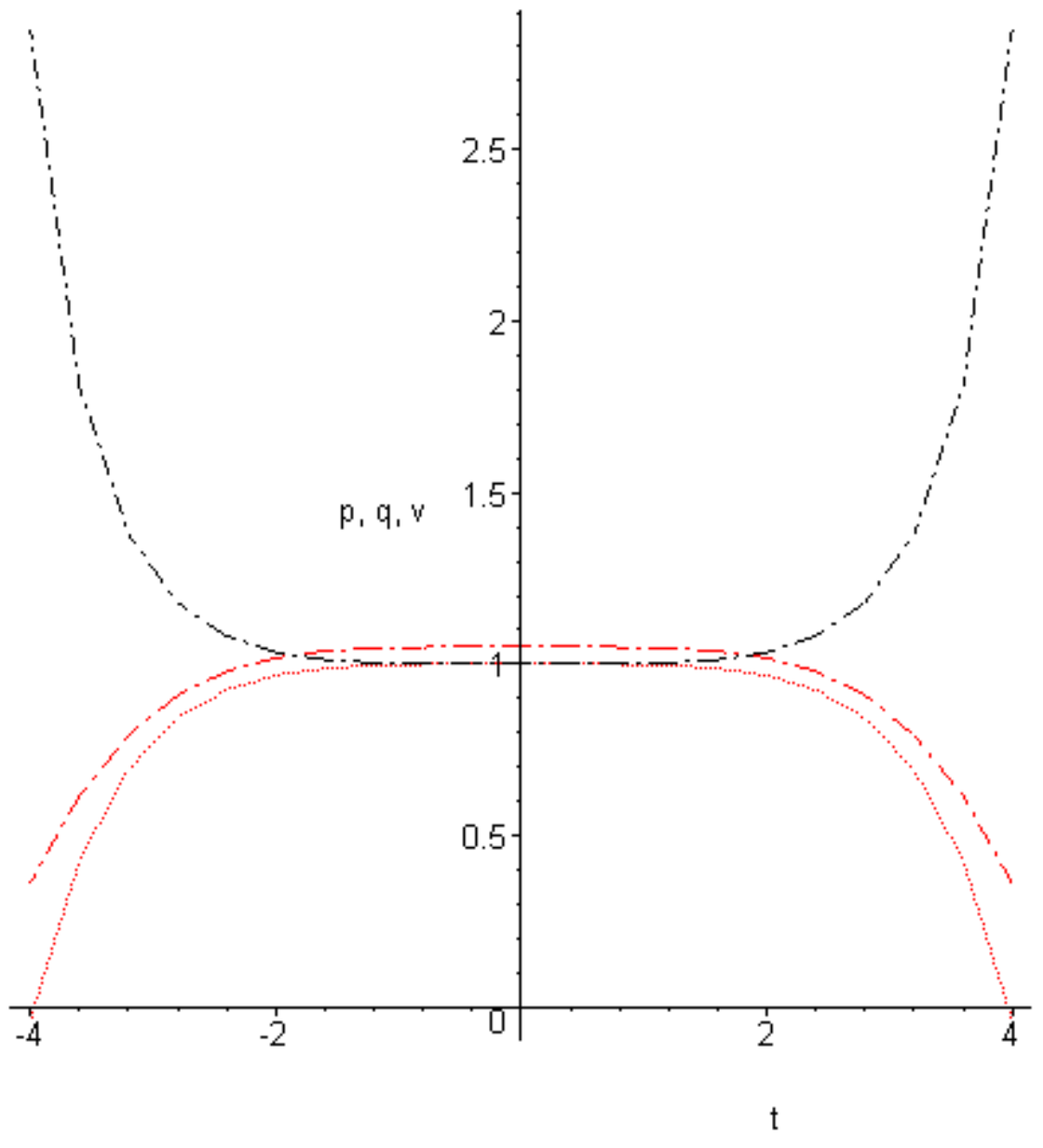

Figure 5. p - pressure $\widetilde{p}_i$, $q$ - pressure $\widetilde{p}_e$, v - self-consistent potential $\widetilde{\varphi}$ in quantum soliton.

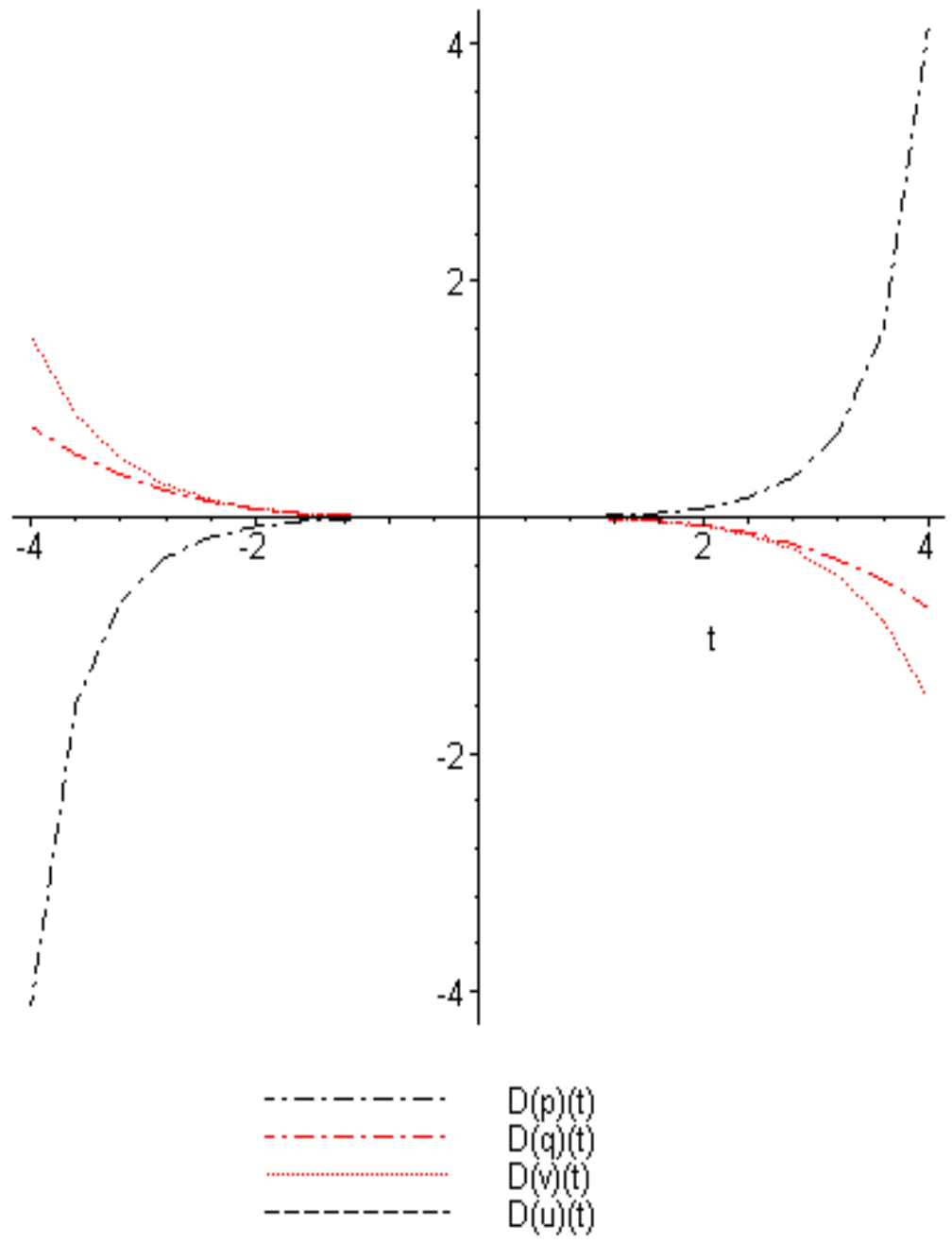

Figure 6. The derivative of pressure of the positive component $\partial\tilde{p}_i/\partial\tilde{\xi}$, the derivative of pressure of negative component $\partial\tilde{p}_e/\partial\tilde{\xi}$, the derivatives of the self-consistent potential $\partial\tilde{\varphi}/\partial\tilde{\xi}$ and velocity $\partial\tilde{u}/\partial\tilde{\xi}$ in quantum soliton.

In following calculations we use the typical ratio of masses $m_i/m_e = 1838$. The initial perturbations in the mixture of heavy positive particles and electrons produce the soliton formation if the pressure $\tilde{p}_i(0)$ of the positive particles is larger than $\tilde{p}_e(0)$ of the negative ones (in Fig. 7, 8 for the following variant (4.3) under consideration **p(0)=1,q(0)=.95**):

**v(0)=1,r(0)=1,s(0)=1/1838,u(0)=1,p(0)=1,q(0)=.95,**
**D(v)(0)=0,D(r)(0)=0,D(s)(0)=0,D(u)(0)=0,D(p)(0)=0,D(q)(0)=0.** (4.3)

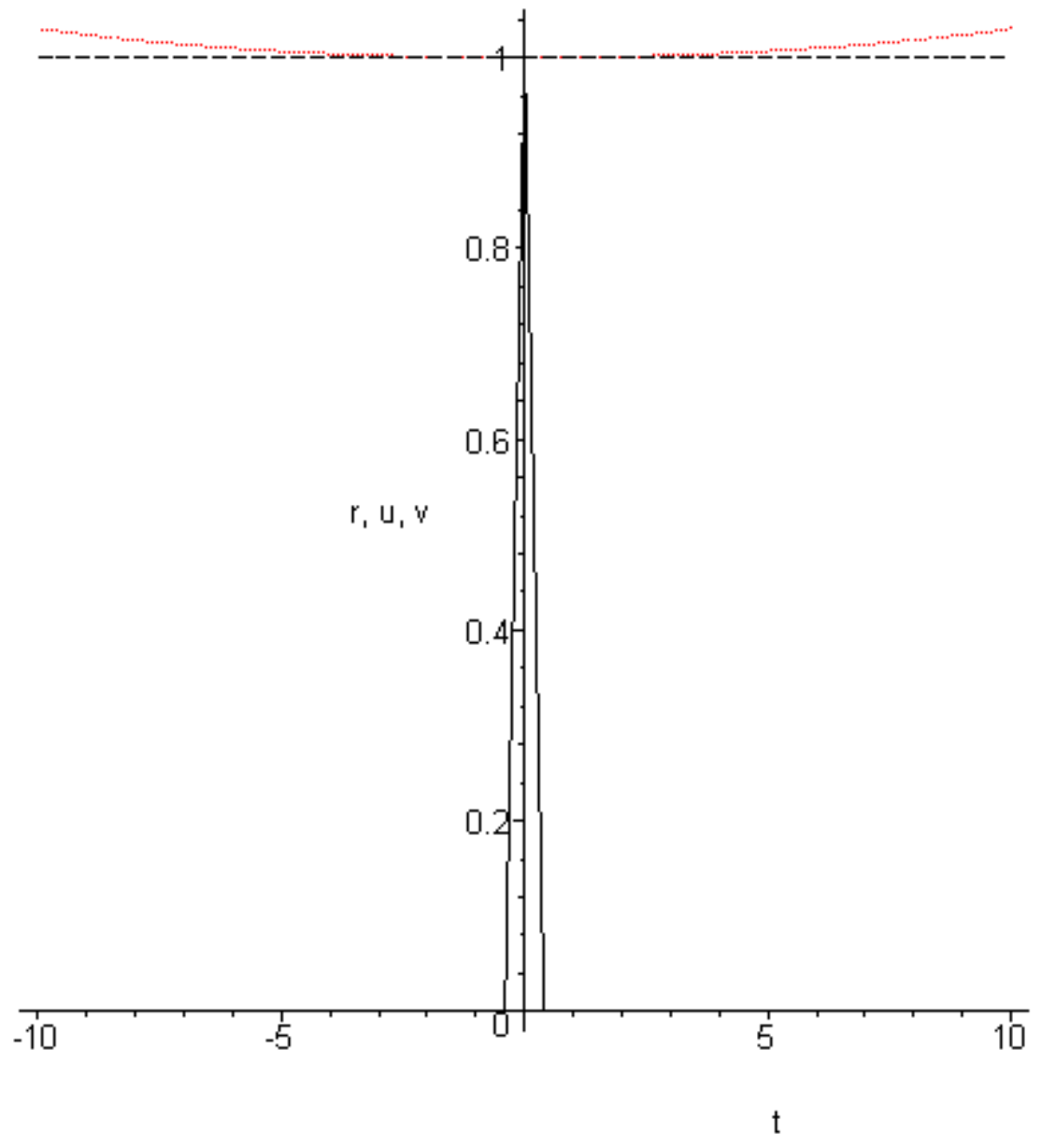

Figure 7. r- density $\tilde{\rho}_i$, u- velocity $\tilde{u}$, v-self-consistent potential $\tilde{\varphi}$ in quantum soliton.

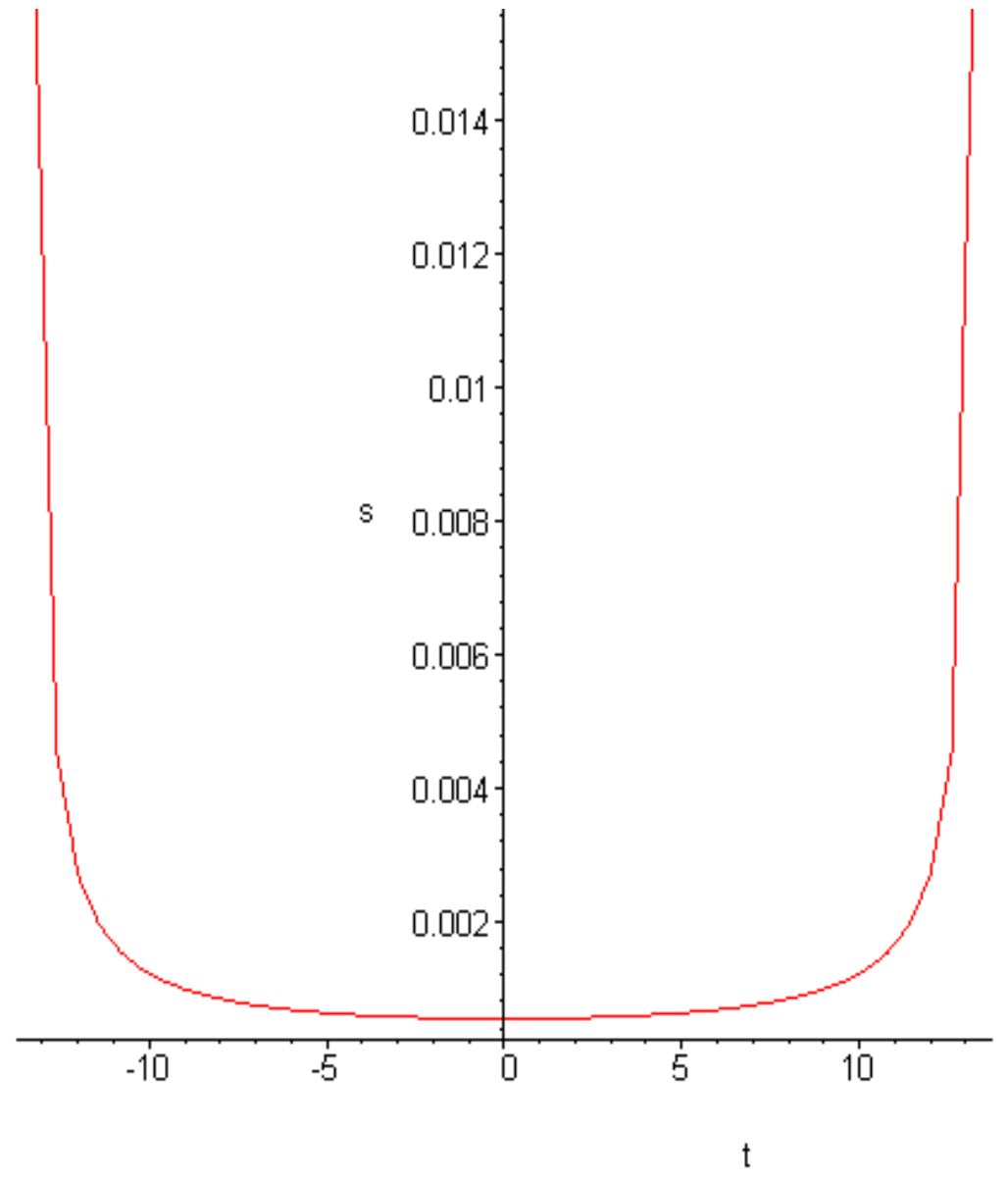

Figure 8. s- density $\tilde{\rho}_e$ in quantum soliton.

In comparison with Figure. 1 – 3 we observe the explicit positive kernel (nuclei) which is typical for atom structures.

Now can be demonstrated the influence of the significant difference in mass of particles $m_i / m_e = 1838$ for the case $\tilde{p}_i(0) < \tilde{p}_e(0)$ in Fig. 9 – 11. We use the Cauchy conditions (4.4)

```
v(0)=1,r(0)=1,s(0)=1/1838,u(0)=1,p(0)=1,q(0)=1.05,
D(v)(0)=0,D(r)(0)=0,D(s)(0)=0,D(u)(0)=0,D(p)(0)=0,D(q)(0)=0.
```
(4.4)

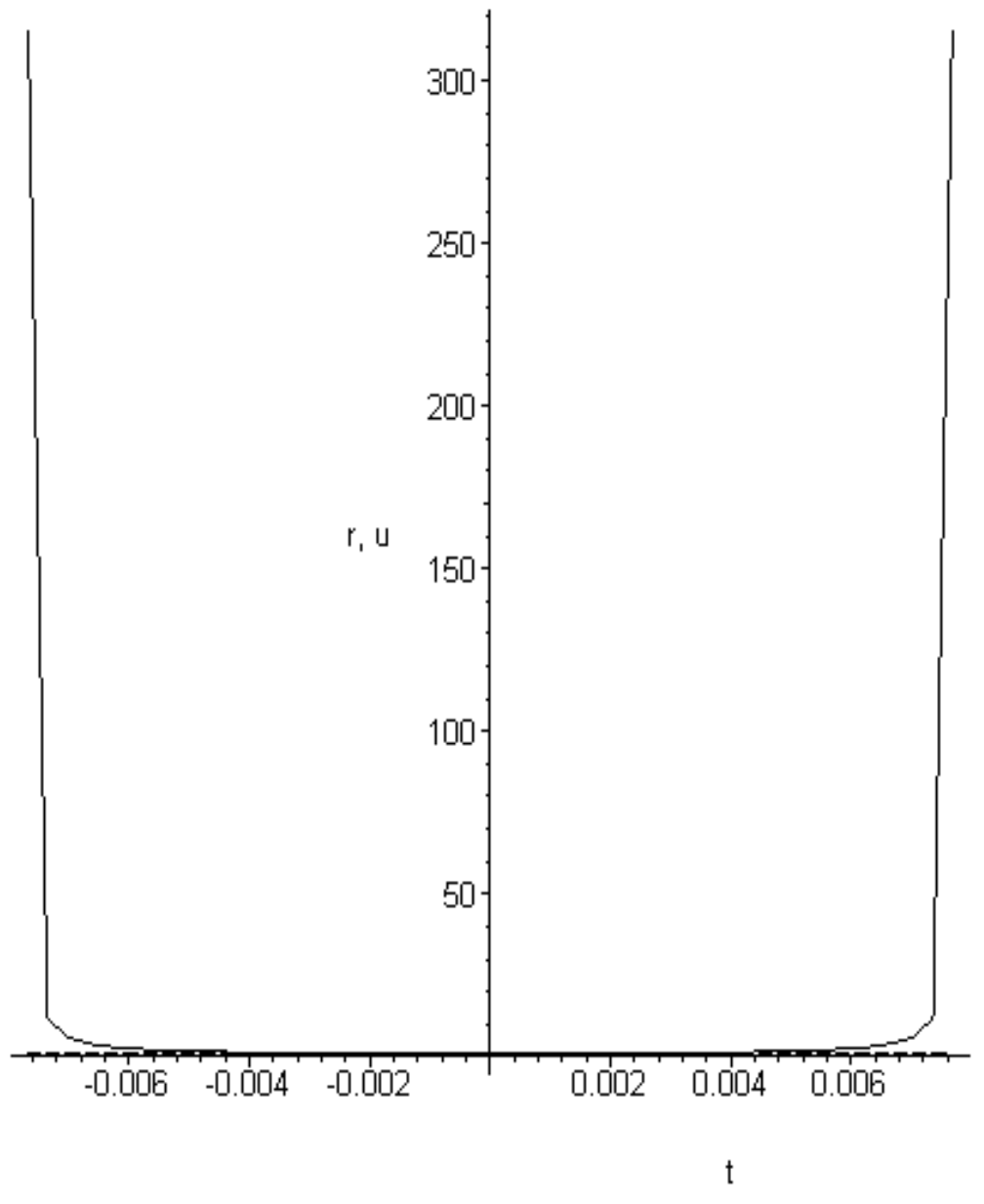

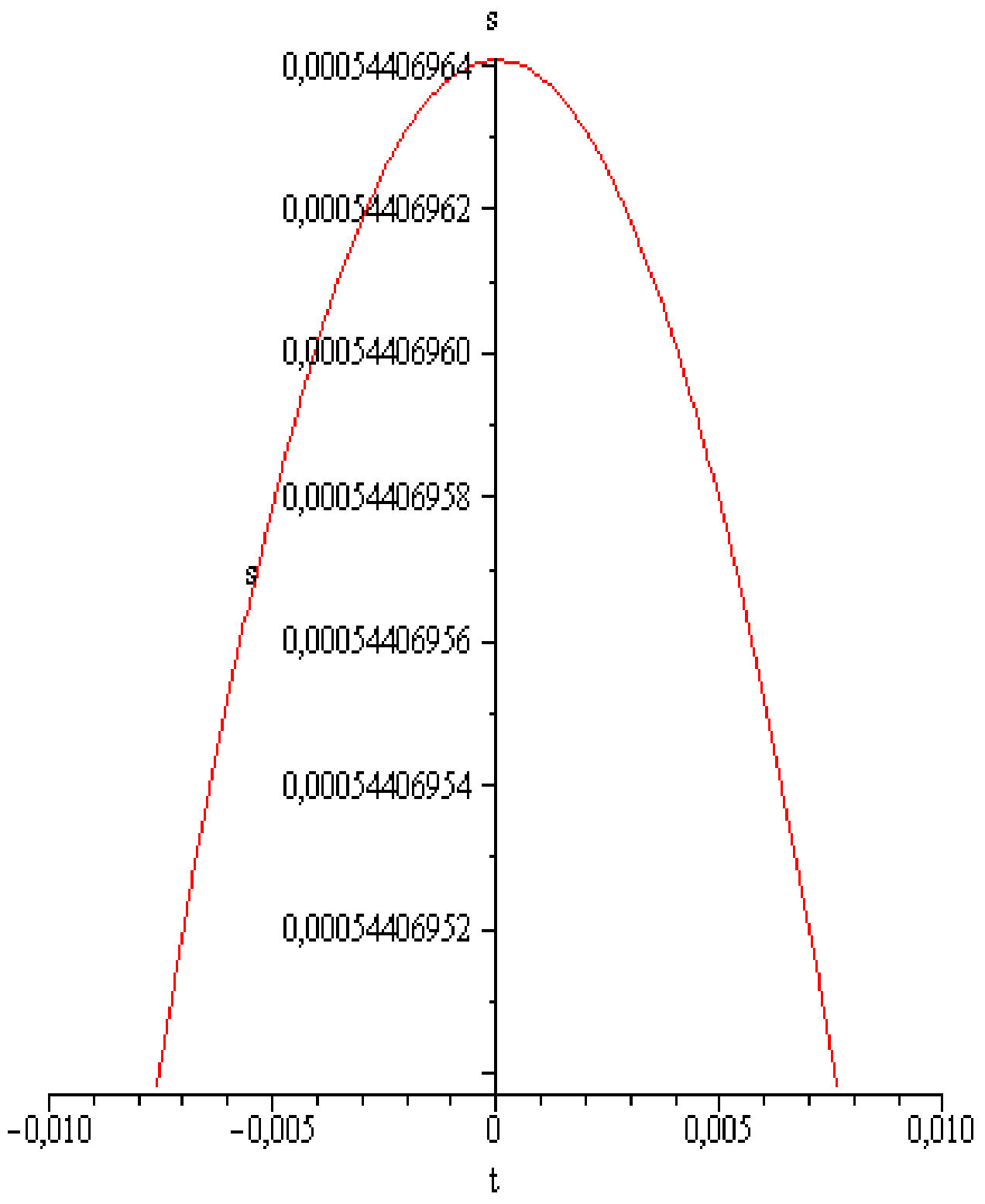

Figure 9. r- density $\tilde{\rho}_i$, u- velocity $\tilde{u}$, Figure 10. s- density $\tilde{\rho}_e$.

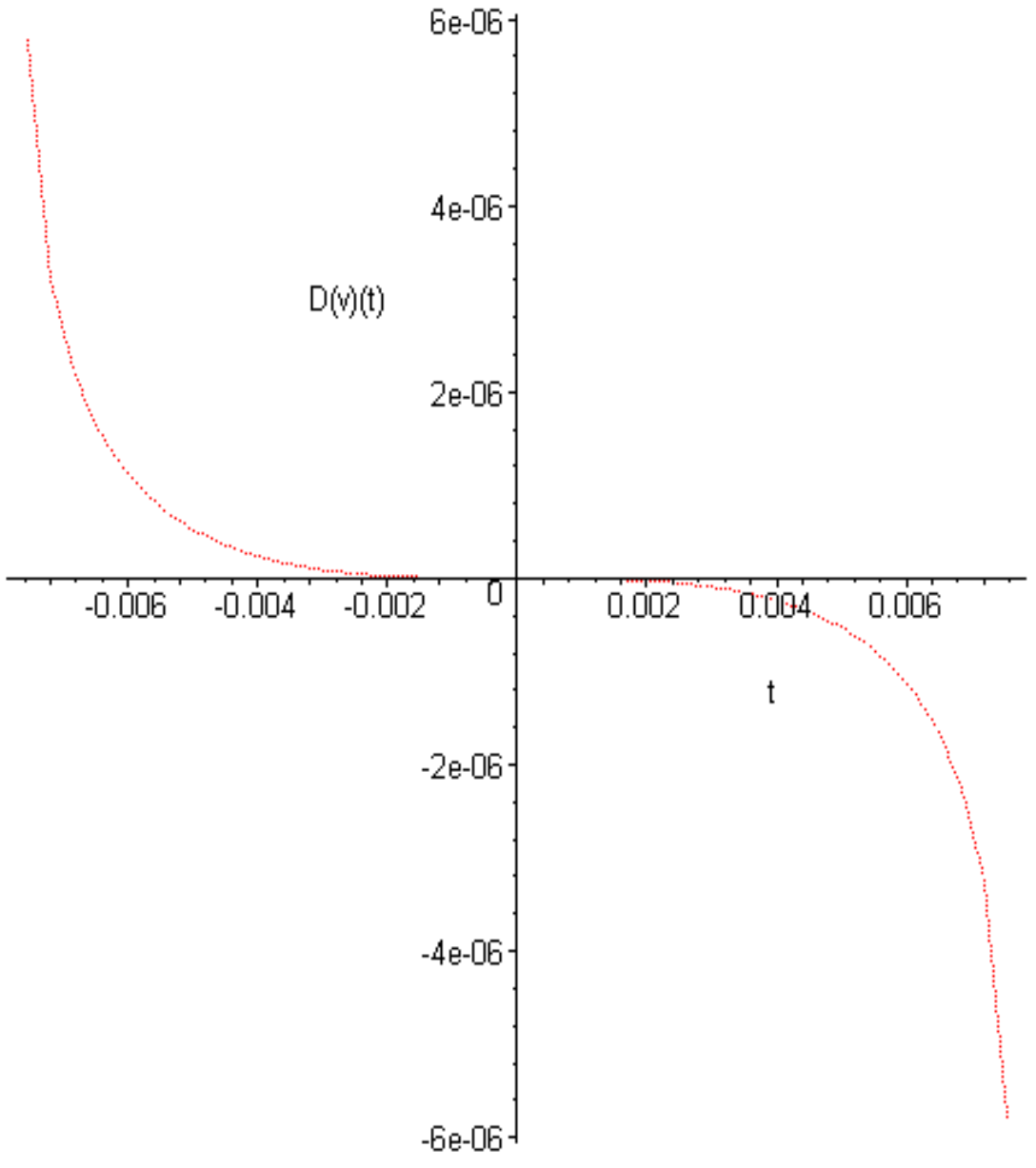

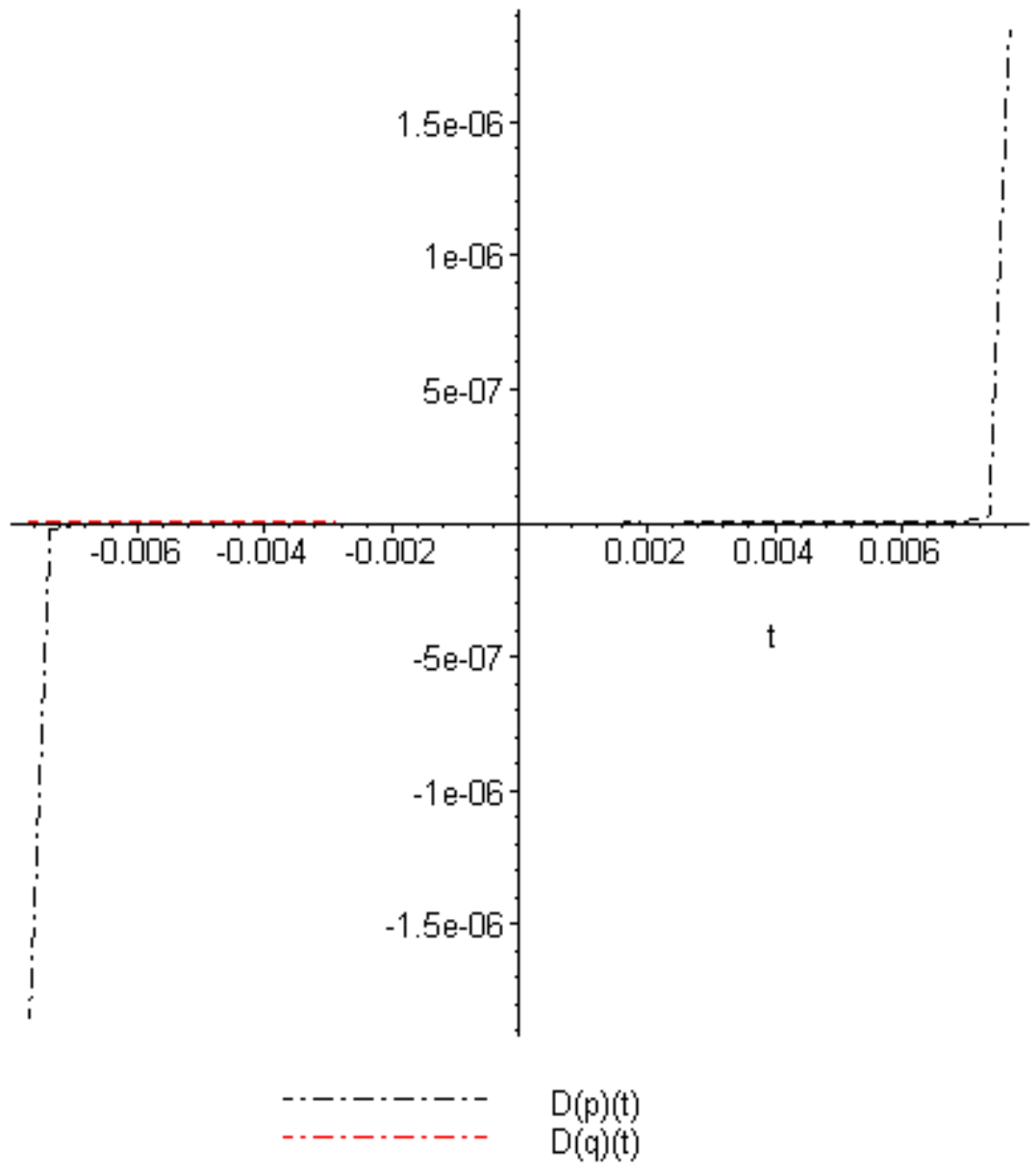

Figure 11.The derivative of the self-consistent potential $\partial\tilde{\varphi}/\partial\tilde{\xi}$ in quantum soliton.

Figure 12. The derivative of pressures $\partial\tilde{p}_i/\partial\tilde{\xi}$ and $\partial\tilde{p}_e/\partial\tilde{\xi}$ in quantum soliton.

We have the “inversed atom structure” for regular matter.

Consider the influence of changing of the rest non-local pressures $\tilde{p}_i(0)$, $\tilde{p}_e(0)$. Figures 13 -15 reflect the following Cauchy conditions (4.5):

```
v(0)=1,r(0)=1,s(0)=1/1838,u(0)=1,p(0)=1,q(0)=0.999,
D(v)(0)=0,D(r)(0)=0,D(s)(0)=0,D(u)(0)=0,D(p)(0)=0,D(q)(0)=0:
```
(4.5)

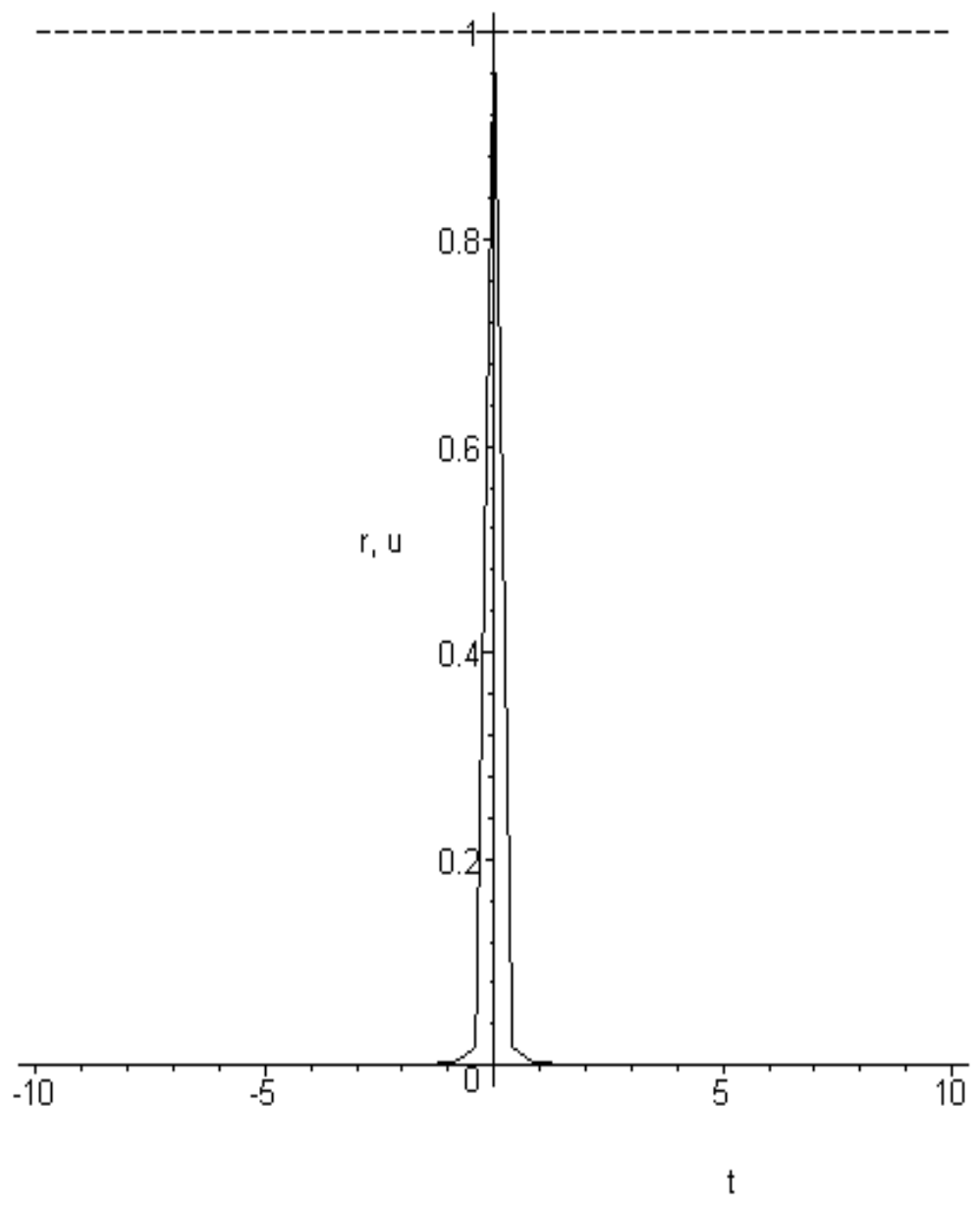

Figure 13. r- density $\tilde{\rho}_i$, u- velocity $\tilde{u}$ in quantum soliton.

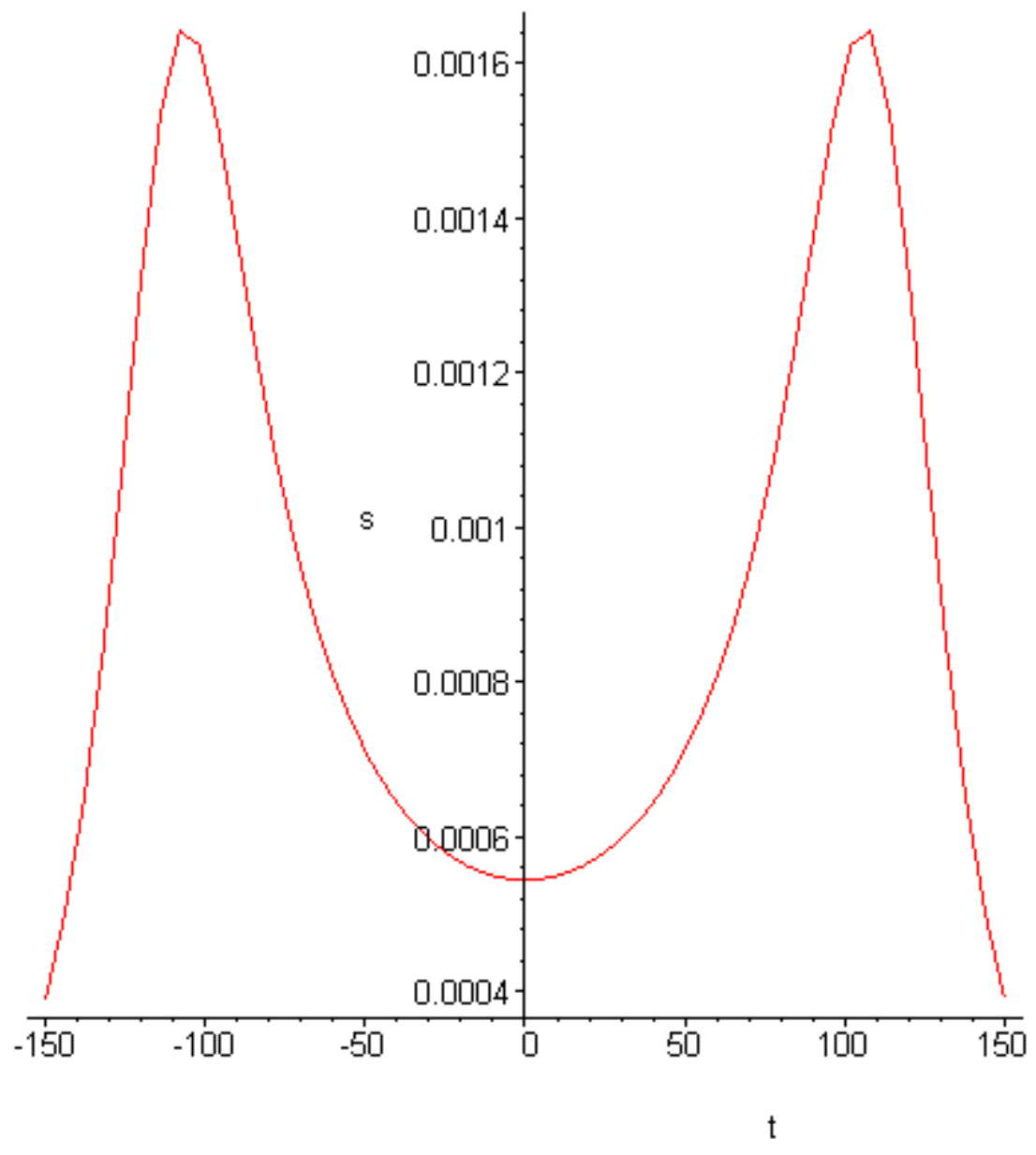

Figure 14. s- density $\tilde{\rho}_e$ in quantum soliton.

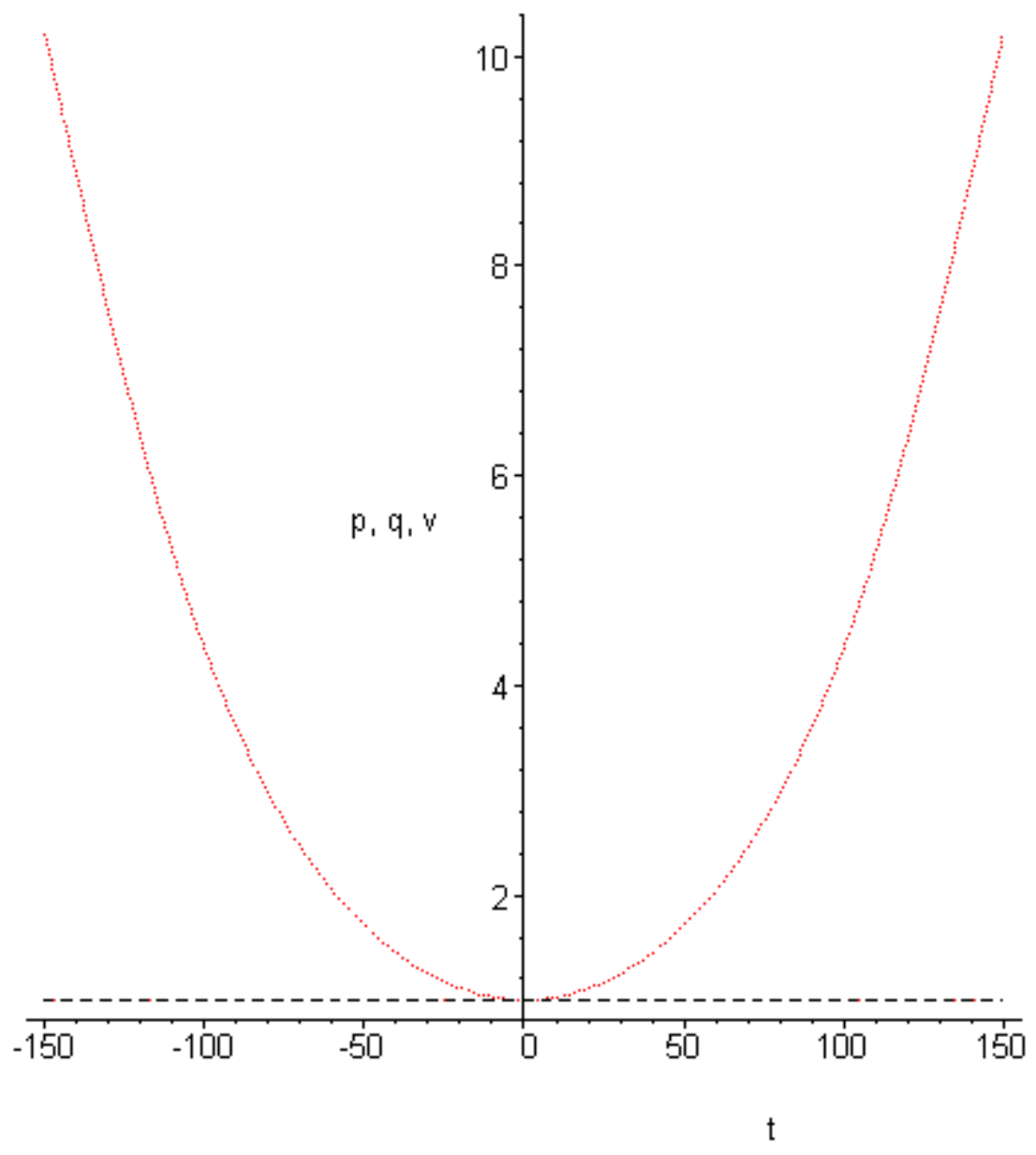

Figure 15. v - self consistent potential $\tilde{\varphi}$ in quantum soliton and pressures $\tilde{p}_i$, $\tilde{p}_e$.

From Figures 7, 8 and 13 - 15 follows that increasing the difference $p_i(0) - p_e(0)$ brings to diminishing of the character domain occupied by soliton. The classical construction with the positive kernel (nuclei) and negative shell is existing if $\tilde{p}_i(0) > \tilde{p}_e(0)$.

But in the opposite case $\tilde{p}_i(0) < \tilde{p}_e(0)$ mathematical modeling for regular matter leads to construction with negative kernel and positive shell for soliton and to diminishing of the linear size of combined soliton in $1.6 \cdot 10^3$ times and cross sections in $\sim 2.6 \cdot 10^6$ times.

Let us demonstrate the possibility to calculate the soliton formations for anti-matter. In following the nomenclatures are used: $s \to \tilde{\rho}_e$ - quantum antiproton density; $q \to \tilde{p}_e$ - rest quantum antiproton pressure; $r \to \tilde{\rho}_i$ - quantum antielectron density; $p \to \tilde{p}_i$ - rest quantum antielectron pressure. Consider the situation when the initial pressure perturbation $\tilde{p}_i(0) > \tilde{p}_e(0)$ and Cauchy conditions (inversed antimatter atom):
The following figures 16 - 21 reflect the calculations of anti-matter atom for conditions (4.6):

$$v(0) = 1, r(0) = 1, s(0) = 1838, u(0) = 1, p(0) = 1, q(0) = 0.95,$$
$$D(v)(0) = 0, D(r)(0) = 0, D(s)(0) = 0, D(u)(0) = 0, D(p)(0) = 0,$$
$$D(q)(0) = 0 \qquad (4.6)$$

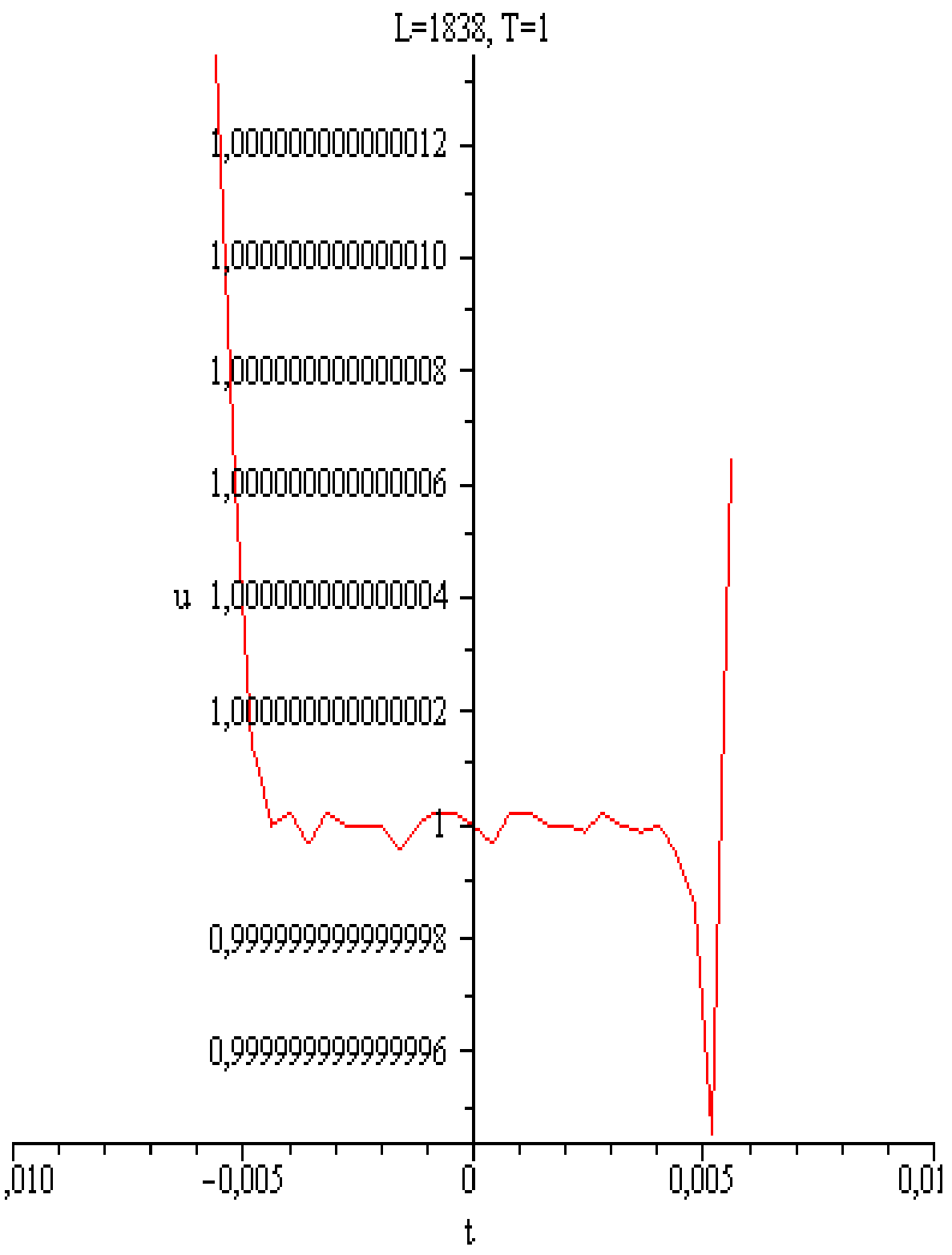

Figure 16. u- velocity $\tilde{u}$.

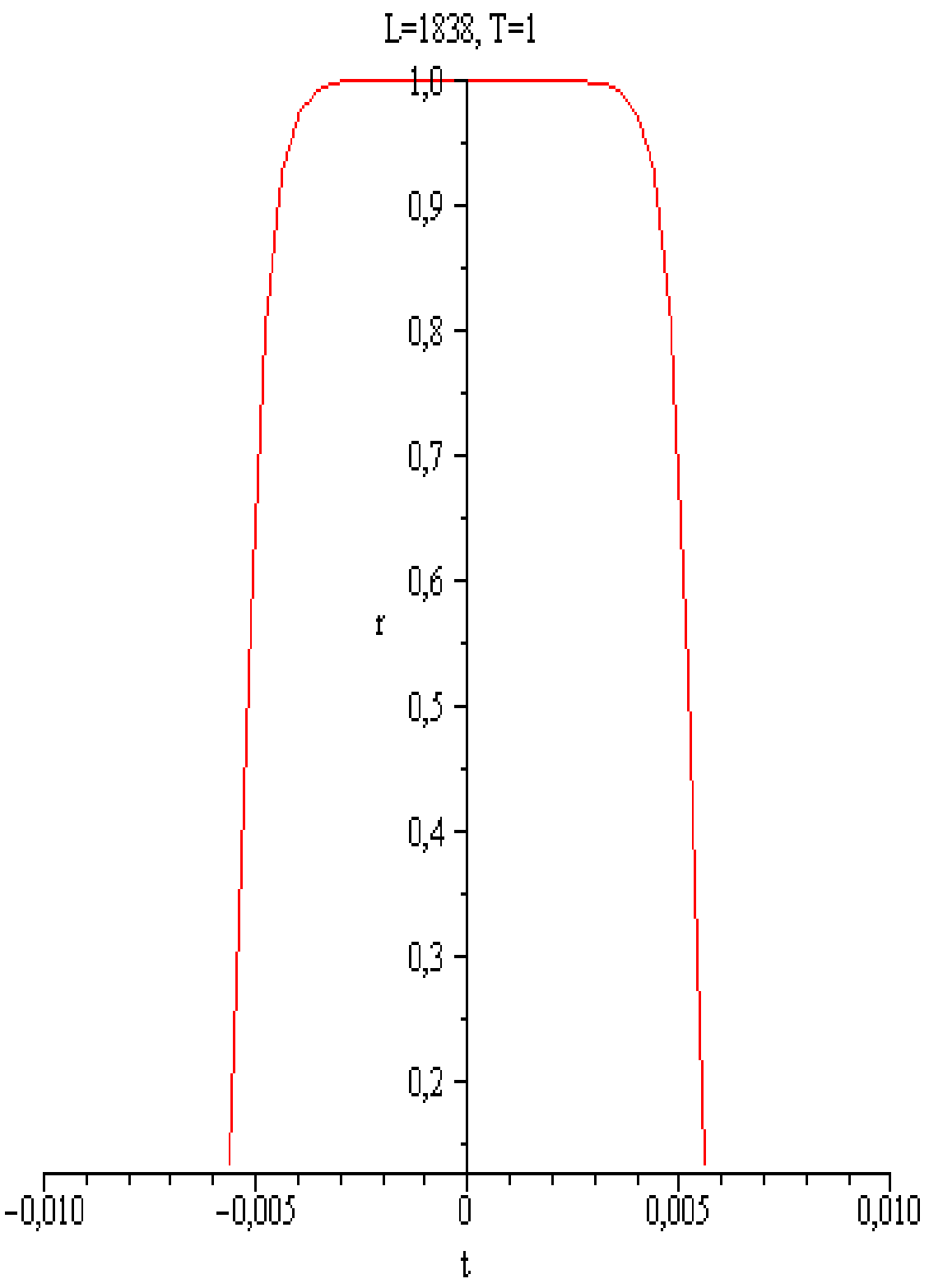

Figure 17. r- density $\tilde{\rho}_i$, in quantum soliton.

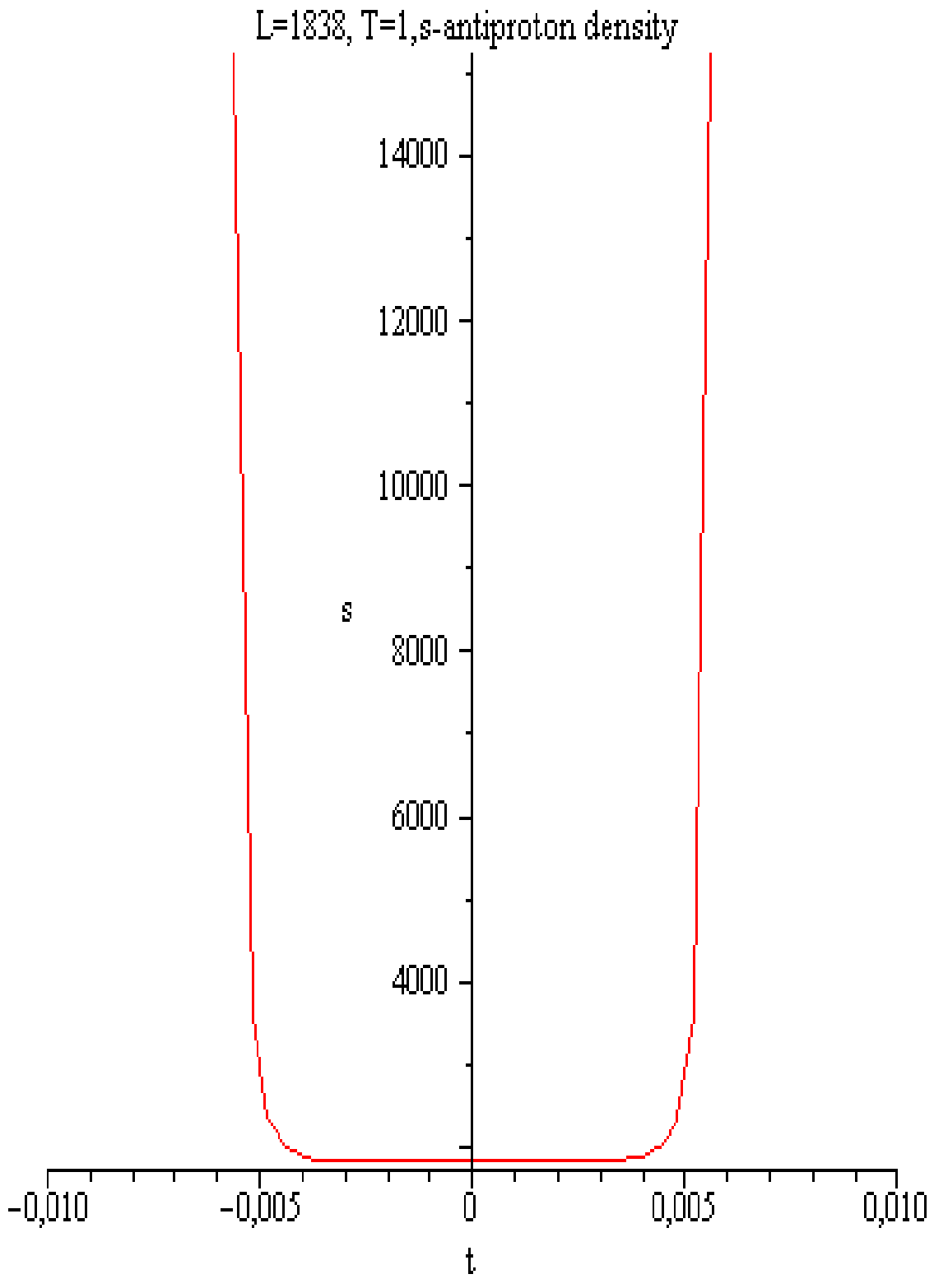

Figure 18. s- density $\tilde{\rho}_e$ in quantum soliton.

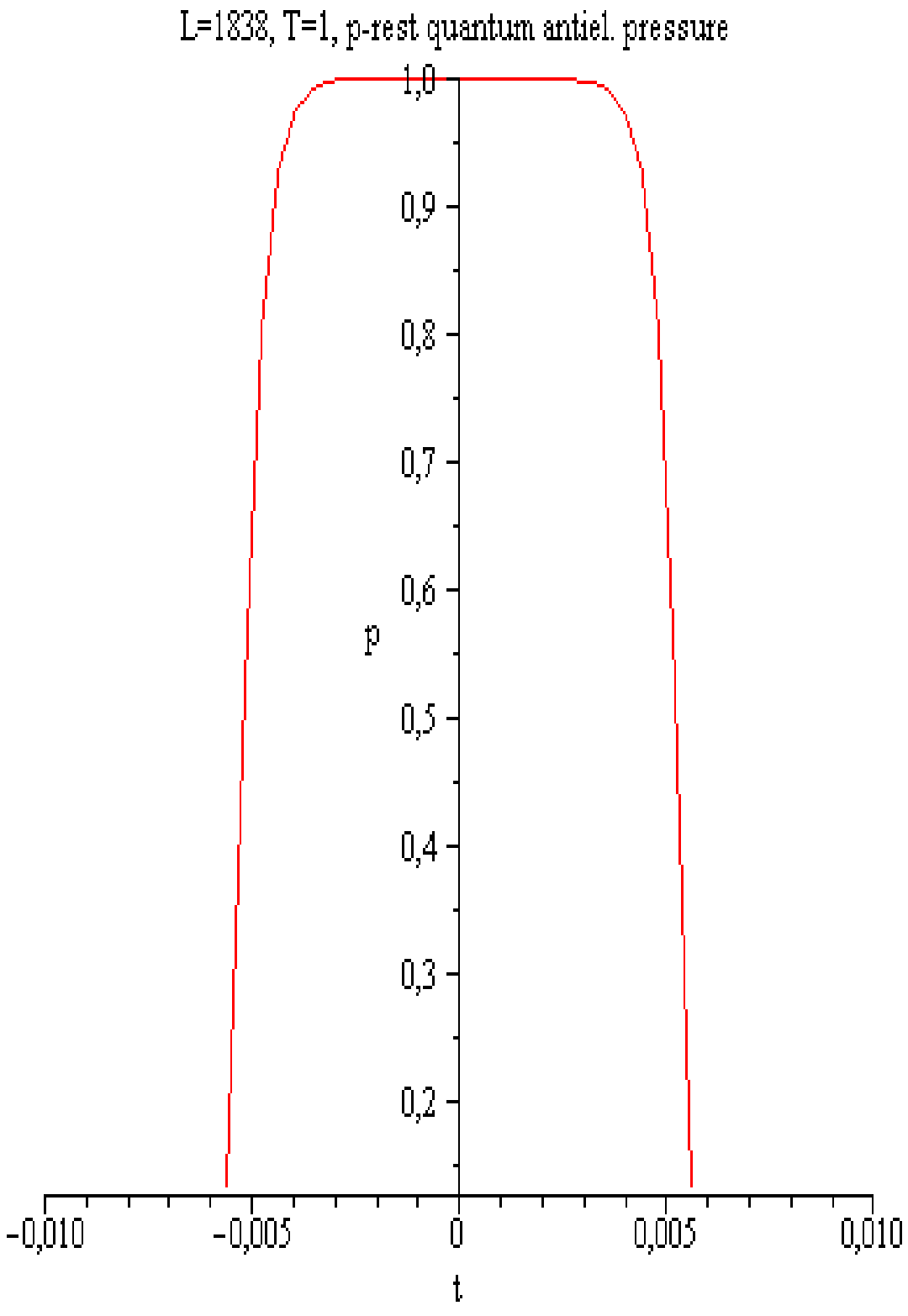

Figure 19. p - pressure $\widetilde{p}_i$ in quantum soliton.

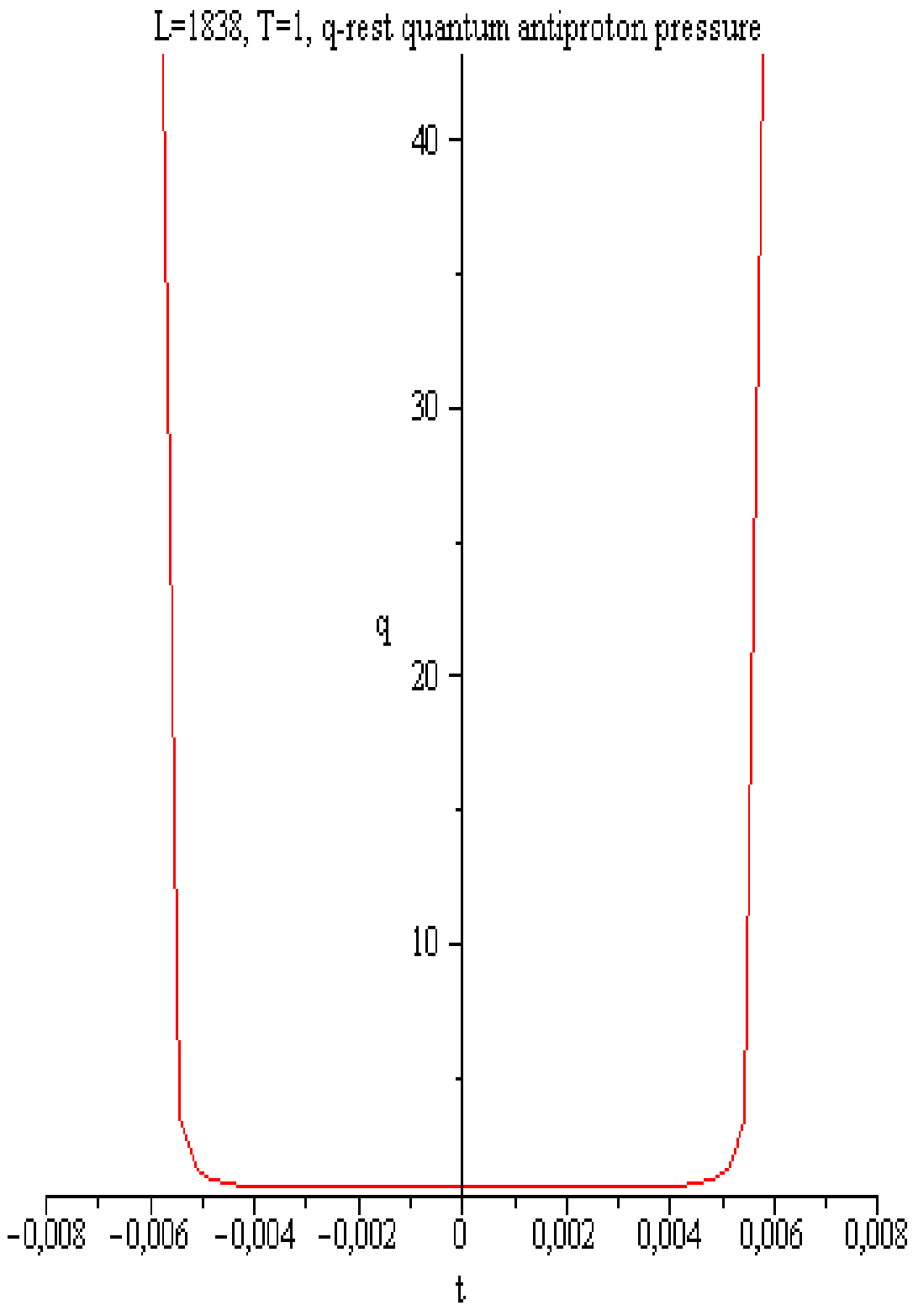

Figure 20. $q$ - pressure $\widetilde{p}_e$ in quantum soliton.

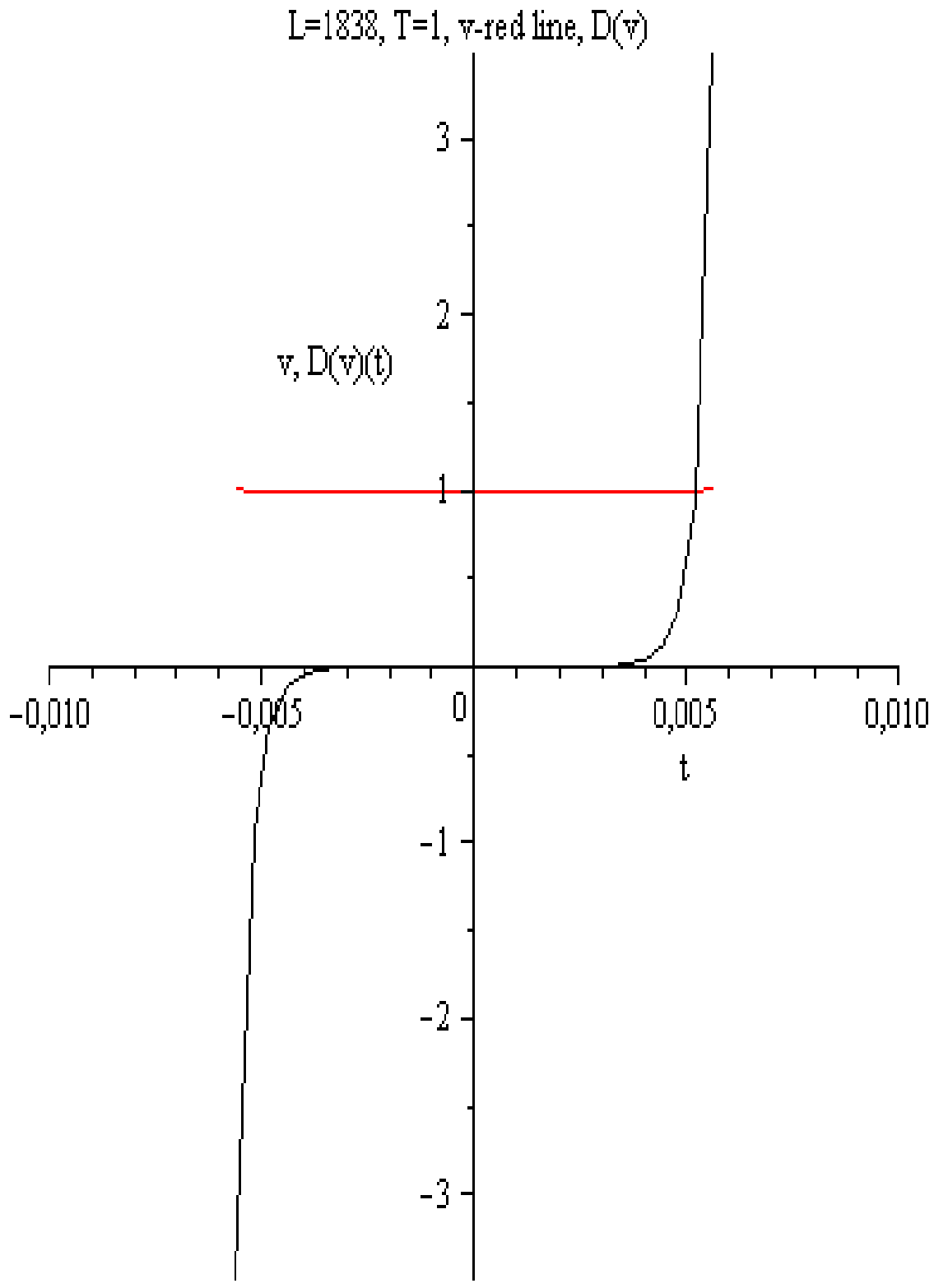

Figure 21. v - self-consistent potential $\widetilde{\varphi}$ and $\partial\widetilde{\varphi}/\partial\widetilde{\xi}$ in quantum soliton.

Figures 22 – 28 belong to the case of anti-matter atom reflected in Cauchy conditions (4.7)

$$v(0) = 1, r(0) = 1, s(0) = 1838, u(0) = 1, p(0) = 1, q(0) = 1.05,$$
$$D(v)(0) = 0, D(r)(0) = 0, D(s)(0) = 0, D(u)(0) = 0, D(p)(0) = 0,$$
$$D(q)(0) = 0 \qquad (4.7)$$

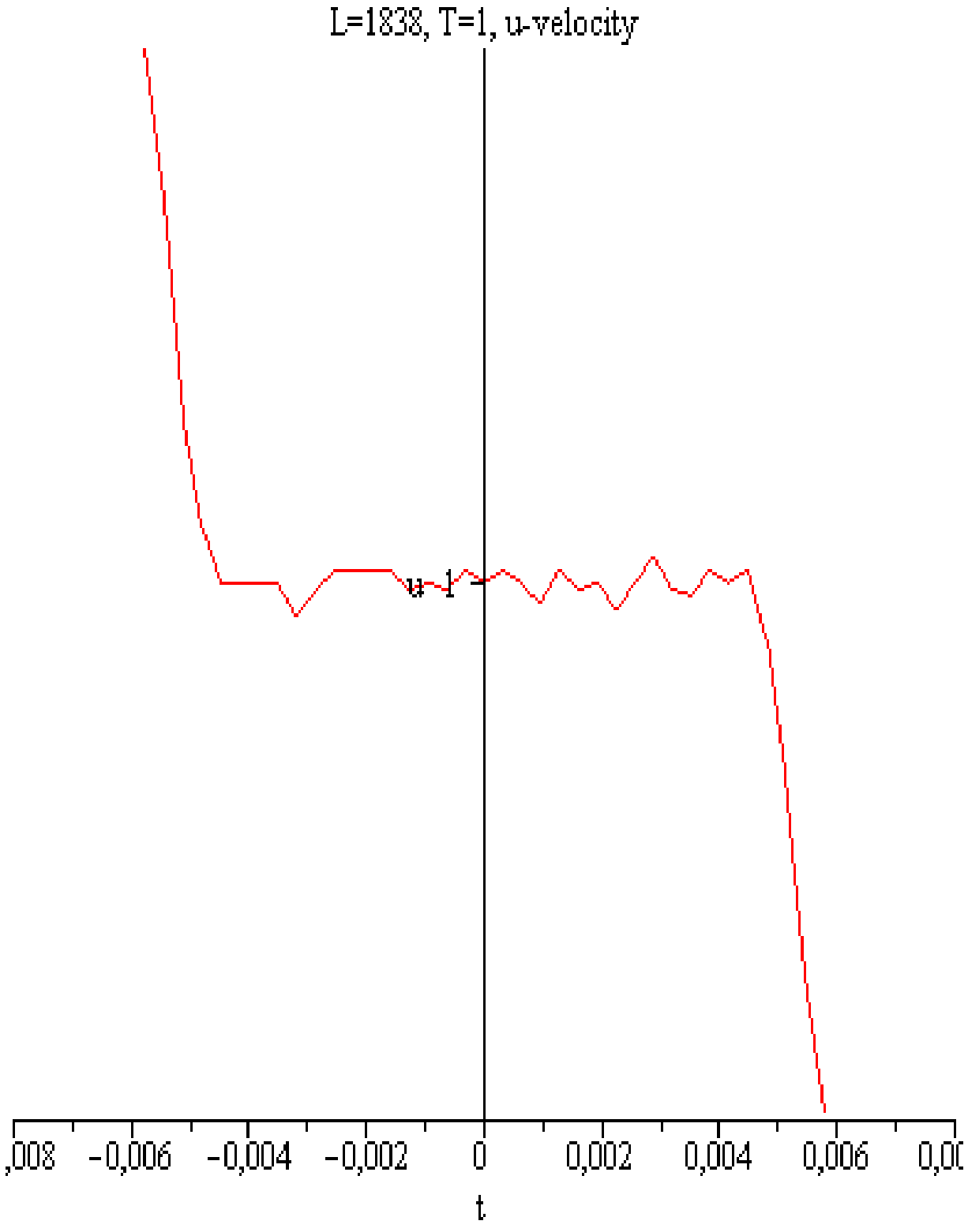

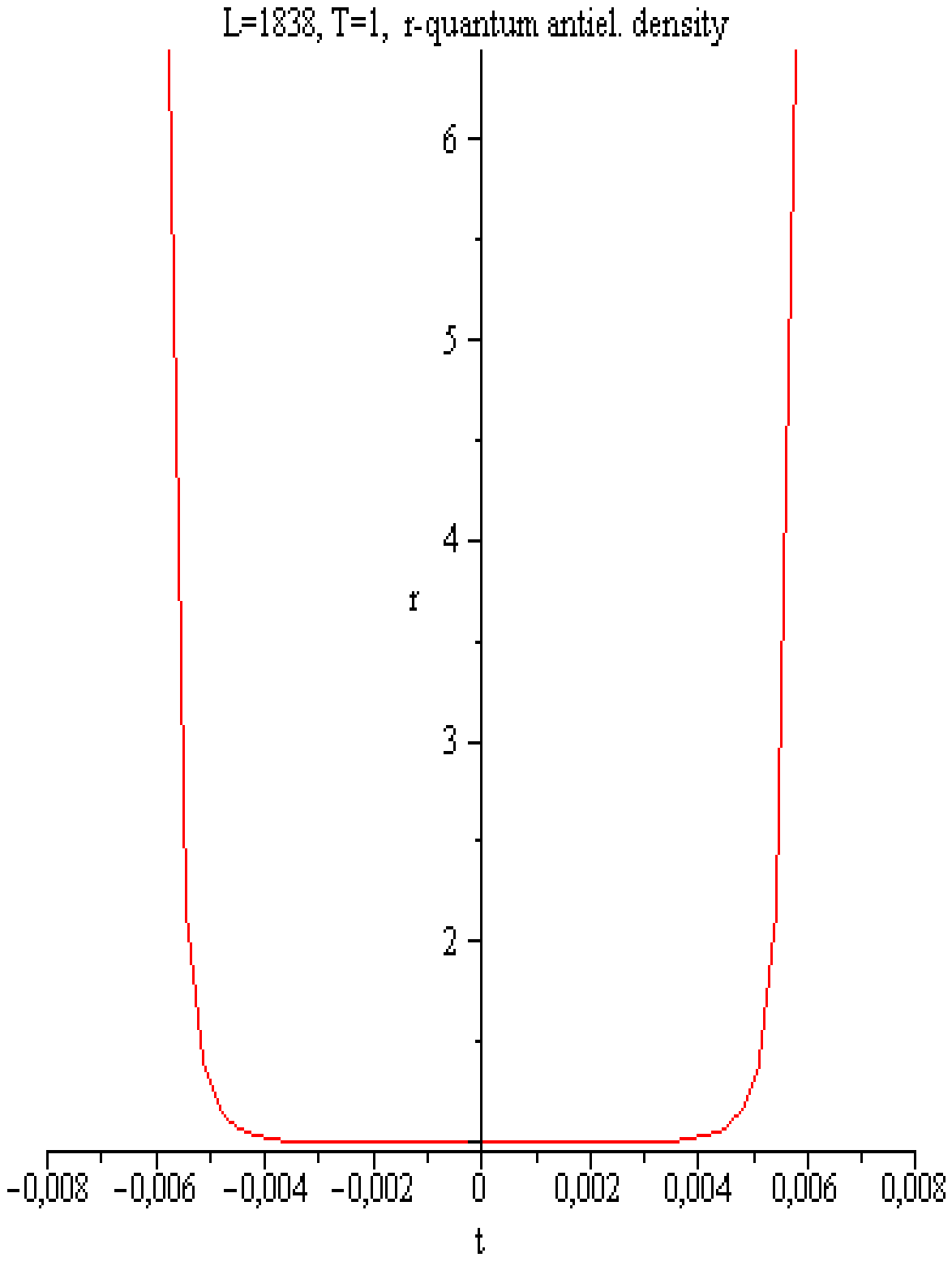

Figure 22. u- velocity $\tilde{u}$ in quantum soliton. Figure 23. r- density $\tilde{\rho}_i$ , in quantum soliton.

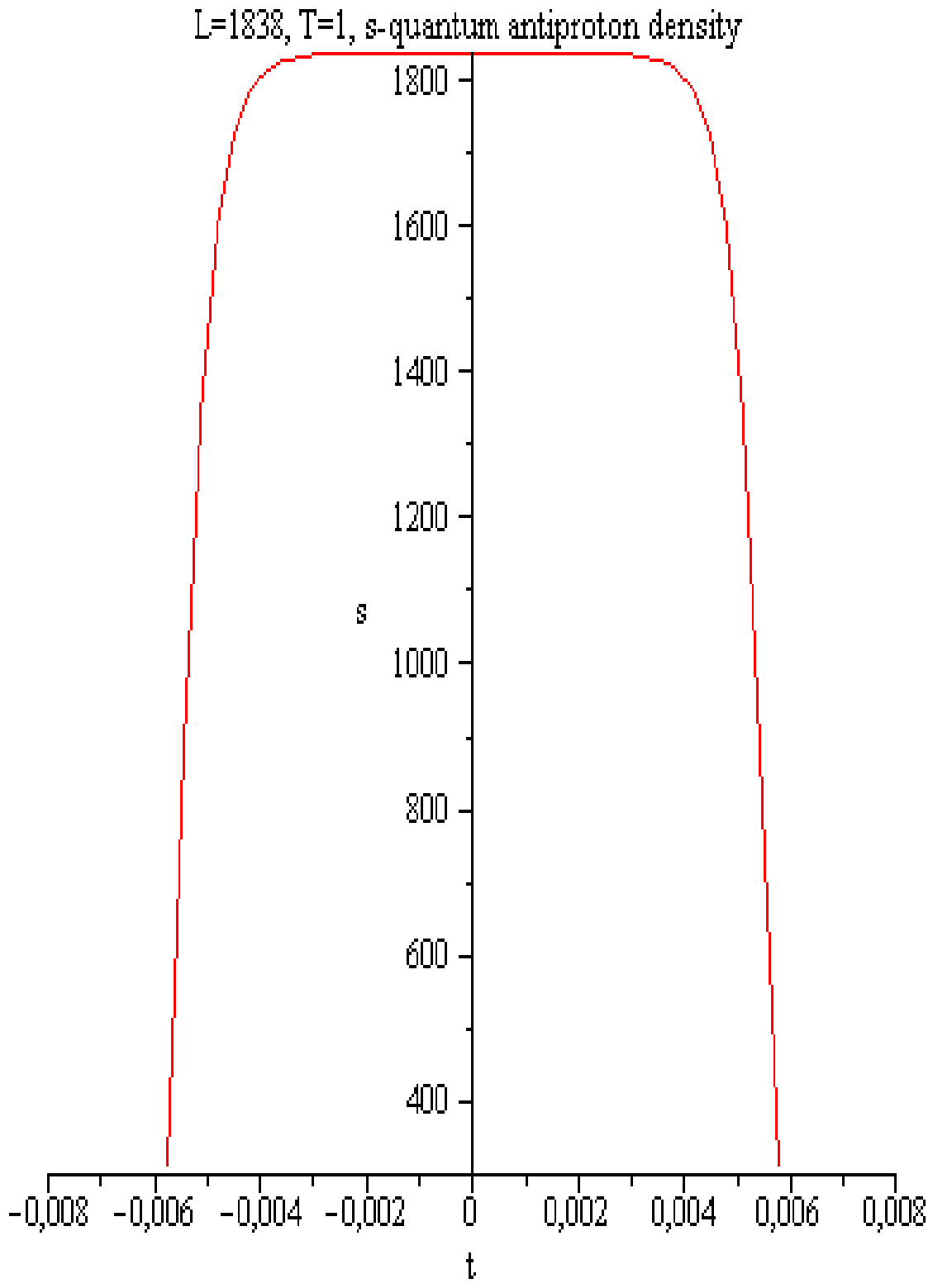

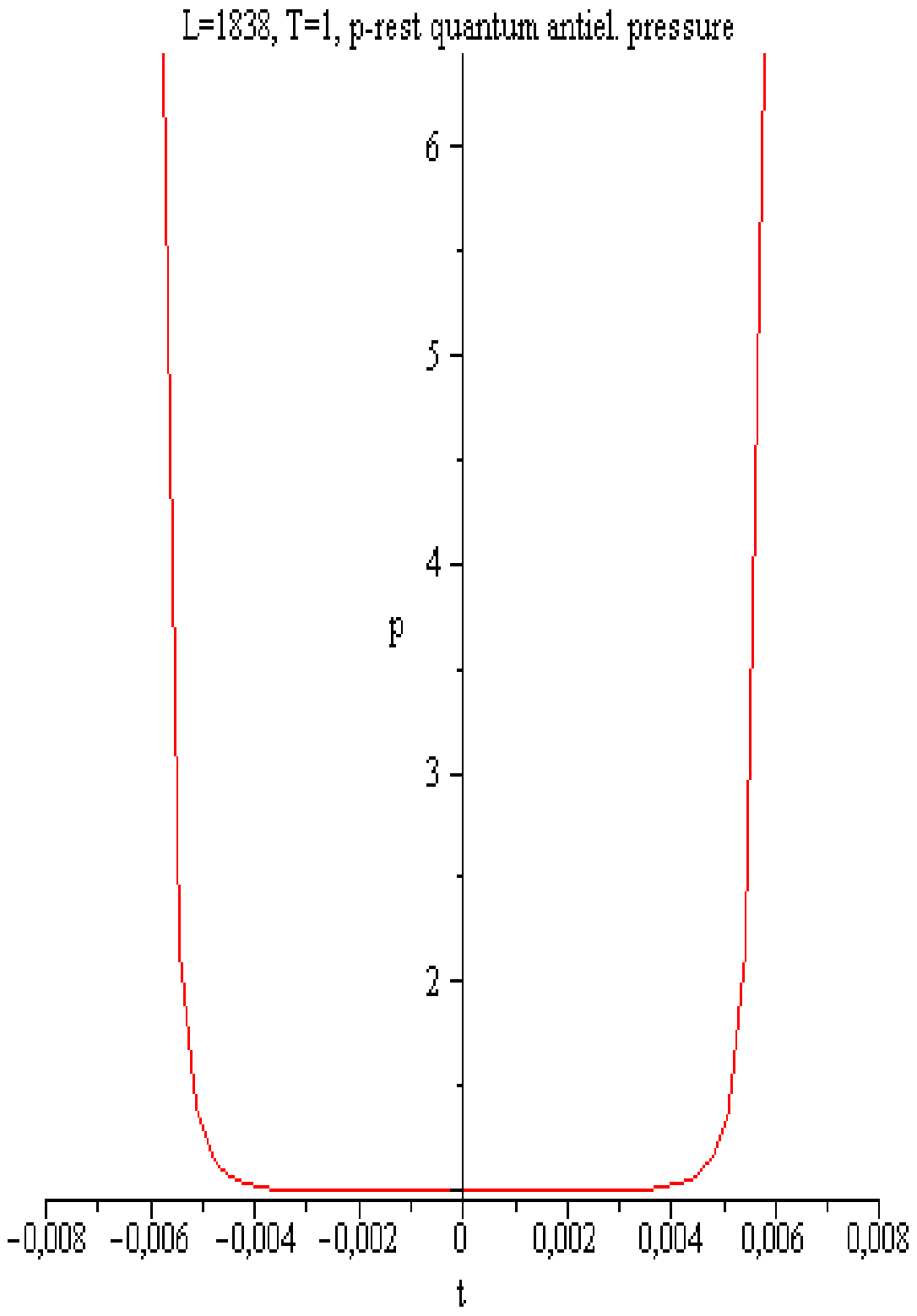

Figure 25. s- density $\tilde{\rho}_e$ - in quantum soliton. Figure 26. p - pressure $\tilde{p}_i$ in quantum soliton.

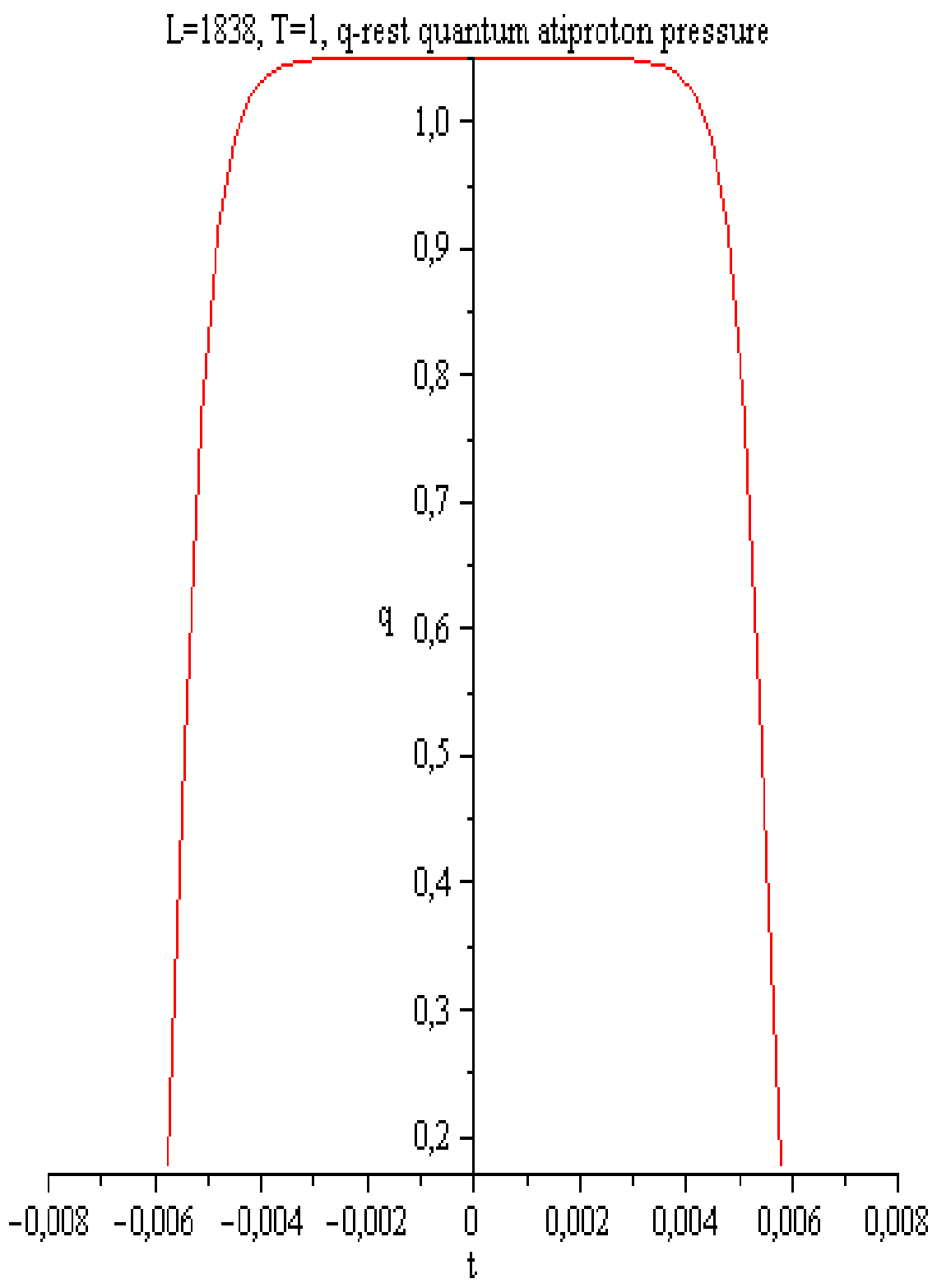

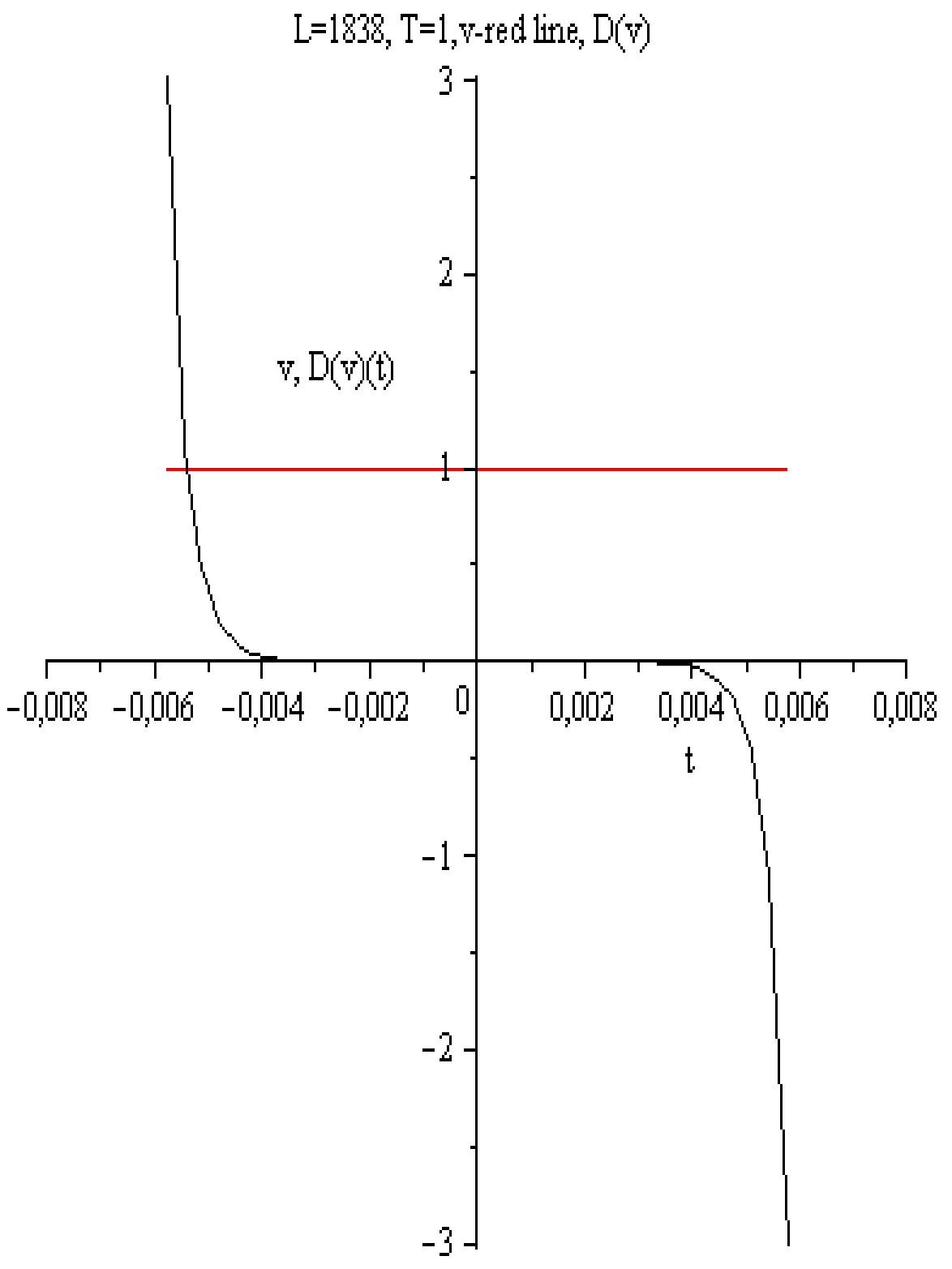

Figure 27. $q$ - pressure $\tilde{p}_e$ in quantum soliton.

Figure 28. v - self-consistent potential $\tilde{\varphi}$ and $\partial\tilde{\varphi}/\partial\tilde{\xi}$ in quantum soliton.

Important to notice that all elements of soliton are moving with the same self-consistent constant velocity if the initial perturbation $\tilde{u}(0)=1$ corresponds to phase velocity.

Quantum solitons are discovered with the help of generalized quantum hydrodynamics (GQH). The solitons have the character of quantum objects (with positive or negative shells) which reach stability as result of equalizing of corresponding pressure of the non-local origin and the self-consistent electric forces. If the initial rest pressures of non-local origin for the positive and negative components are equal each other, the stable soliton does not exist.

Four cases of the principal significance are considered – creations of atoms and anti-atoms for the cases $\tilde{p}_i(0)<\tilde{p}_e(0)$ and $\tilde{p}_i(0)>\tilde{p}_e(0)$. For both cases mathematical modeling for anti-matter leads to diminishing of the linear size of combined soliton in $2\cdot10^3$ times and cross sections in ~ $4\cdot10^6$ times in comparison with atoms of regular matter.

Now the conclusion of the principal significance. Matter and anti-matter atoms after creation in the Big Bang activity are involved in the process of the collisional relaxation. But the cross sections of anti-matter atoms and inversed atoms of regular matter have so small cross sections that they are leaving the physical system. Now anti-matter forms the outer edge of visible Universe.

Several teams of physicists (including the CERN command) have taken a big step towards making the first spectroscopic measurements on a beam of antihydrogen atoms. Precise measurements could be used to study a fundamental quantum transformation known as the charge-parity-time (CPT) operation. The formulated results of delivered generalized quantum hydrodynamics should be taken into account.

But regular matter takes part in the Hubble expansion, the corresponding theory is considered in the next item from position of the non-local physics.

## 5. Plasma – gravitational analogy in the generalized theory of Landau damping. Hubble expansion.

My aim consists in the application of plasma – gravitational analogy for the effect of Hubble expansion using the generalized theory of Landau damping and the generalized Boltzmann physical kinetics developed by me (Alexeev 2009ab). The collisionless damping of electron plasma waves was predicted by Landau in 1946 (Landau, 1946) and later was confirmed experimentally. Landau damping plays a significant role in many electronics experiments and belongs to the most well known phenomenon in statistical physics of ionized gases. The physical origin of the collisionless Landau wave damping is simple. Really, if individual electron moves in the periodic electric field, this electron can diminish its energy (electron velocity larger than phase velocity of wave) or receive additional energy from the wave (electron velocity less than phase velocity of wave). Then the total energy balance for a swarm of electrons depends on quantity of "cold" and "hot" electrons. For the Maxwellian distribution function, the quantity of "cold" electrons is more than quantity of "hot" electrons. This fact leads to, so-called, the collisionless Landau damping of the electric field perturbation. In spite of transparent physical sense, the effect of Landau damping has continued to be of great interest to theorist as well. Much of this interest is connected with counterintuitive nature of result itself coupled with the rather abstruse mathematical nature of Landau's original derivation (including so-called Landau's rule of complex integral calculation). Moreover, for these reasons there were publications containing some controversy over the reality of the phenomenon. In papers (Alexeev 2009ab) the difficulties originated by Landau's derivation were clarified. The mentioned consideration leads to another solution of Vlasov - Landau equation, these ones in agreement with data of experiments. The problem solved in this article consists in consideration of the generalized theory of Landau damping in gravitating systems from viewpoint of Generalized Boltzmann Physical Kinetics and non-local physics. The influence of the particle collisions is taken into account.

Plasma – gravitational analogy is well-known and frequently used effect in physical kinetics. The origin of analogy is simple and is connected with analogy between Coulomb law and Newtonian law of gravitation. From other side electrical charges can have different signs whereas there is just one kind of "gravitational charge" (i.e. masses of particles) corresponding to the force of attraction. This fact leads to the extremely important distinctions in formulation of the generalized theory of Landau damping in gravitational media. In the following, we intend to use the classical non-relativistic Newtonian law of gravitation

$$\mathbf{F}_{21} = \gamma_N \frac{m_1 m_2}{r_{12}^2} \frac{\mathbf{r}_{12}}{r_{12}}, \tag{5.1}$$

where $\mathbf{F}_{21}$ is the force acting on the particle "1" from the particle "2", $\mathbf{r}_{12}$ is vector directed from the center-of-mass of the particle "1" to the particle "2", $\gamma_N$ is gravitational constant $\gamma_N = 6.6 \cdot 10^{-8} cm^3 /(g \cdot s^2)$; the corresponding force $\mathbf{g}_{21}$ per mass unit is

$$\mathbf{g}_{21} = \mathbf{F}_{21} / m_1. \tag{5.2}$$

The flux

$$\Phi = \int_S g_n dS \tag{5.3}$$

for closed surface $S$ can be calculated using (5.2); one obtains

$$\int_S g_n dS = -4\pi\gamma_N \int_V \rho^a dV, \tag{5.4}$$

where $\rho^a$ is density *inside* of volume $V$ bounded by the surface $S$. As usual, Eq. (3.4) can be rewritten as

$$\int_V \left(div\ \mathbf{g} + 4\pi\gamma_N \rho^a\right) dV = 0. \tag{5.5}$$

The definite integral (5.5) is equal to zero for arbitrary volume $V$, then

$$div\ \mathbf{g} = -4\pi\gamma_N \rho^a , \qquad (5.6)$$

and after introduction the gravitational potential $\Psi$

$$\mathbf{g} = -\partial\Psi / \partial\mathbf{r} \qquad (5.7)$$

we reach the known Poisson equation

$$\Delta\Psi = 4\pi\gamma_N \rho^a . \qquad (5.8)$$

Generalized Boltzmann physical kinetics leads to possibility to calculate the density $\rho^a$ using the density $\rho$ (obtained with the help of the one particle DF $f$) and the fluctuation term $\rho^{fl}$. All fluctuation terms in the GBE theory were tabulated (Alexeev,       ) and for $\rho^{fl}$ we have

$$\rho^{fl} = \tau\left(\frac{\partial\rho}{\partial t} + \frac{\partial}{\partial\mathbf{r}}\cdot\rho\mathbf{v}_0\right), \qquad (5.9)$$

where $\mathbf{v}_0$ is hydrodynamic velocity. After substitution of $\rho^{fl}$ in (3.8) one obtains

$$\Delta\Psi = 4\pi\gamma_N\left[\rho - \tau\left(\frac{\partial\rho}{\partial t} + \frac{\partial}{\partial\mathbf{r}}\cdot\rho\mathbf{v}_0\right)\right]. \qquad (5.10)$$

From equations (5.6) - (5.9) follow that classical Newtonian field equation

$$\Delta\Psi = 4\pi\gamma_N \rho \qquad (5.11)$$

valid only for situation when the fluctuations terms can be omitted and then

$$\rho = \rho^a . \qquad (5.12)$$

This condition can be considered as the simplest closure condition, but in the general case, the other hydrodynamic equations should be involved into consideration because Eq. (5.10) contains hydrodynamic velocity $\mathbf{v}_0$. As result, one obtains the system of moment equations, i.e. gravitation equation

$$\frac{\partial}{\partial\mathbf{r}}\cdot\mathbf{g} = -4\pi\gamma_N\left[\rho - \tau\left(\frac{\partial\rho}{\partial t} + \frac{\partial}{\partial\mathbf{r}}\cdot\rho\mathbf{v}_0\right)\right], \qquad (5.13)$$

and generalized continuity, motion and energy equations which can be further applied to the theory of the rotation curves of spiral galaxies.

The following mathematical transformations will be obtained on the level of the generalized theory of Landau damping based on the generalized Boltzmann equation (GBE), and need in some preliminary additional explanations from viewpoint of so-called dark energy and dark matter.

As it was mentioned above the accelerated cosmological expansion was discovered in direct astronomical observations. For explanation of this acceleration new idea was introduced in physics about existing of a force with the opposite sign which is called universal antigravitation. In the simplest interpretation, dark energy is related usually to the Einstein cosmological constant. In review (Chernin 2008) the modified Newton force is written as

$$F(r) = -\frac{\gamma_N M}{r^2} + \frac{8\pi\gamma_N}{3}\rho_{\mathrm{v}} r , \qquad (5.14)$$

where $\rho_{\mathrm{v}}$ is the Einstein – Gliner vacuum density introduced also in (Gliner 2002). The problem can be solved without the ideology of the Einstein – Gliner vacuum. However for us it is interesting the interpretation of the modified law (5.14). In the limit of large distances, the influence of central mass $M$ becomes negligibly small and the field of forces is determined only by the second term in the right side of (5.14). It follows from relation (5.14) that there is a "equilibrium" distance $r_{\mathrm{v}}$, at which the sum of the gravitation and antigravitation forces is equal to zero. In other words $r_{\mathrm{v}}$ is "the zero-gravitational radius". For so called Local Group of galaxies estimation of $r_{\mathrm{v}}$ is about 1Mpc, (Chernin 2008). There are no theoretical methods of the density $\rho_{\text{v}}$ calculation. Obviously, the second term in relation (5.14) should be defined as result of solution of the self-consistent gravitational problem.

Let us return now to the formulation of plasma-gravitational analogy in the frame of generalized theory of Landau damping. I intend to apply the GBE model with the aim to obtain the dispersion relation for one component gas placed in the self-consistent gravitational field and to consider effect antigravitation in the frame of the Newton theory of gravitation.

With this aim let us admit now that there is a gravitational perturbation $\delta\Psi$ in the system of particles connected with the density perturbation $\delta\rho$. These perturbations connected with the perturbation of DF in the system, which was before in the local equilibrium. In doing so, we will make the additional assumptions for simplification of the problem, namely:

(a) Consideration of the self-consistent gravitational field correspond to the area of the large distance $r$ (see relation (5.14)) from the central mass $M$ where the first term is not significant and in particular the problem correspond to the plane case. As mentioned above the second term should be defined as a self-consistent force of the Newtonian origin

$$F = -\frac{\partial \delta\Psi}{\partial x}. \tag{5.15}$$

(b) The integral collision term is written in the Bhatnagar - Gross - Krook (BGK) form

$$J = -\frac{f - f_0}{v_r^{-1}} \tag{5.16}$$

into the right-hand side of the GBE. Here, $f_0$ and $v_r^{-1} = \tau_r$ are respectively a certain equilibrium distribution function and the relaxation time.

(c) The evolution of particles in a self-consistent gravitational field corresponds to a non-stationary one-dimensional model, $u$ is the velocity component along the $x$ axis.

(d) The distribution function $f$ deviate little from its equilibrium counterpart $f_0$.

$$f = f_0(u) + \delta f(x,u,t). \tag{5.17}$$

(e) A wave number $k$ and a complex frequency $\omega$ $(\omega = \omega' + i\omega'')$are appropriate to the wave mode considered;

$$\delta f = \langle \delta f \rangle e^{i(kx-\omega t)}, \tag{5.18}$$

$$\delta\Psi = \langle \delta\varphi \rangle e^{i(kx-\omega t)}. \tag{5.19}$$

(f) The quadratic GBE terms determining the deviation from the equilibrium DF are neglected.

Under these assumptions listed above, the GBE is written as follows:

$$\frac{\partial f}{\partial t} + u\frac{\partial f}{\partial x} + F\frac{\partial f}{\partial u} - \tau\left\{\frac{\partial^2 f}{\partial t^2} + 2u\frac{\partial^2 f}{\partial t\partial x} + u^2\frac{\partial^2 f}{\partial x^2} + 2F\frac{\partial^2 f}{\partial t\partial u} + \right.$$
$$\left. + \frac{\partial F}{\partial t}\frac{\partial f}{\partial u} + F\frac{\partial f}{\partial x} + u\frac{\partial F}{\partial x}\frac{\partial f}{\partial u} + F^2\frac{\partial^2 f}{\partial u^2} + 2uF\frac{\partial^2 f}{\partial u\partial x}\right\} = -v_r\delta f \tag{5.20}$$

where the relations take place for the corresponding terms in Eq. (5.20)

$$\frac{\partial f}{\partial t} = -i\omega\delta f, \quad u\frac{\partial f}{\partial x} = iku\delta f, \quad F\frac{\partial f}{\partial u} = -\frac{\partial\delta\Psi}{\partial x}\frac{\partial f_0}{\partial u}, \quad \frac{\partial^2 f}{\partial t^2} = -\omega^2\delta f, \quad 2u\frac{\partial^2 f}{\partial t\partial x} = 2\omega uk\delta f,$$
$$u^2\frac{\partial^2 f}{\partial x^2} = -u^2k^2\delta f, \ 2F\frac{\partial^2 f}{\partial t\partial u} = 0, \frac{\partial F}{\partial t}\frac{\partial f}{\partial u} = -\frac{\partial}{\partial t}\frac{\partial\delta\Psi}{\partial x}\frac{\partial f}{\partial u} = -\omega k\delta\Psi\frac{\partial f_0}{\partial u}, \ F\frac{\partial f}{\partial x} = 0, \tag{5.21}$$
$$u\frac{\partial f}{\partial u}\frac{\partial F}{\partial x} = -u\frac{\partial f}{\partial u}\frac{\partial}{\partial x}\frac{\partial\delta\Psi}{\partial x} = k^2u\delta\Psi\frac{\partial f_0}{\partial u}, \ F^2\frac{\partial^2 f}{\partial u^2} = 0, \ \frac{\partial^2 f}{\partial u\partial x}2uF = 0.$$

We are concerned with developing (within the GBE framework) the dispersion relation for gravitational field, and substitution of (5.21) into Eq. (5.20) yields

$$\left\{i(ku-\omega) + v_r + \tau(ku-\omega)^2\right\}\langle\delta f\rangle - \langle\delta\Psi\rangle\frac{\partial f_0}{\partial u}k\{i + \tau(ku-\omega)\} = 0. \tag{5.22}$$

For the physical system under consideration, the influence of the collision term $v_r\langle\delta f\rangle$ is rather small. Using for this case the Poisson equation in the form (5.11) and then the relation

$$k^2\langle\delta\Psi\rangle = -4\pi\gamma_N\langle\delta n\rangle, \tag{5.23}$$

one obtains from (5.22), (5.23)

$$\langle\delta f\rangle = \frac{4\pi\gamma_N m}{k}\frac{\left[i-\tau(\omega-ku)\right]\frac{\partial f_0}{\partial u}}{i(\omega-ku)-\tau(\omega-ku)^2-v_r}\langle\delta n\rangle. \tag{5.24}$$

After integration over all $u$ we arrive at the dispersion relation

$$1 = \frac{4\pi\gamma_N m}{k}\int_{-\infty}^{+\infty}\frac{\frac{\partial f_0}{\partial u}\left[i-\tau(\omega-ku)\right]}{i(\omega-ku)-\tau(\omega-ku)^2-v_r}du. \tag{5.25}$$

Let us suppose that the velocity depending part of DF $f_0$ corresponds to the Maxwell DF. Then after differentiating in Eq. (5.25) under the sign of integral and some transformations we obtain the integral dispersion equation

$$1-\frac{1}{r_A^2k^2}\left[1-\sqrt{\frac{m}{2\pi k_BT}}\int_{-\infty}^{+\infty}\frac{\{[i-\tau(\omega-ku)]\omega-v_r\}e^{-mu^2/2k_BT}}{i(\omega-ku)-\tau(\omega-ku)^2-v_r}du\right]=0, \tag{5.26}$$

where

$$r_A = \sqrt{\frac{k_BT}{4\pi\gamma_N m^2 n}}. \tag{5.27}$$

Poisson equation (5.11) has the structure like the Poisson equation for the electrical potential, as result the relation for $r_A$ is analogous to the Debye - Hueckel radius $r_D = \sqrt{k_BT/(4\pi e^2 n)}$.

Introducing now the dimensionless variables

$$\hat{u} = u\sqrt{\frac{m}{2k_BT}},\ \hat{\omega} = \omega\frac{1}{k}\sqrt{\frac{m}{2k_BT}},\ \hat{v}_r = v_r\frac{1}{k}\sqrt{\frac{m}{2k_BT}},\ \hat{\tau} = \tau k\sqrt{\frac{2k_BT}{m}} \tag{5.28}$$

we can rewrite Eq (3.26) in the form

$$1-\frac{1}{r_A^2k^2}\left[1-\frac{1}{\sqrt{\pi}}\int_{-\infty}^{+\infty}\frac{\{[i-\hat{\tau}(\hat{\omega}-\hat{u})]\hat{\omega}-\hat{v}_r\}e^{-\hat{u}^2}}{i(\hat{\omega}-\hat{u})-\hat{\tau}(\hat{\omega}-\hat{u})^2-\hat{v}_r}d\hat{u}\right]=0. \tag{5.29}$$

Now consider a situation in which the denominator of the complex integrand in Eq. (5.29) becomes zero. The quadratic equation

$$\hat{\tau}y^2 - iy + \hat{v}_r = 0,\ y = \hat{\omega}-\hat{u} \tag{5.30}$$

has the roots

$$y_1 = \frac{i}{2\hat{\tau}}\left(1+\sqrt{1+4\hat{\tau}\hat{v}_r}\right),\ y_2 = \frac{i}{2\hat{\tau}}\left(1-\sqrt{1+4\hat{\tau}\hat{v}_r}\right). \tag{5.31}$$

Hence, Eqn (5.29) can be rewritten as

$$1-\frac{1}{r_A^2k^2}\left[1+\frac{1}{\tau\sqrt{\pi}}\int_{-\infty}^{+\infty}\frac{\{[i+\hat{\tau}(\hat{u}-\hat{\omega})]\hat{\omega}-\hat{v}_r\}e^{-\hat{u}^2}}{(\hat{u}-\hat{u}_1)(\hat{u}-\hat{u}_2)}d\hat{u}\right]=0 \tag{5.32}$$

where

$$\hat{u}_1 = \hat{\omega}-y_1,\ \hat{u}_2 = \hat{\omega}-y_2. \tag{5.33}$$

Let us transform equation (5.32) to the following one:

$$1-\frac{1}{r_A^2 k^2}\left\{1+\frac{1}{\sqrt{\pi}}\left[\left(\frac{i\widehat{v}_r+0.5\widehat{\omega}}{\sqrt{1+4\widetilde{\tau}\widehat{v}_r}}-0.5\widehat{\omega}\right)\int_{-\infty}^{+\infty}\frac{e^{-\widehat{u}^2}}{(\widehat{u}_1-\widehat{u})}d\widehat{u}-\right.\right.$$

$$\left.\left.-\left(\frac{i\widehat{v}_r+0.5\widehat{\omega}}{\sqrt{1+4\widetilde{\tau}\widehat{v}_r}}+0.5\widehat{\omega}\right)\int_{-\infty}^{+\infty}\frac{e^{-\widehat{u}^2}}{(\widehat{u}_2-\widehat{u})}d\widehat{u}\right]\right\}=0 \qquad (5.34)$$

Equation (5.34) contains improper Cauchy type integrals. From the theory of complex variables is known Cauchy's integral formula: if the function $f(z)$ is analytic inside and on a simple closed curve *C*, and $z_0$ is any point inside C, then

$$f(z_0)=-\frac{1}{2\pi i}\oint_C\frac{f(z)}{z_0-z}dz \qquad (5.35)$$

where *C* is traversed in the positive (counterclockwise) sense.

Let *C* be the boundary of a simple closed curve placed in lower half plane (for example a semicircle of radius *R*) with the corresponding element of real axis, $z_0$ is an interior point. As usual after adding to this semicircle a cross-cut connecting semicircle *C* with the interior circle (surrounding $z_0$) of the infinite small radius for analytic $f(z)$ the following formula obtains

$$\oint_C\frac{f(z)}{z_0-z}dz=-\int_{-R}^{R}\frac{f(\widetilde{x})}{z_0-\widetilde{x}}d\widetilde{x}+\int_{C_R}\frac{f(z)}{z_0-z}dz+2\pi i f(z_0), \qquad (5.36)$$

because the integrals along cross-cut cancel each other, ($z=\widetilde{x}+i\widetilde{y}$).

Analogous for upper half plane

$$\oint_C\frac{f(z)}{z_0-z}dz=\int_{-R}^{R}\frac{f(\widetilde{x})}{z_0-\widetilde{x}}d\widetilde{x}+\int_{C_R}\frac{f(z)}{z_0-z}dz+2\pi i f(z_0). \qquad (5.37)$$

The formulae (5.36), (5.37) could be used for calculation (including the case $R\to\infty$) of the integrals along the real axis with the help of the residual theory *for arbitrary* $z_0$ if analytical function $f(z)$ satisfies the special conditions of decreasing by $R\to\infty$.

Let us consider now integral $\int_{C_R}\frac{e^{-z^2}}{z_0-z}dz$. Generally speaking, for function $f(z)=e^{-z^2}$ Cauchy's conditions are not satisfied. Really for a point $z=\widetilde{x}+i\widetilde{y}$ this function is $f(z)=e^{\widetilde{y}^2-\widetilde{x}^2}[\cos(2\widetilde{x}\widetilde{y})-i\sin(2\widetilde{x}\widetilde{y})]$ and $f(z)$ is growing by $|\widetilde{y}|>|\widetilde{x}|$ for this part of $C_R$.

*But from physical point of view in* ***the linear problem*** *of interaction of individual particles* ***only*** *with waves of potential self-consistent gravitational field the natural assumption can be introduced that solution depends* ***only*** *of concrete* $z_0=\widehat{\omega}'+i\widehat{\omega}''$, *but does not depend of another possible modes of oscillations in physical system.*

It can be realized only if the calculations do not depend of choosing of contour $C_R$. This fact leads to the additional conditions, for lower half plane

$$\int_{-\infty}^{\infty}\frac{f(\widetilde{x})}{z_0-\widetilde{x}}d\widetilde{x}=2\pi i f(z_0), \qquad (5.38)$$

and for upper half plane

$$\int_{-\infty}^{\infty}\frac{f(\widetilde{x})}{z_0-\widetilde{x}}d\widetilde{x}=-2\pi i f(z_0). \qquad (5.39)$$

As it is shown in (Alexeev 2009ab) Landau approximation for improper integral contains in implicit form restrictions (valid only for close vicinity of $\tilde{x}$ -axis) for the choice of contour $C$; these restrictions lead to the continuous spectrum. The question arises, is it possible to find solutions of the equation (5.34) by the restrictions (5.38), (5.39)? In the following will be shown that the conditions (5.38), (5.39) together with (5.34) lead to the discrete spectrum of $z_0 = \hat{\omega}' + i\hat{\omega}''$ and from physical point of view conditions (5.38), (5.39) can be considered as condition of quantization.

The relations (5.38), (5.39) are the additional conditions which physical sense consists in the extraction of independent oscillations – oscillations which existence does not depend on presence of other oscillations in considering physical system.

Then equation (5.34) produces the dispersion relation, which admits a damped gravitational wave solution $(\hat{\omega}'' < 0)$ for small influence of the collision integral (see also (Alexeev 2009b)):

$$\mp e^{\hat{u}_2^2} \frac{1 - r_A^2 k^2}{2\sqrt{\pi}} = \frac{\hat{v}_r}{\sqrt{1 + 4\hat{\tau}\hat{v}_r}} - \frac{i\hat{\omega}}{2}\left(1 + \frac{1}{\sqrt{1 + 4\hat{\tau}\hat{v}_r}}\right). \tag{5.40}$$

where

$$\hat{u}_2^2 = \hat{\omega}'^2 - \hat{\omega}''^2 - \hat{\omega}'' \frac{\sqrt{1 + 4\hat{\tau}\hat{v}_r} - 1}{\hat{\tau}} - \frac{1 + 2\hat{\tau}\hat{v}_r - \sqrt{1 + 4\hat{\tau}\hat{v}_r}}{2\hat{\tau}^2} + i\left(2\hat{\omega}'' + \frac{\sqrt{1 + 4\hat{\tau}\hat{v}_r} - 1}{\hat{\tau}}\right)\hat{\omega}'. \tag{5.41}$$

The time of the collision relaxation $\tau_{rel} = v_r^{-1}$ for gravitational physical system can be estimated in terms of the mean time $\tau$ between close collisions and the Coulomb logarithm:

$$\tau\, v_r = \Lambda,\ \hat{\tau}\, \hat{v}_r = \Lambda. \tag{5.42}$$

We separate the real and imaginary parts in equation (5.40). One obtains for the real part

$$\begin{aligned}
&\mp \frac{1 - r_a^2 k^2}{2\sqrt{\pi}} \exp\left\{\hat{\omega}'^2 - \hat{\omega}''^2 - \hat{\omega}''\hat{v}_r \frac{\sqrt{1 + 4\Lambda} - 1}{\Lambda} - \hat{v}_r^2 \frac{1 + 2\Lambda - \sqrt{1 + 4\Lambda}}{2\Lambda^2}\right\} = \\
&= \left[\frac{\hat{v}_r}{\sqrt{1 + 4\Lambda}} + 0.5\hat{\omega}'' + \frac{0.5\hat{\omega}''}{\sqrt{1 + 4\Lambda}}\right] \cos\left[\hat{\omega}'\left(2\hat{\omega}'' + \hat{v}_r \frac{\sqrt{1 + 4\Lambda} - 1}{\Lambda}\right)\right] - \\
&- 0.5\hat{\omega}'\left[1 + \frac{1}{\sqrt{1 + 4\Lambda}}\right] \sin\left[\hat{\omega}'\left(2\hat{\omega}'' + \hat{v}_r \frac{\sqrt{1 + 4\Lambda} - 1}{\Lambda}\right)\right].
\end{aligned} \tag{5.43}$$

Similarly, for the imaginary part we find

$$\begin{aligned}
&0.5\hat{\omega}'\left[1 + \frac{1}{\sqrt{1 + 4\Lambda}}\right] \cos\left[\hat{\omega}'\left(2\hat{\omega}'' + \hat{v}_r \frac{\sqrt{1 + 4\Lambda} - 1}{\Lambda}\right)\right] + \\
&+ \left[\frac{\hat{v}_r}{\sqrt{1 + 4\Lambda}} + 0.5\hat{\omega}'' + \frac{0.5\hat{\omega}''}{\sqrt{1 + 4\Lambda}}\right] \sin\left[\hat{\omega}'\left(2\hat{\omega}'' + \hat{v}_r \frac{\sqrt{1 + 4\Lambda} - 1}{\Lambda}\right)\right] = 0.
\end{aligned} \tag{5.44}$$

Coulomb logarithm $\Lambda$ is large for such objects like galaxies, the typical value $\Lambda \sim 200$ and the system of equations (5.43), (5.44) for the large Coulomb logarithm $\Lambda$ simplifies to

$$\mp \frac{1 - r_a^2 k^2}{\sqrt{\pi}} e^{\hat{\omega}'^2 - \hat{\omega}''^2} = \hat{\omega}'' \cos(2\hat{\omega}'\hat{\omega}'') - \hat{\omega}' \sin(2\hat{\omega}'\hat{\omega}''), \tag{5.45}$$

$$\hat{\omega}' \cos(2\hat{\omega}'\hat{\omega}'') + \hat{\omega}'' \sin(2\hat{\omega}'\hat{\omega}'') = 0. \tag{5.46}$$

Let us introduce the notation

$$\alpha = 2\hat{\omega}'\hat{\omega}'', \ \beta = 1 - r_a^2 k^2, \tag{5.47}$$

we obtain the universal equation

$$-e^{\sigma \ \cot\sigma} \sin\sigma = \frac{\pi}{2\beta^2}\sigma, \tag{5.48}$$

where $\sigma = -2\alpha = -4\hat{\omega}'\hat{\omega}''$. This equation does not depend on the sign in front of parameter $\beta$ in (5.45). The exact solution of equation (5.48) can be found with the help of the $W$-function of Lambert

$$\sigma_n = \mathrm{Im}\left[W_n\left(\frac{2\beta^2}{\pi}\right)\right], \tag{5.49}$$

frequencies $\hat{\omega}_n', \hat{\omega}_n''$ are (see also (5.18), (5.28))

$$\omega_n' = k\sqrt{-\frac{k_B T}{2m}\sigma_n \tan\frac{\sigma_n}{2}}\ ,.\ \omega_n'' = -k\sqrt{-\frac{k_B T}{2m}\sigma_n \cot\frac{\sigma_n}{2}} \tag{5.50}$$

In asymptotic for large entire positive $n$ (singular point $r_A k = 1$ is considered further in this section)

$$\sigma_n = \left(n + \frac{1}{2}\right)\pi,\ \hat{\omega}_n' = \frac{\sqrt{\sigma_n}}{2} = \frac{1}{2}\sqrt{\pi\left(n + \frac{1}{2}\right)},\ \hat{\omega}_n'' = -\frac{\sqrt{\sigma_n}}{2} = -\frac{1}{2}\sqrt{\pi\left(n + \frac{1}{2}\right)}. \tag{5.51}$$

The exact solution for the $n$ – th discrete solution from the spectrum of oscillations follows from (5.49), (5.50):

$$\hat{\omega}_n = \frac{1}{2}\sqrt{-\mathrm{Im}\left[W_n\left(\frac{2\left(1 - r_A^2 k^2\right)^2}{\pi}\right)\right]\tan\left[\frac{1}{2}\mathrm{Im}\left[W_n\left(\frac{2\left(1 - r_A^2 k^2\right)^2}{\pi}\right)\right]\right]} -$$
$$- \frac{i}{2}\sqrt{-\mathrm{Im}\left[W_n\left(\frac{2\left(1 - r_A^2 k^2\right)^2}{\pi}\right)\right]\cot\left[\frac{1}{2}\mathrm{Im}\left[W_n\left(\frac{2\left(1 - r_A^2 k^2\right)^2}{\pi}\right)\right]\right]}. \tag{5.52}$$

The square of the oscillation frequency of the longitudinal gravitational waves $\hat{\omega}_n'^2$ is proportional to the wave energy. Hence, the energy of waves is quantized, and as $n$ grows one obtains the asymptotic expression analogous to quantum levels of quantum oscillator in one dimension

$$\hat{\omega}_n'^2 = \frac{\pi}{4}\left(n + \frac{1}{2}\right), \tag{5.53}$$

the squares of possible dimensionless frequencies become equally spaced:

$$\hat{\omega}_{n+1}'^2 - \hat{\omega}_n'^2 = \frac{\pi}{4}. \tag{5.54}$$

or

$$\omega_{n+1}'^2 - \omega_n'^2 = \frac{\pi}{2}k^2 \frac{k_B T}{m}. \tag{5.55}$$

This difference can be connected with energy of Newtonian graviton. Figures 29 and 30 reflect the result of calculations for 200 discrete levels for the case of the large Coulomb logarithm $\Lambda$. For high levels this spectrum contains many very close equidistant curves with partly practically straight lines, which human eyes can perceive as background. Moreover plotter from the technical point of view has no possibility to reflect the small curvature of lines approximating this curvature as a long step. My suggestion is to turn this shortcoming into merit in explication of topology of high quantum levels in quantum systems.

Really, extremely interesting that this (from the first glance) grave shortcoming of plotters lead to the automatic construction of approximation for derivatives $d(r_D k)/d\hat{\omega}'$ and $d(r_D k)/d\hat{\omega}''$.

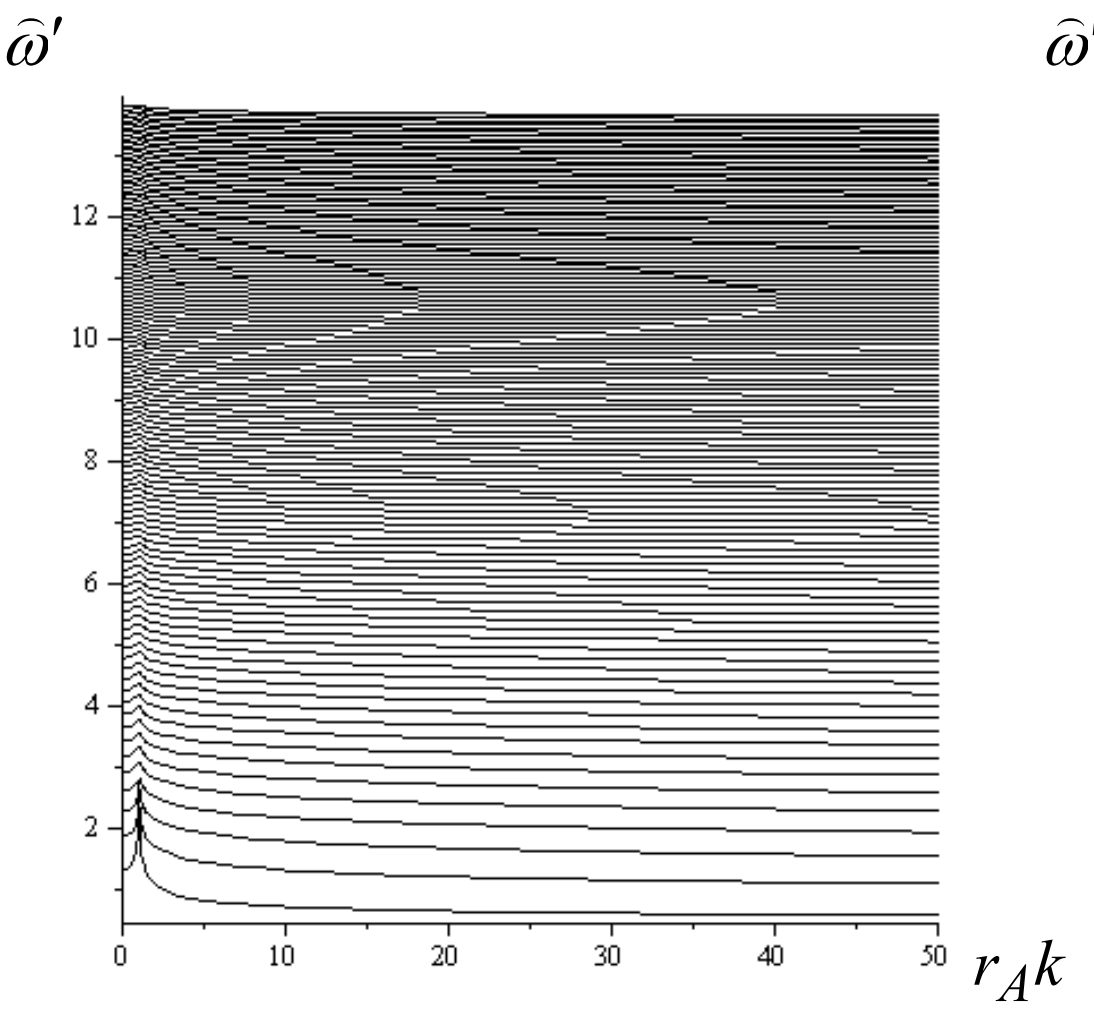

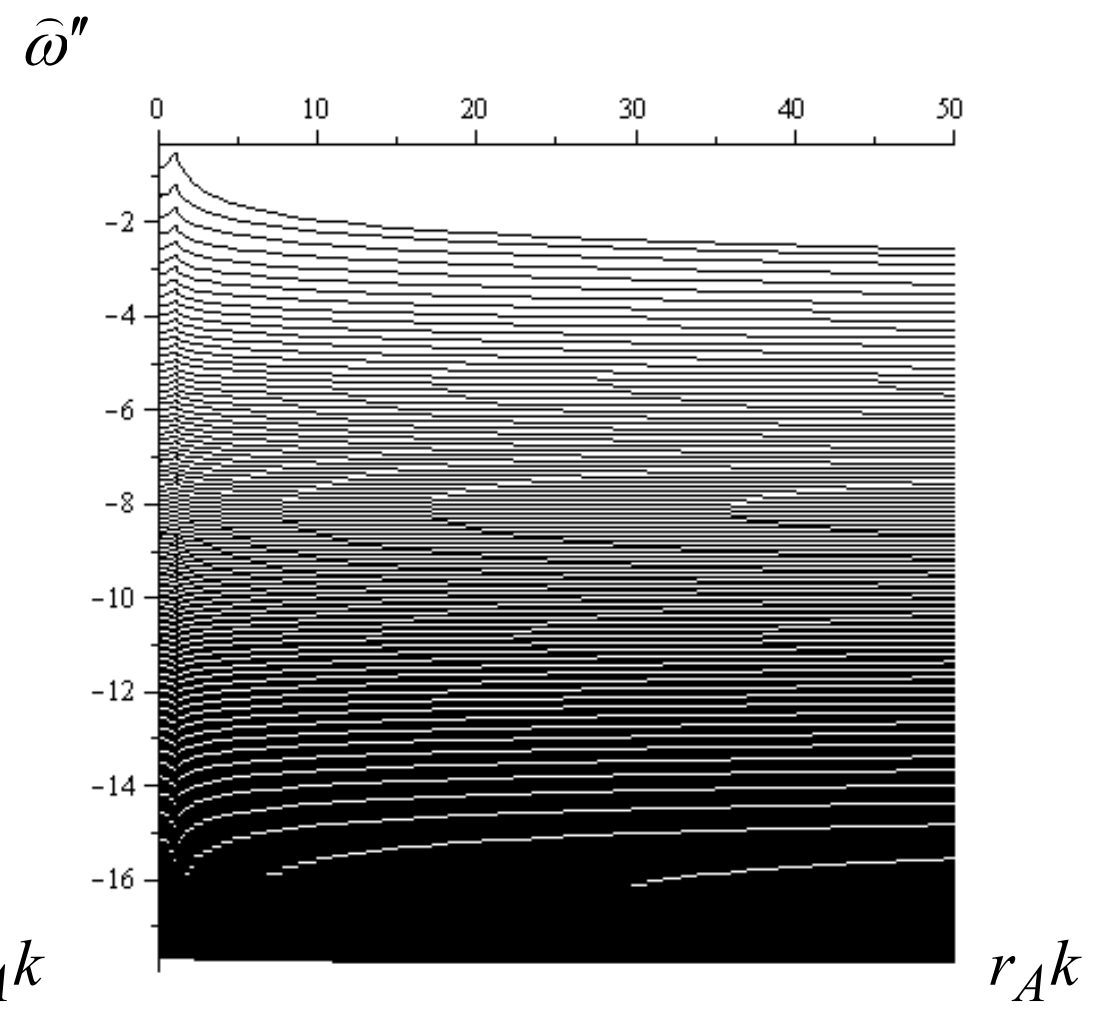

Fig. 29. The dimensionless frequency $\hat{\omega}'$ versus parameter $r_A k$, (left).

Fig. 30. The dimensionless frequency $\hat{\omega}''$ versus parameter $r_A k$, (right).

You can see this very complicated topology of curves in Figs. 29, 30 including the discrete spectrum of the bell-like curves in the mentioned figures. This singularity is connected with the existence of generalized derivatives $d(r_A k)/d\hat{\omega}'$, $d(r_A k)/d\hat{\omega}''$ for discontinuous functions. This effect has no attitude to the mathematical programming and looks in the definite sense like effect of “shroud of Christ” – self-organization of visible information in the human conscience. Enlarging of scaling shows that the complicated curves topology exists also in the black domain. Then figures 29, 30 can be used for understanding of the future development of events in physical system after the initial linear stage. For example Fig. 29 shows the discrete set of frequencies which vicinity corresponds to passing over from abnormal to normal dispersion (for example, by $\hat{\omega}' \sim 7$) for discrete systems of $r_A k$. Of course the non-linear stage needs the special investigation with using of another methods including the method of direct mathematical modeling It seems that the curves of high levels have different topology in comparison with the low levels. Nevertheless, it is far from reality, the high-level frequencies have the same character features as low frequencies. The highest levels are placed so close to each other that we see dark background on on figures 30 – 32.

It is of interest to investigate the singular point where

$$r_A k = 1. \tag{5.56}$$

Note the solution $\sigma \to \pi + 0$ and therefore $\hat{\omega}' \to \infty$ and $\hat{\omega}'' \to 0$. But phase velocity of wave $u_\phi = \omega' r_a$ and phase velocity of gravitational wave turns into infinity (in the frame of non-relativistic theory) and damping is equal to zero. In vicinity of $r_A k = 1$ one obtains “gravitational window” with increasing of frequency $\hat{\omega}'_n$ and decreasing of damping; the corresponding wave lengths $\lambda_A$ is

$$\lambda_A = 2\pi r_A . \qquad (5.57)$$

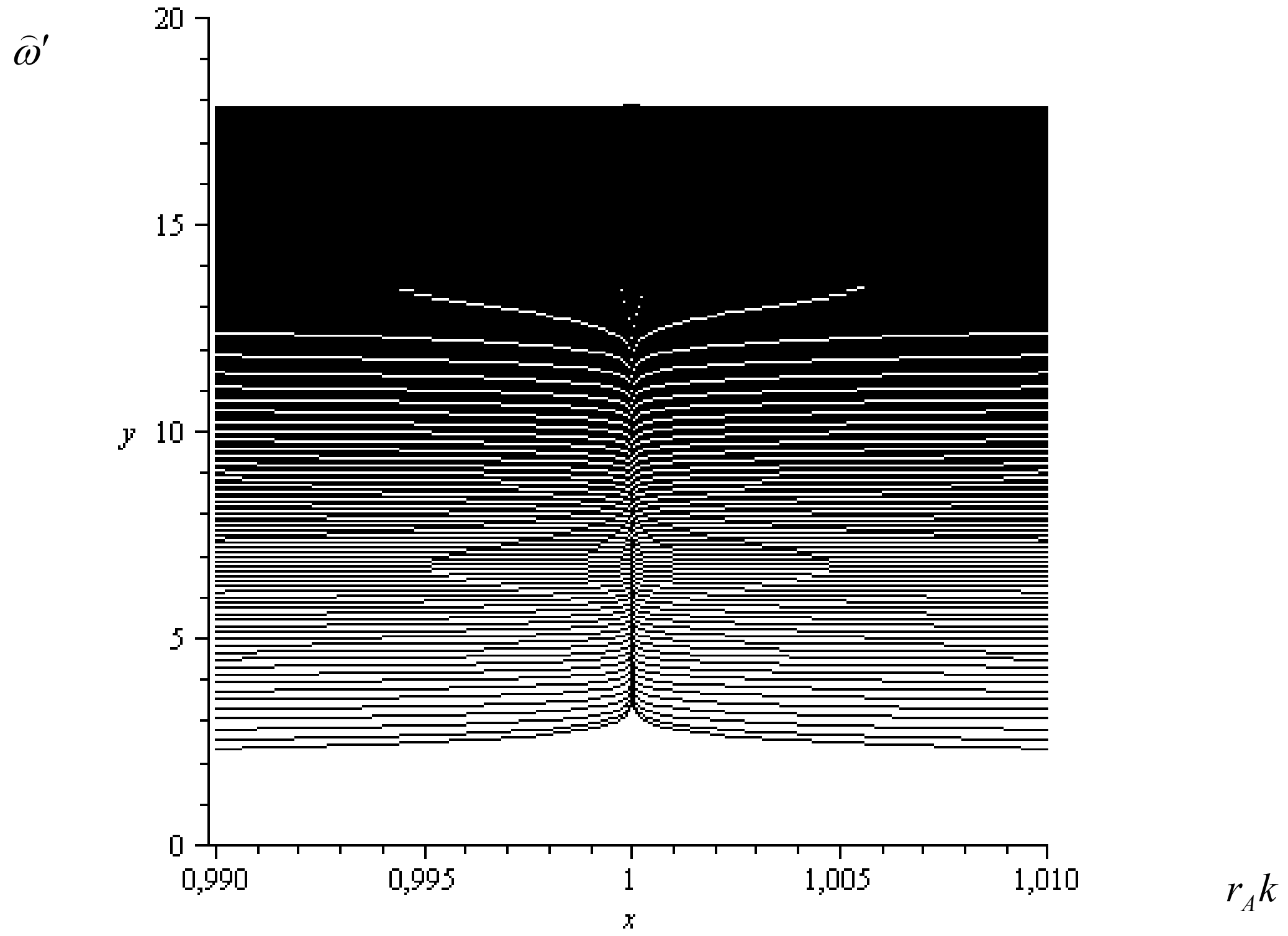

Figure 31. Topology of the dispersion curves $\widehat{\omega}'$ in the vicinity of the gravitational window.

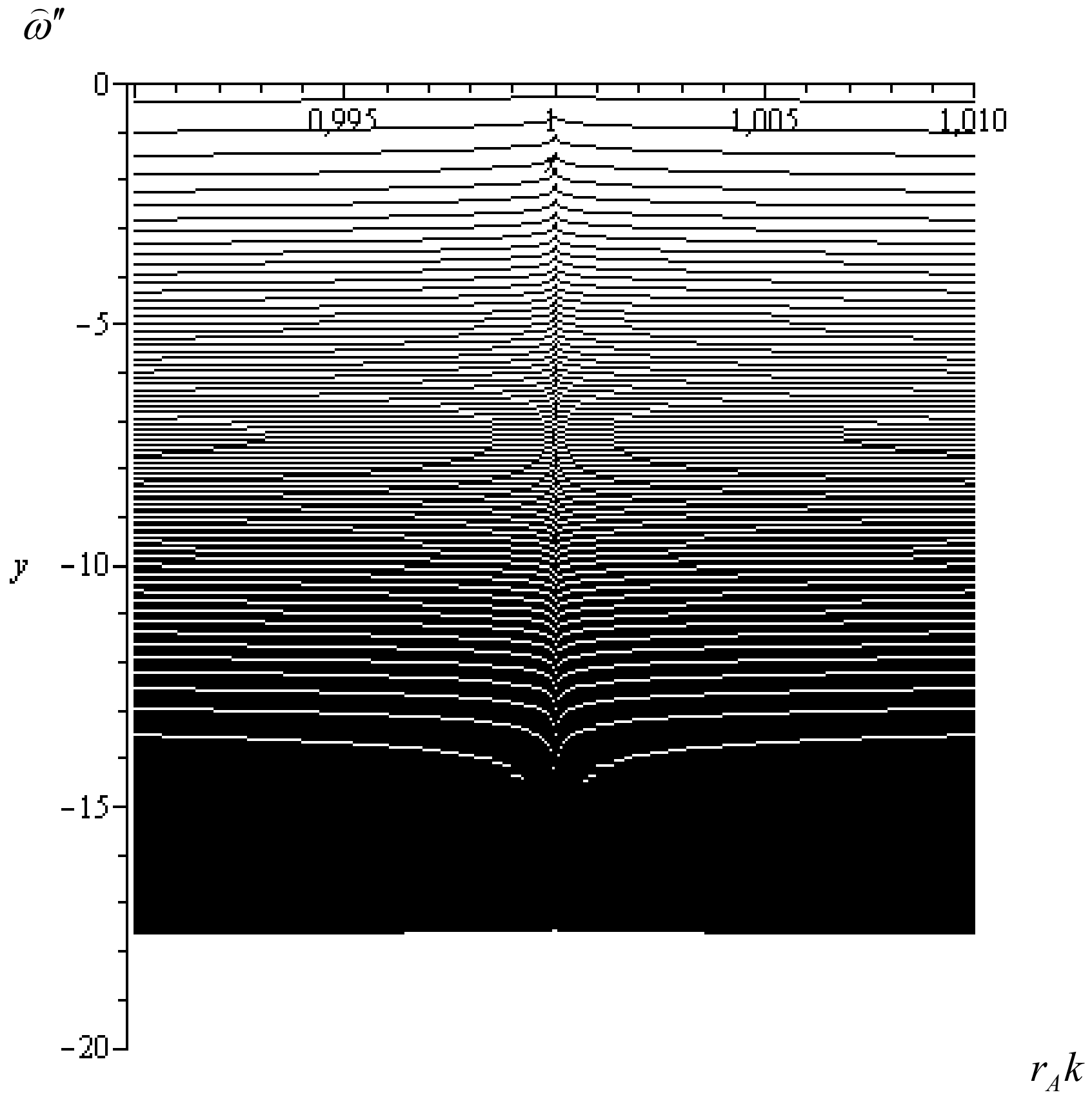

Figure 32. Topology of the dispersion curves $\widehat{\omega}''$ in the vicinity of the gravitational window.

Figures 31, 32 reflect the topology of the dispersion curves in the vicinity of the gravitational window.

Let us make some estimates. The mean density (calculated via luminous matter in stars and galaxies) $\rho \approx (0.01 - 0.02)\rho_c$, where $\rho_c = 0.47 \cdot 10^{-29}$ g/cm$^3$. Using also for estimates $T = 3$ K, $n = 0.3 \cdot 10^{-7}$ cm$^{-3}$, $m = 1.6 \cdot 10^{-24} g$, one obtains from relation (5.57) $r_A = 0.82 \cdot 10^{23}$cm $= 0.027$ Mpc and $\lambda_A = 0.17$ Mpc.

Now we can create the physical picture leading to the Hubble flow. *The main origin of Hubble effect (including the matter expansion with acceleration) is self – catching of expanding matter by the self – consistent Newtonian gravitational field in conditions of weak influence of central massive bodies.*

The relation (5.57) is the condition of this self – catching as result of explosion with appearance of waves for which the wave lengths is of about $\lambda_A$. Gravitational self – catching takes place for Big Bang having given birth to the global expansion of Universe, but also for Little Bang (Chernin 2008) in so-called Local Group of galaxies. The gravitationally bound system of the Local Group can exist only within region $r < r_v$. In this case $r_v$ need not to be connected with the modified Newton force and can be considered as character value where gravitation of the central mass is not significant. Outside the Group at distances $r > r_v$ the Hubble flow of Galaxies starts. This "no reentering radius" was found as result of direct observations of the Local Group: $r_v$ is of order (or less) 1 Mpc. This value is in good coincidence with the length $\lambda_A = 0.17$ Mpc. Some important remarks should be done:

1. Effects of gravitational self-catching should be typical for Universe. The existence of "Hubble boxes" discussed in review (Chernin 2008) as typical blocks of the nearby Universe.
2. As it follows from figures 29-32 effect of gravitational self-catching exists *in finite region* close to $r_A k = 1$, the phase velocity is defined by discrete spectrum $u_{\phi,n} = \tilde{\omega}'_n \sqrt{2k_B T / m}$.
3. Gravitational self-catching can be significant in the Earth conditions.

The last remark needs to be explained. Gravitational self-catching can be essential in the physical systems which character lengths correlates with $r_A$ *in conditions of weak influence of central massive bodies.* The corresponding conditions are realizing in tsunami waves. For water by the earth conditions
$T = 300$ K, $\rho = 1$ g/cm$^3$; $\gamma_N = 6.6 \cdot 10^{-8}$cm$^3/(\text{g} \cdot \text{s}^2)$, the lengths $r_A = 407.43$ km and $\lambda_A = 2558.66$ km. For close collisions $r_c \sim 10^{-7}$ cm and Coulomb logarithm $\Lambda_a = \ln \frac{r_A}{r_c} \sim 10^2$.
The delivered theory can be applied in the Earth conditions if the influence of central mass can be excluded from consideration. This situation realizes in the tsunami motion because the direction to the Earth center supposes perpendicular to the direction of additional self-consistent gravitational force acting in the tsunami wave. In essence, the catching of water mass is realizing by longitudinal self-consistent gravitational wave. I don't intend here to go into details, but the origin of effects of the small attenuation can be qualitatively explained from position of kinetic theory. Let us calculate the mean velocity $\bar{u}_+$ of particles moving in a chosen direction. If this direction is considering as the positive ones, then $u > 0$ and for the Maxwellian function $f_0$

$$\bar{u}_+ = \sqrt{\frac{m}{2\pi k_B T}} \int_0^{+\infty} e^{-mu^2/2k_B T} u du = \sqrt{\frac{k_B T}{2\pi m}}. \tag{5.58}$$

Kinetic energy, connected with this motion is

$$m\bar{u}_+^{\ 2}/2 = k_B T/(4\pi). \qquad (5.59)$$

From relations (5.27), (5.59) follow

$$\bar{u}_+ = r_A\sqrt{2\rho\gamma_N} \qquad (5.60)$$

Therefore, if the selected direction is opposite to the direction of the wave motion, energy of gravitational field $E_a = \gamma_N m\rho r_a^2$ (per particle) should be applied for exclusion of such kind of particles. For water in considered estimation one obtains $E_a = \gamma_N m\rho r_A^2 = 3.293\cdot 10^{-15}\,\mathrm{erg}$, $\bar{u}_+ = r_A\sqrt{2\rho\gamma_N} = 533$ km/hour- the typical value of tsunami in ocean. Otherwise, the wave expansion leads to the energy dissipation of directional motion in the form of the chaotic heat motion. But in the case if the forces of gravitation attraction counteract (or keep to a minimum) these losses, the wave is moving without attenuation.

## 6. Disk galaxy rotation curves and the problem of dark matter.

About forty years after Zwicky's initial observations, in the late 1960s and early 1970s, Vera Rubin, astronomer at the Department of Terrestrial Magnetism at the Carnegie Institution of Washington presented findings based on a new sensitive spectrograph that could measure the velocity curve of edge-on spiral galaxies to a greater degree of accuracy than had ever before been achieved. Together with Kent Ford, Rubin (Rubin *et al.* 1970, 1980) announced at a 1975 meeting of the American Astronomical Society the astonishing discovery that most stars in spiral galaxies orbit at roughly the same speed reflected schematically on figure 32. The following extensive radio observations determined the detailed rotation curve of spiral disk galaxies to be flat (as the curve B), much beyond as seen in the optical band. Obviously the trivial balance between the gravitational and centrifugal forces leads to relation between orbital speed $V$ and galactocentric distance $r$ as $V^2 = \gamma_N M/r$ beyond the physical extent of the galaxy of mass $M$ (the curve A). The obvious contradiction with the velocity curve B having a 'flat' appearance out to a large radius, was explained by introduction of a new physical essence – dark matter because for spherically symmetric case the hypothetical density distribution $\rho(r)\sim 1/r^2$ leads to $V = const$. The result of this activity is known – undetectable dark matter which does not emit radiation, inferred solely from its gravitational effects. But it means that upwards of 50% of the mass of galaxies was contained in the dark galactic halo.

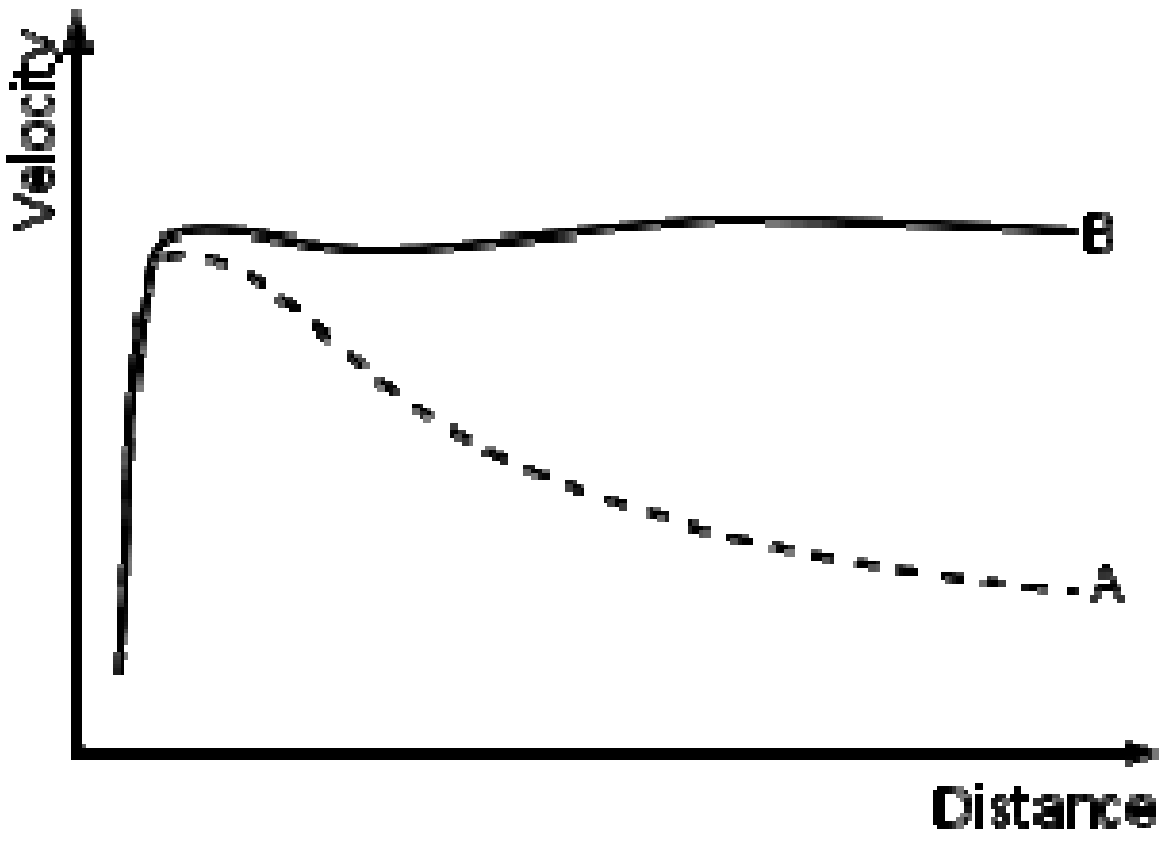

Figure 32. Rotation curve of a typical spiral galaxy: predicted (**A**) and observed (**B**).

In the following I intend to show that the character features reflected on figure 32 can be explained in the frame of Newtonian gravitation law and the non-local kinetic description created by me. With this aim let us consider the formation of the soliton's type of solution of the generalized hydrodynamic equations for gravitational media like galaxy in the self consistent gravitational field. Our aim consists in calculation of the self-consistent hydrodynamic moments of possible formation like gravitational soliton. In the first approximation for spiral galaxies the problem can be considered in the frame of the non-stationary 1D formulation. It is sufficient for our aims, introduction of spherical coordinate system or generalized relativistic description (Alexeev & Ovchinnikova 2010ab) do not lead to changing of following conclusions. Then the system of GHE consist from the generalized Poisson equation reflecting the effects of the density and the density flux perturbations, continuity equation, motion and energy equations. The GHE derivation can be found in (Alexeev 1994, 2004). This system of four equations for non-stationary 1D case is written in the form (see also (1.9)-(1.12) and (2.1)-(2.6)):

(Poisson equation)

$$\frac{\partial^2 \Psi}{\partial x^2} = 4\pi\gamma_N \left[ \rho - \tau \left( \frac{\partial \rho}{\partial t} + \frac{\partial}{\partial x}(\rho u) \right) \right], \tag{6.1}$$

(continuity equation)

$$\frac{\partial}{\partial t}\left\{ \rho - \tau\left[ \frac{\partial \rho}{\partial t} + \frac{\partial}{\partial x}(\rho v_0) \right] \right\} + \frac{\partial}{\partial x}\left\{ \rho v_0 - \tau\left[ \frac{\partial}{\partial t}(\rho v_0) + \right.\right.$$
$$\left.\left. + \frac{\partial}{\partial x}(\rho v_0^2) + \frac{\partial p}{\partial x} + \rho\frac{\partial \Psi}{\partial x} \right] \right\} = 0, \tag{6.2}$$

(motion equation)

$$\frac{\partial}{\partial t}\left\{ \rho u - \tau\left[ \frac{\partial}{\partial t}(\rho u) + \frac{\partial}{\partial x}(\rho u^2) + \frac{\partial p}{\partial x} + \rho\frac{\partial \Psi}{\partial x} \right] \right\} + \frac{\partial \Psi}{\partial x}\left[ \rho - \tau\left( \frac{\partial \rho}{\partial t} + \frac{\partial}{\partial x}(\rho u) \right) \right] +$$
$$+ \frac{\partial}{\partial x}\left\{ \rho u^2 + p - \tau\left[ \frac{\partial}{\partial t}(\rho u^2 + p) + \frac{\partial}{\partial x}(\rho u^3 + 3pu) + 2\rho u\frac{\partial \Psi}{\partial x} \right] \right\} = 0, \tag{6.3}$$

(energy equation)

$$\frac{\partial}{\partial t}\left\{ \rho u^2 + 3p - \tau\left[ \frac{\partial}{\partial t}(\rho u^2 + 3p) + \frac{\partial}{\partial x}(\rho u^3 + 5pu) + 2\rho u\frac{\partial \Psi}{\partial x} \right] \right\} +$$
$$+ \frac{\partial}{\partial x}\left\{ \rho u^3 + 5pu - \tau\left[ \frac{\partial}{\partial t}(\rho u^3 + 5pu) + \frac{\partial}{\partial x}\left(\rho u^4 + \right.\right.\right.$$
$$\left.\left.\left. + 8pu^2 + 5\frac{p^2}{\rho}\right) + \frac{\partial \Psi}{\partial x}(3\rho u^2 + 5p) \right] \right\} + \tag{6.4}$$
$$+ 2\frac{\partial \Psi}{\partial x}\left\{ \rho u - \tau\left[ \frac{\partial}{\partial t}(\rho u) + \frac{\partial}{\partial x}(\rho u^2 + p) + \rho\frac{\partial \Psi}{\partial x} \right] \right\} = 0,$$

where $u$ is translational velocity of the one species object, $\Psi$ - self consistent gravitational potential, $\rho$ is density and $p$ is pressure, $\tau$ is non-locality parameter.

Equations (6.1) (6.4) have remarkable structure from positions of existing now many attempts to correct results of the mathematical modeling of galactic halo by introducing $\rho_D$ - dark matter density, which has different approximations from the observation data. As you see from (6.1) $\rho_D$ coincide with $-\tau\left[ \frac{\partial \rho}{\partial t} + \frac{\partial}{\partial x}(\rho v_0) \right]$. Other analogical terms in equations (6.1) – (6.4) could be associated with the flux of dark matter, energy density of dark matter, energy flux of dark matter and so on. Obviously this activity has no physical sense.

Let us introduce the coordinate system moving along the positive direction of $x$- axis in ID space with velocity $C = u_0$ equal to phase velocity of considering object

$$\xi = x - Ct. \qquad (6.5)$$

Taking into account the De Broglie relation we should wait that the group velocity $u_g$ is equal $2u_0$. In moving coordinate system all dependent hydrodynamic values are function of $(\xi, t)$.We investigate the possibility of the object formation of the soliton type. For this solution there is no explicit dependence on time for coordinate system moving with the phase velocity $u_0$. Write down the system of equations (6.1) - (6.4) in the dimensionless form, where dimensionless symbols are marked by tildes. For the scales $\rho_0$, $u_0, x_0 = u_0 t_0, \Psi_0 = u_0^2$, $\gamma_{N0} = \frac{u_0^2}{\rho_0 x_0^2}$ $p_0 = \rho_0 u_0^2$ and conditions $\tilde{C} = C/u_0 = 1$, the equations take the form:

(generalized Poisson equation)

$$\frac{\partial^2 \tilde{\Psi}}{\partial \tilde{\xi}^2} = 4\pi\tilde{\gamma}_N \left[ \tilde{\rho} - \tilde{\tau}\left( -\frac{\partial \tilde{\rho}}{\partial \tilde{\xi}} + \frac{\partial}{\partial \tilde{\xi}}(\tilde{\rho}\tilde{u}) \right) \right], \qquad (6.6)$$

(continuity equation)

$$\frac{\partial \tilde{\rho}}{\partial \tilde{\xi}} - \frac{\partial \tilde{\rho}\tilde{u}}{\partial \tilde{\xi}} + \frac{\partial}{\partial \tilde{\xi}}\left\{ \tilde{\tau}\left[ \frac{\partial}{\partial \tilde{\xi}}\left[ \tilde{p} + \tilde{\rho}\tilde{u}^2 + \tilde{\rho} - 2\tilde{\rho}\tilde{u} \right] + \tilde{\rho}\frac{\partial \tilde{\Psi}}{\partial \tilde{\xi}} \right] \right\} = 0, \qquad (6.7)$$

(motion equation)

$$\frac{\partial}{\partial \tilde{\xi}}\left( \tilde{\rho}\tilde{u}^2 + \tilde{p} - \tilde{\rho}\tilde{u} \right) + \frac{\partial}{\partial \tilde{\xi}}\left\{ \tilde{\tau}\left[ \frac{\partial}{\partial \tilde{\xi}}\left( 2\tilde{\rho}\tilde{u}^2 - \tilde{\rho}\tilde{u} + 2\tilde{p} - \tilde{\rho}\tilde{u}^3 - 3\tilde{p}\tilde{u} \right) + \tilde{\rho}\frac{\partial \tilde{\Psi}}{\partial \tilde{\xi}} \right] \right\} +$$

$$+ \frac{\partial \tilde{\Psi}}{\partial \tilde{\xi}}\left\{ \tilde{\rho} - \tilde{\tau}\left[ -\frac{\partial \tilde{\rho}}{\partial \tilde{\xi}} + \frac{\partial}{\partial \tilde{\xi}}(\tilde{\rho}\tilde{u}) \right] \right\} - 2\frac{\partial}{\partial \tilde{\xi}}\left\{ \tilde{\tau}\tilde{\rho}\tilde{u}\frac{\partial \tilde{\Psi}}{\partial \tilde{\xi}} \right\} = 0, \qquad (6.8)$$

(energy equation)

$$\frac{\partial}{\partial \tilde{\xi}}\left( \tilde{\rho}\tilde{u}^2 + 3\tilde{p} - \tilde{\rho}\tilde{u}^3 - 5\tilde{p}\tilde{u} \right) -$$

$$- \frac{\partial}{\partial \tilde{\xi}}\left\{ \tilde{\tau}\frac{\partial}{\partial \tilde{\xi}}\left( 2\tilde{\rho}\tilde{u}^3 + 10\tilde{p}\tilde{u} - \tilde{\rho}\tilde{u}^2 - 3\tilde{p} - \tilde{\rho}\tilde{u}^4 - 8\tilde{p}\tilde{u}^2 - 5\frac{\tilde{p}^2}{\tilde{\rho}} \right) \right\} +$$

$$+ \frac{\partial}{\partial \tilde{\xi}}\left\{ \tilde{\tau}\left( 3\tilde{\rho}\tilde{u}^2 + 5\tilde{p} \right)\frac{\partial \tilde{\Psi}}{\partial \tilde{\xi}} \right\} - 2\tilde{\rho}\tilde{u}\frac{\partial \tilde{\Psi}}{\partial \tilde{\xi}} - 2\frac{\partial}{\partial \tilde{\xi}}\left\{ \tilde{\tau}\tilde{\rho}\tilde{u}\frac{\partial \tilde{\Psi}}{\partial \tilde{\xi}} \right\} + \qquad (6.9)$$

$$+ 2\tilde{\tau}\frac{\partial \tilde{\Psi}}{\partial \tilde{\xi}}\left[ -\frac{\partial}{\partial \tilde{\xi}}(\tilde{\rho}\tilde{u}) + \frac{\partial}{\partial \tilde{\xi}}\left( \tilde{\rho}\tilde{u}^2 + \tilde{p} \right) + \tilde{\rho}\frac{\partial \tilde{\Psi}}{\partial \tilde{\xi}} \right] = 0,$$

Some comments to the system of four ordinary non-linear equations (6.6) – (6.9):

1. Every equation from the system is of the second order and needs two conditions. The problem belongs to the class of Cauchy problems.
2. In comparison for example, with the Schrödinger theory connected with behavior of the wave function, no special conditions are applied for dependent variables including the domain of the solution existing. This domain is defined automatically in the process of the numerical solution of the concrete variant of calculations.

3. From the introduced scales $\rho_0$, $u_0, x_0 = u_0 t_0, \Psi_0 = u_0^2$, $\gamma_{N0} = \frac{u_0^2}{\rho_0 x_0^2}$, $p_0 = \rho_0 u_0^2$, only three parameters are independent, namely, $\rho_0$, $u_0, x_0$.

4. Approximation for the dimensionless non-local parameter $\tilde{\tau}$ should be introduced (see (1.7), (1.8), (2.7)-(2.10)). In the definite sense it is not the problem of the hydrodynamic level of the physical system description (like the calculation of the kinetic coefficients in the classical hydrodynamics). Interesting to notice that quantum GHE were applied with success for calculation of atom structure (Alexeev 2008ab), which is considered as two species charged $e,i$ mixture. The corresponding approximations for non-local parameters $\tau_i$, $\tau_e$ and $\tau_{ei}$ are proposed in (Alexeev 2008b). In the theory of the atom structure (Alexeev 2008b) after taking into account the Balmer's relation, (1.8) transforms into

$$\tau_e = {n\hbar}\big/{m_e u^2}, \quad (6.10)$$

where $n = 1,2,...$ is principal quantum number. As result the length scale relation was written as $x_0 = \frac{H}{m_e u_0} = \frac{n\hbar}{m_e u_0}$. But the value $\nu^{qu} = \hbar / m_e$ has the dimension $[cm^2 / s]$ and can be titled as quantum viscosity, $\nu^{qu} = 1.1577 \ cm^2 / s.$ Then

$$\tau_e = {n\nu^{qu}}\big/{u^2}. \quad (6.11)$$

Introduce now the quantum Reynolds number

$$\mathrm{Re}^{qu} = \frac{u_0 x_0}{\nu^{qu}}. \quad (6.12)$$

As result from (6.10), (6.11) follows the condition of quantization for $\mathrm{Re}^{qu}$. Namely

$$\mathrm{Re}^{qu} = n, \quad n = 1,2,... \quad (6.13)$$

5. Taking into account the previous considerations I introduce the following approximation for the dimensionless non-local parameter

$$\tilde{\tau} = \frac{1}{\tilde{u}^2}, \quad (6.14)$$

or

$$\tau = \frac{1}{u^2} u_0 x_0 = \frac{\nu_0^k}{u^2}, \quad (6.15)$$

where the scale for the kinematical viscosity is introduced $\nu_0^k = u_0 x_0$, (compare with (6.11)). Then we have the physically transparent result – non-local parameter is proportional to the kinematical viscosity and in inverse proportion to the square of velocity.

Now we are ready to display the results of the mathematical modeling realized with the help of Maple (the versions Maple 9 or more can be used). The system of generalized hydrodynamic equations (6.6) – (6.9) have the great possibilities of mathematical modeling as result of changing of eight Cauchy conditions describing the character features of initial perturbations which lead to the soliton formation.

The following Maple notations on figures are used: r- density $\tilde{\rho}$, u- velocity $\tilde{u}$ ( solid black line), p - pressure $\tilde{p}$ (black dashed line) and v - self consistent potential $\tilde{\Psi}$. Explanations placed under all following figures, Maple program contains Maple's notations – for example the expression $D(u)(0) = 0$ means in the usual notations $\frac{\partial \tilde{u}}{\partial \tilde{\xi}}(0) = 0$, independent variable $t$ responds to $\tilde{\xi}$. We begin with investigation of the problem of principle significance – is it possible after a perturbation (defined by Cauchy conditions) to obtain the gravitational object of the soliton's kind as result of the self-organization of the matter? With this aim let us consider the initial perturbations

**u(0)=1,p(0)=1,r(0)=1,D(u)(0)=0,D(p)(0)=0,D(r)(0)=0,D(v)(0)=0, v(0)=1**

The following figures reflect the result of solution of equations. (6.6) – (6.9) with the choice of scales leading to $\tilde{\gamma}_N = 1$. Figures 33 – 36 correspond to the approximation of the non-local parameter $\tilde{\tau}$ in the form (4.14).

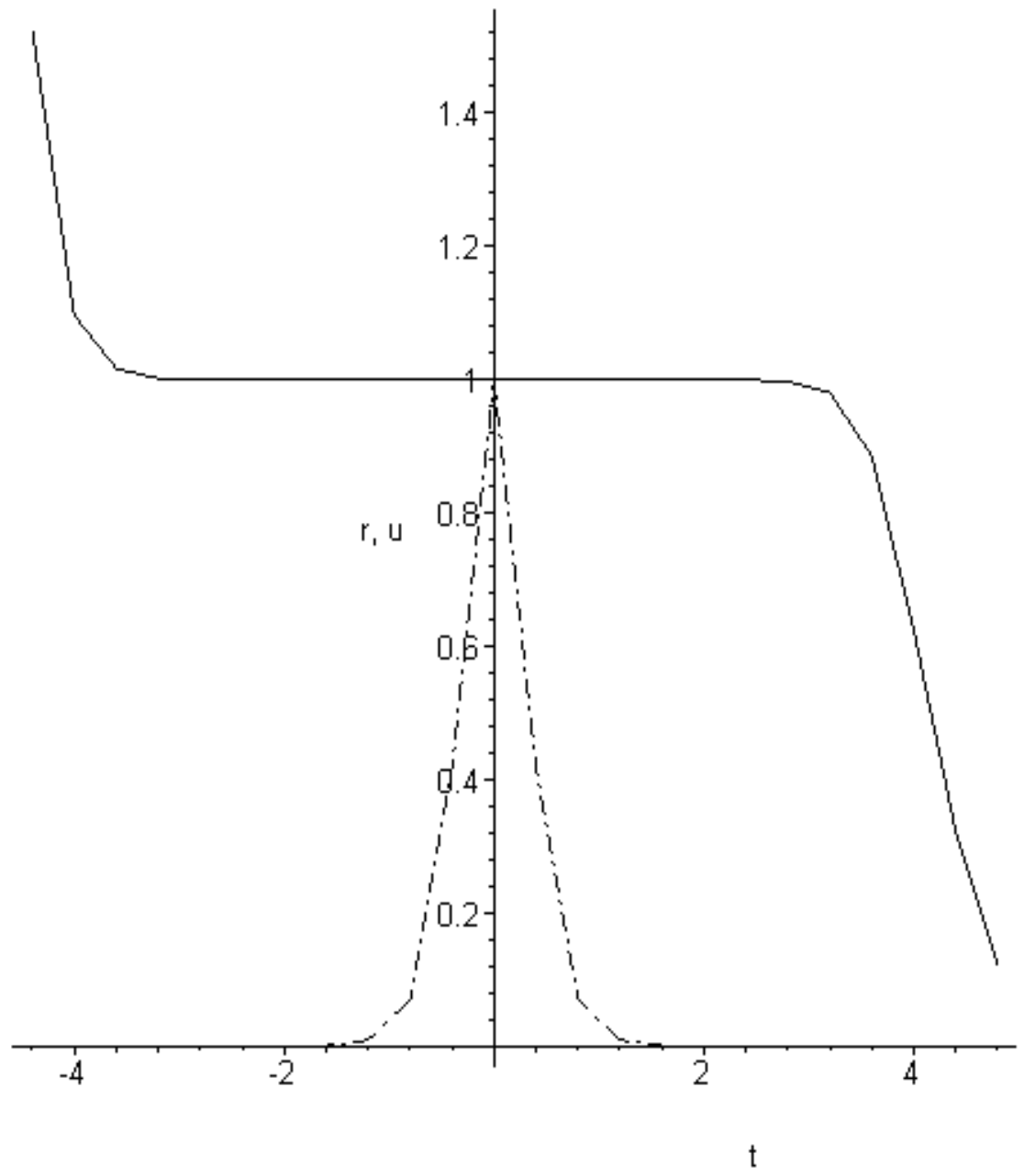

Figure 33. r - density $\tilde{\rho}$ (black dash dotted line), u - velocity $\tilde{u}$ in gravitational soliton.

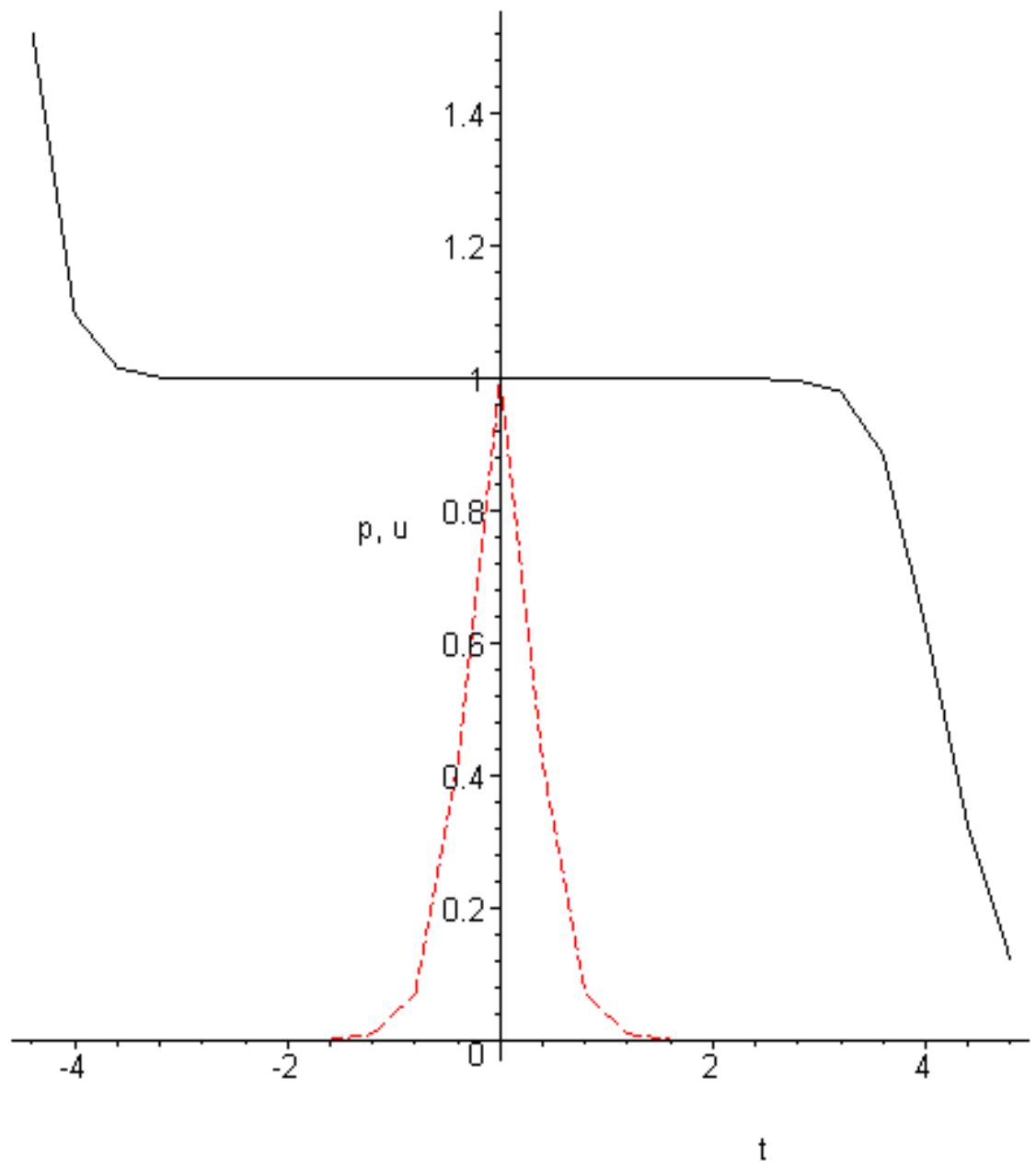

Figure 34. p - pressure $\tilde{p}$, u- velocity $\tilde{u}$ in gravitational soliton.

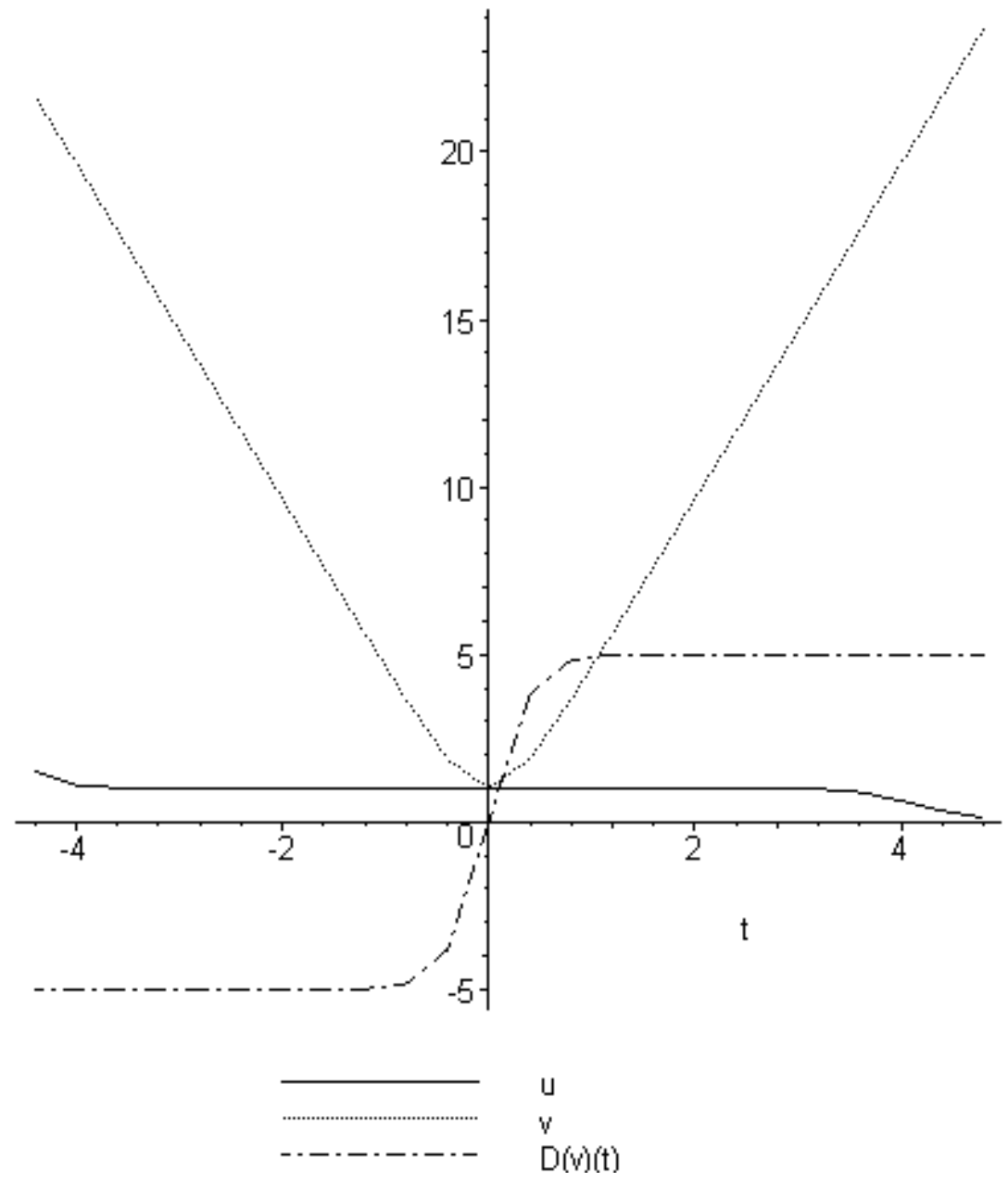

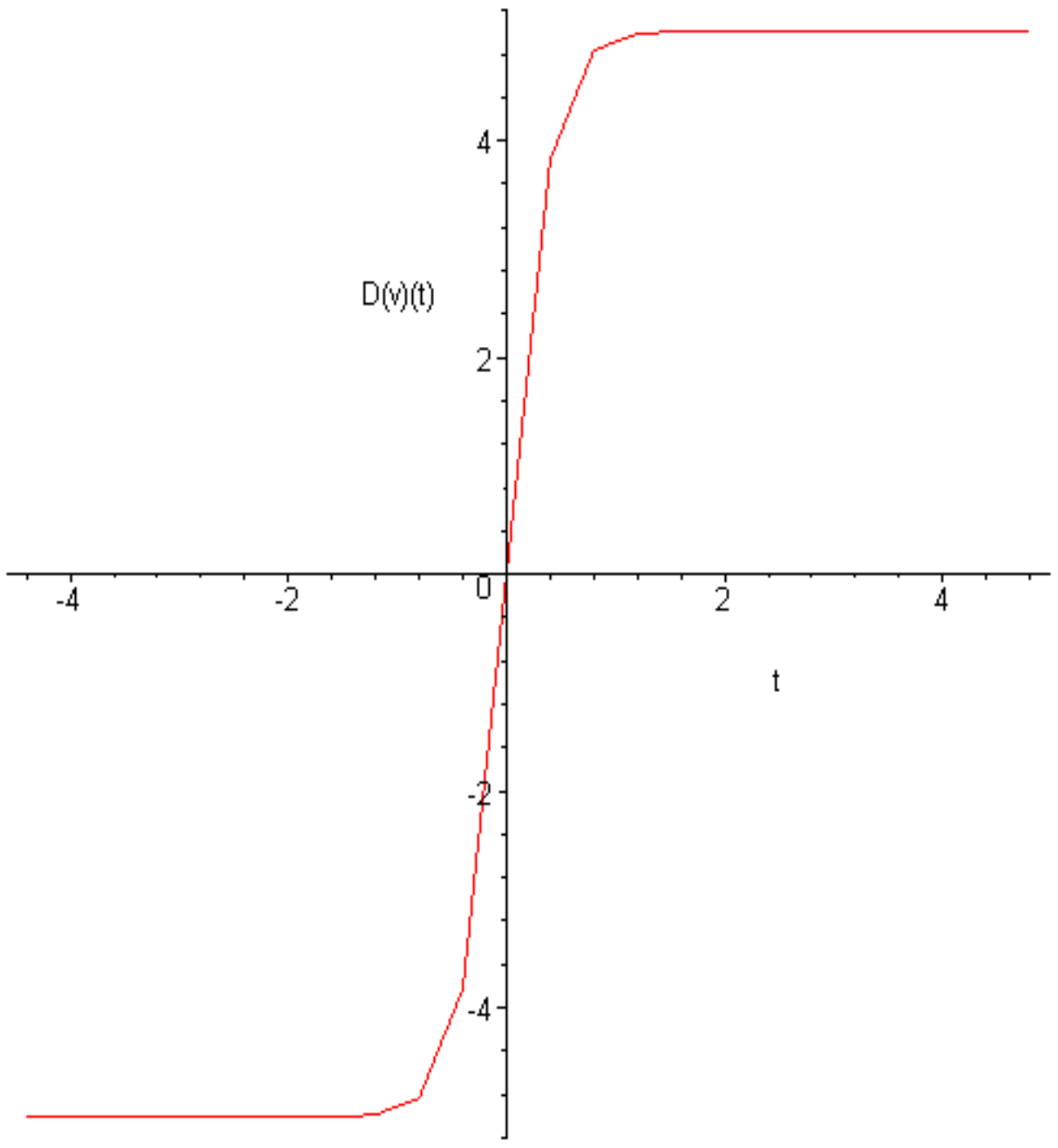

Figure 35. u- velocity $\tilde{u}$, v - self consistent potential $\tilde{\Psi}$, $D(\mathrm{v})(t) = \partial\tilde{\Psi}/\partial\tilde{\xi}$ in quantum soliton, (left).

Figure 36. $D(\mathrm{v})(t) = \partial\tilde{\Psi}/\partial\tilde{\xi}$ in quantum soliton, (right).

Figure 33 displays the gravitational object placed in bounded region of 1D space, all parts of this object are moving with the same velocity. Important to underline that no special boundary conditions were used for this and all following cases. Then this soliton is product of the self-organization of gravitational matter. Figures 34, 35 contain the answer for formulated above question about stability of the object. The derivative (see figure 36) $\frac{\partial \tilde{\Psi}}{\partial \tilde{\xi}} = \frac{\partial \Psi}{\partial \xi} \frac{x_0}{u_0^2} = -g(\xi) \Big/ \left(u_0^2 / x_0\right) = -\tilde{g}(\xi)$ is proportional to the self-consistent gravitational force acting on the soliton and in its vicinity. Therefore the stability of the object is result of the self-consistent influence of the gravitational potential and pressure.

Extremely important that the self-consistent gravitational force has the character of the flat area which exists in the vicinity of the object. This solution exists only in the restricted area of space; the corresponding character length is defined automatically as result of the numerical solution of the problem.

The non-local parameter $\tilde{\tau}$, in the definite sense plays the role analogous to kinetic coefficients in the usual Boltzmann kinetic theory. The influence on the results of calculations is not too significant. The same situation exists in the generalized hydrodynamics.

Really, let us use the another approximation for $\tilde{\tau}$ in the simplest possible form, namely

$$\tilde{\tau} = 1. \tag{6.16}$$

The following figures 37 – 40 reflect the results of solution of equations. (6.6) – (6.9) with the choice of scales leading to $\tilde{\gamma}_N = 1$, but with the approximation of the non-local parameter $\tilde{\tau}$ in the form (6.16).

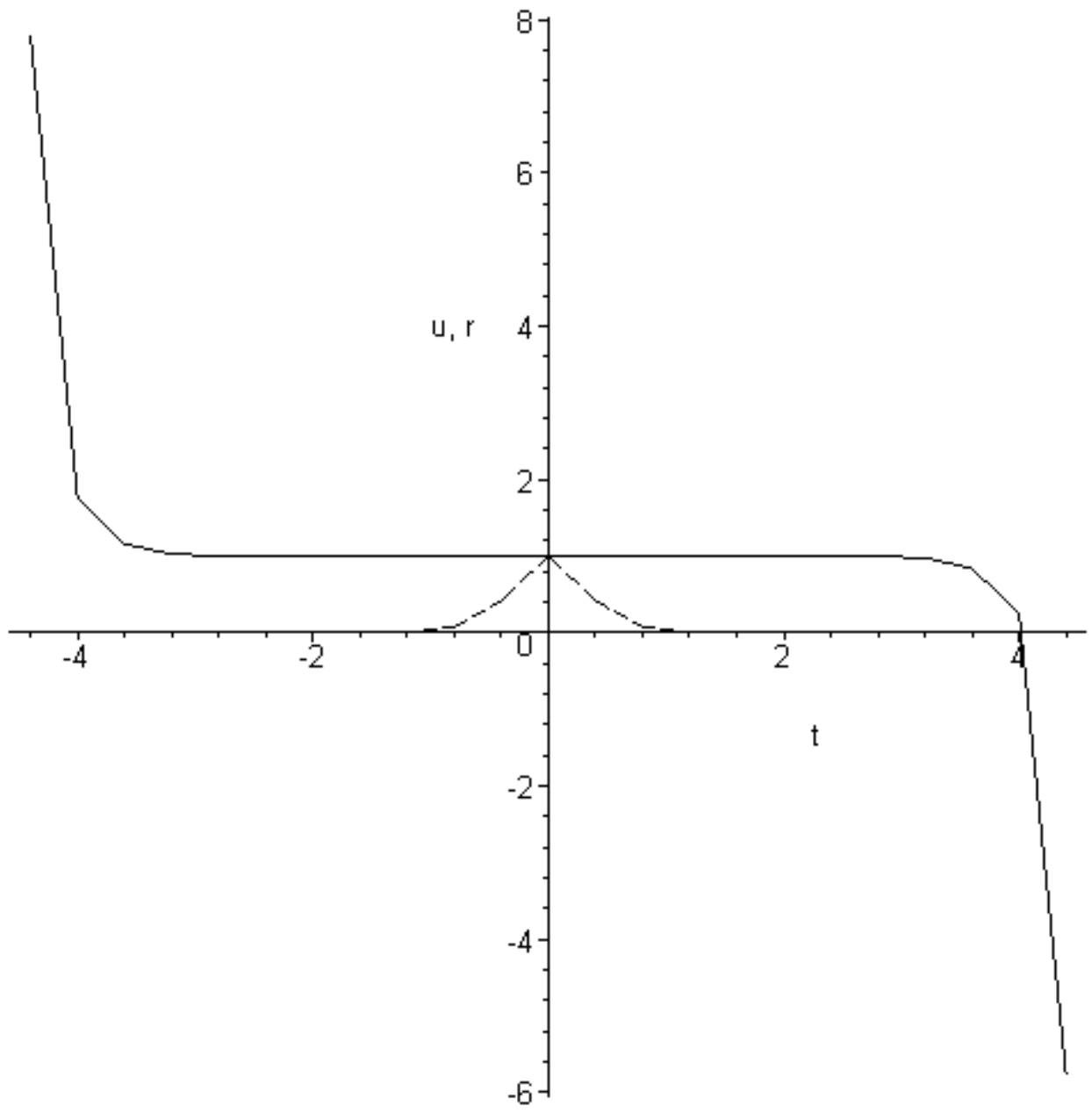

Figure 37. r - density $\tilde{\rho}$ (black dashed line), u - $\tilde{u}$ in gravitational soliton.

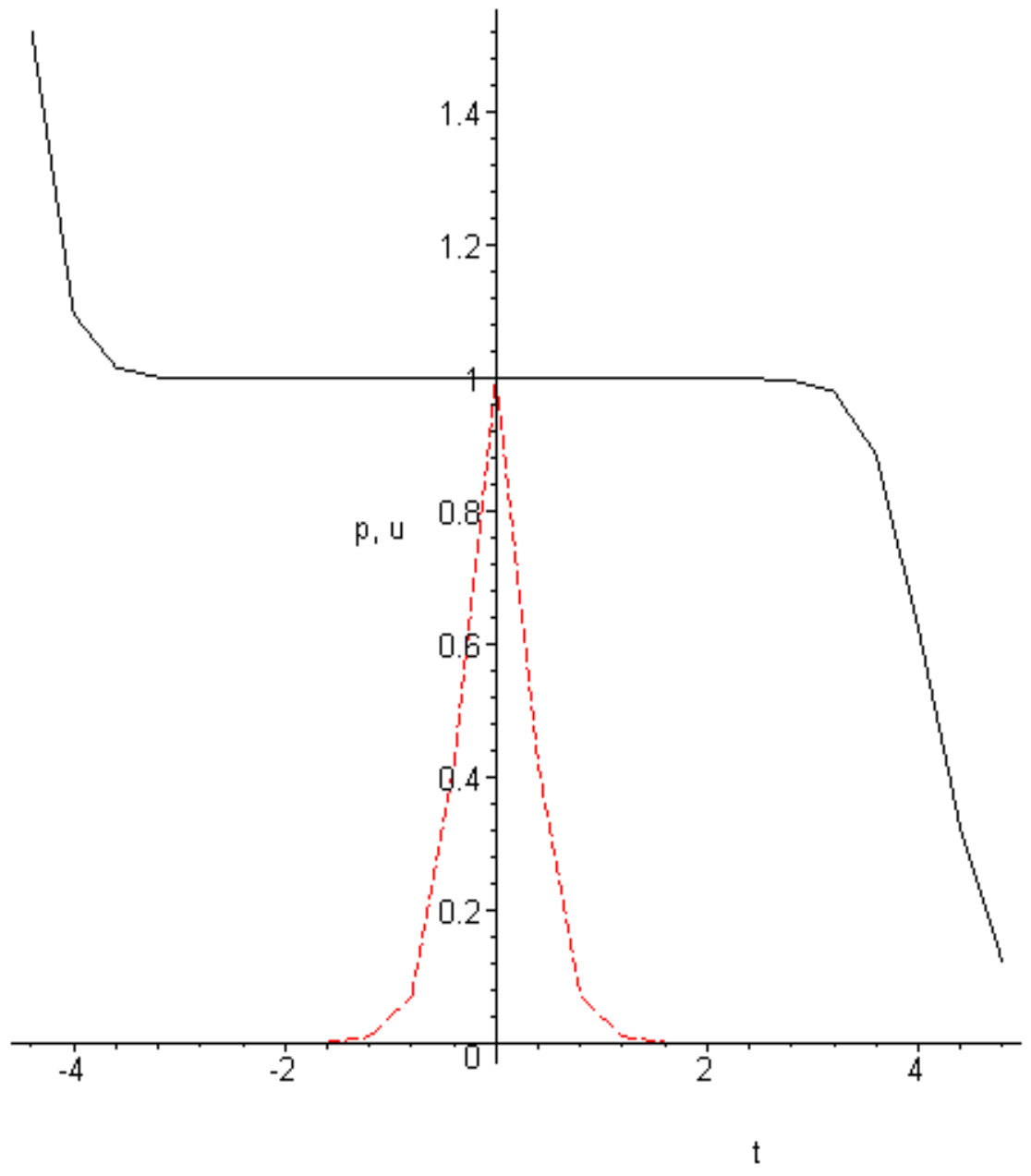

Figure 38. p - pressure $\tilde{p}$ (dashed line), u-velocity $\tilde{u}$ in gravitational soliton.

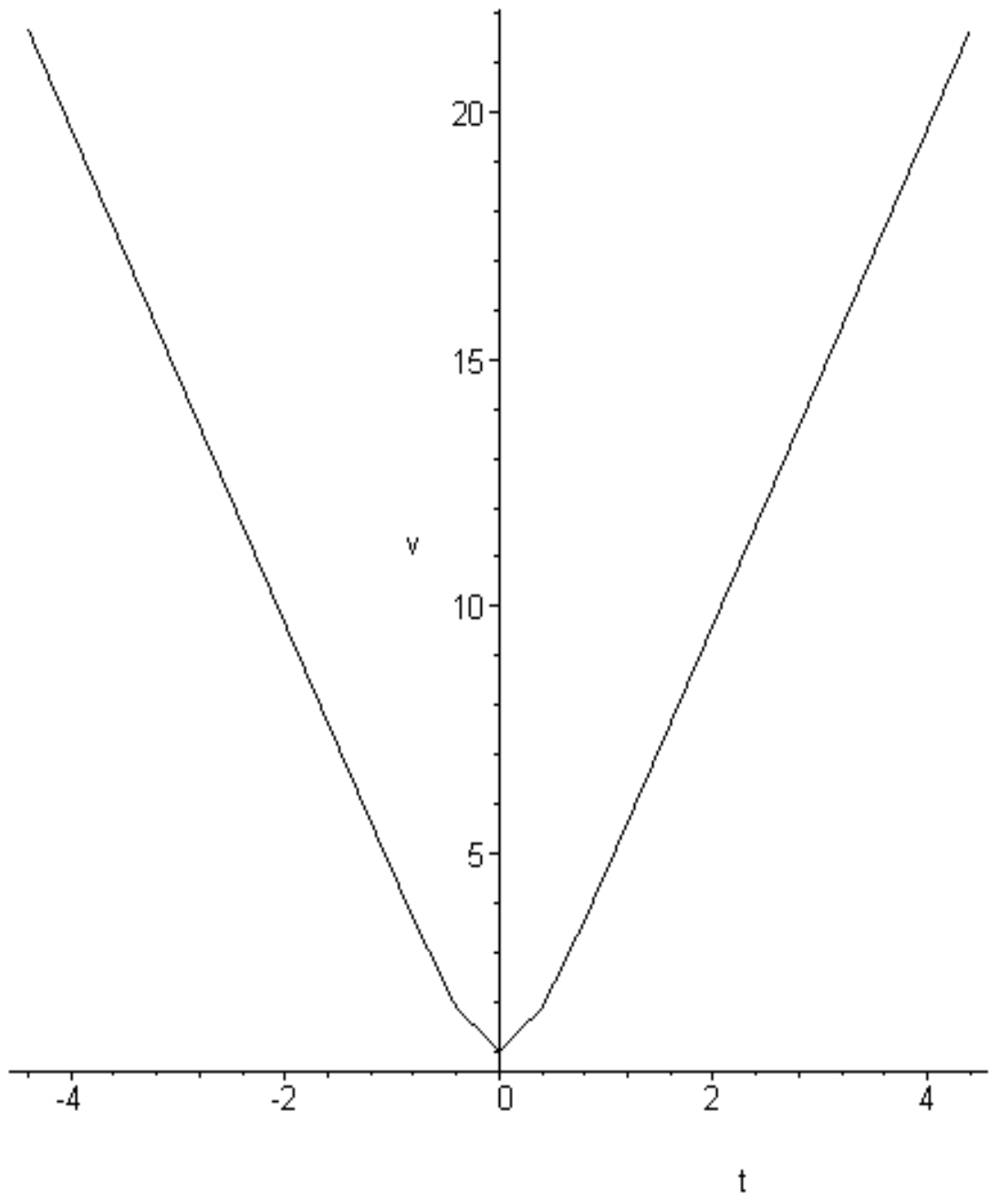

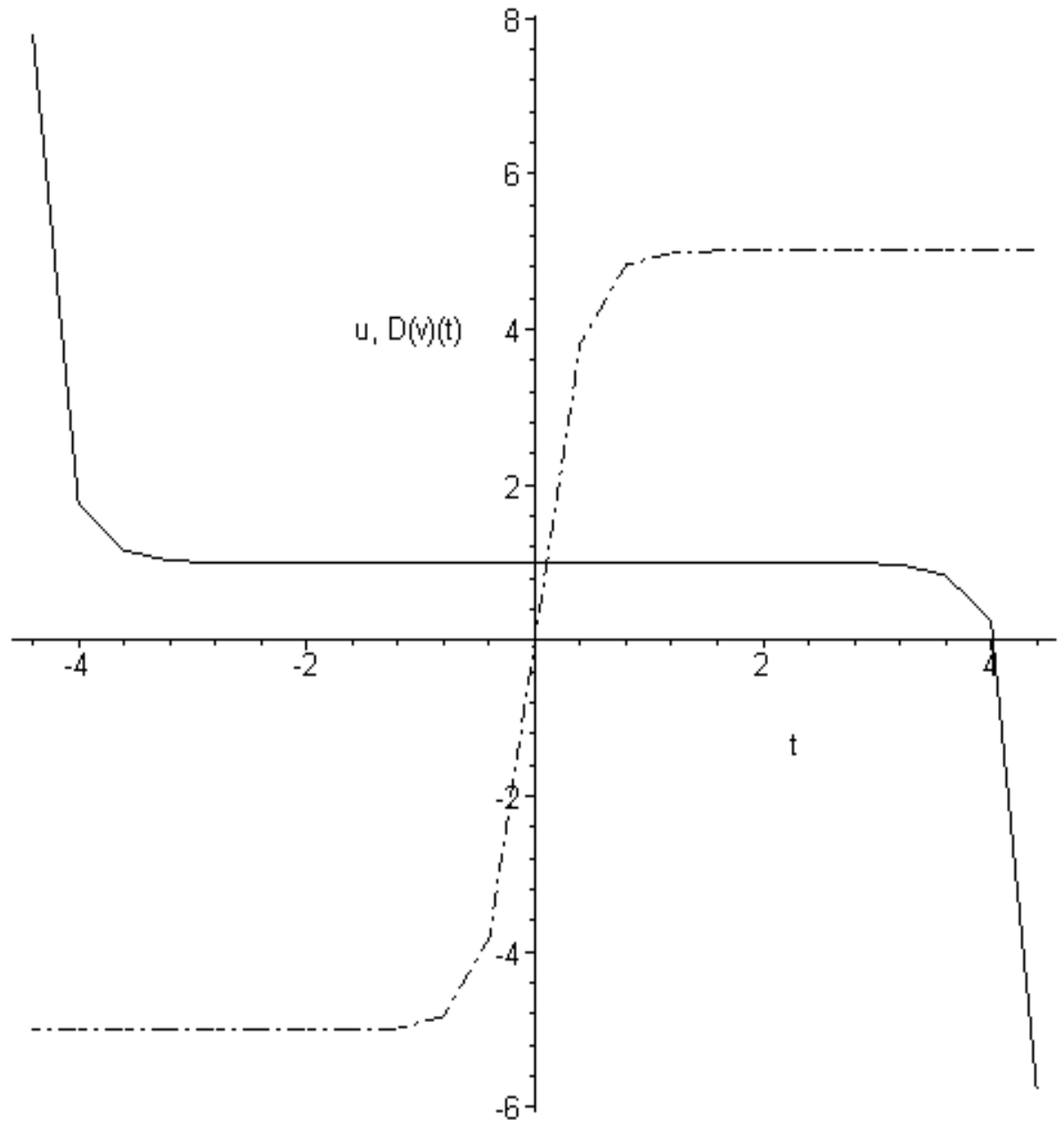

Figure 39. v – self consistent potential $\tilde{\Psi}$ in gravitational soliton, (left).

Figure 40. u- velocity $\tilde{u}$ (black solid line), $D(\mathrm{v})(t) = \frac{\partial \tilde{\Psi}}{\partial \tilde{\xi}}$ in quantum soliton, (right).

As it is follows from figures 33 – 40 in the vicinity of the central massive galaxy kernel – gravitational soliton, exists the domain with the constant gravitational force acting on the unit of mass. As result rotation curve of a typical spiral galaxy follows the curve like (B) instead of (A) on figure 32. These peculiar features of the halo movement can be explained without new concepts like “dark matter”.

Taking into account all obtained results of mathematical modeling including small cross sections of the antimatter particles and Hubble effect, it is possible to wait for the formation of, so to speak, the double matter – antimatter layer close, as a rule, to the edge of visible Universe. “Discharge” in the mentioned double layer - annihilation of matter and antimatter on the level of stars and “anti-stars”, galaxies and “anti-galaxies” should lead to often bursts in the form of gamma rays accompanied by an optical counterpart that peaked at a visible magnitude. Moreover this model leads to the isotropic distribution in visible space of mentioned events of explosions. “Discharge” in the mentioned double layer should lead to gamma-ray bursts in the form of highly focused explosions, with most of the explosion energy collimated into a narrow jet. But Gamma-ray bursts (GRBs) - flashes of gamma rays associated with extremely energetic explosions in distant galaxies – were really detected in 1967 by the Vela satellites, a series of satellites designed to detect covert nuclear weapons tests. All mentioned above effects are really observed now including narrow jets traveling at speeds close to the speed of light. Gamma-ray bursts are very bright as observed from Earth despite typical immense distances. GRBs have extremely luminous optical counterparts as well. GRB 080319B, for example, was accompanied by an optical counterpart that peaked at a visible magnitude comparable to that of the dimmest naked-eye stars despite the burst's distance of 7.5 billion light years (Racusin, J.L. *et al.* 2008).

## 7. Conclusion

The problems of anti-matter, antigravitation, dark energy, dark matter reflects the crisis of the local transport kinetic theory. The origin of difficulties consists in Oversimplification following from principles of local physics.

In other words the problem of Oversimplification is not "trivial" simplification of the important problem. The situation is much more serious – Oversimplification based on principles of local physics, and obvious crisis, we see in astrophysics, simply reflects the general shortenings of the local kinetic transport theory. The aim of investigations undertaken by me here and before consists in creation of the unified non-local theory of transport processes. It is not a new idea that physics is unified construction but not the collection of inconsistent facts. I hope this paper demonstrate the validity of this conception.

## References

Alekseev, B.V. 1982 *Matematicheskaya Kinetika Reagiruyushchikh Gazov (Mathematical theory of reacting gases)* Moscow, Nauka.

Alekseev, B.V. 2000 Physical principles of the generalized Boltzmann kinetic theory of gases. *Physics-Uspekhi*, **170** (6) 601-629.

Alekseev, B.V. 2003 Physical principles of the generalized Boltzmann kinetic theory of ionized gases. *Physics-Uspekhi*, **173** (2) 139-167.

Alexeev, B.V. 1994 The generalized Boltzmann equation, generalized hydrodynamic equations and their applications. *Phil. Trans. Roy. Soc. Lond.* **349** p.417-443.

Alexeev, B. V. 1995 The generalized Boltzmann equation. Physica A. **216.** 459-468.

Alexeev, B.V. 2004 *Generalized Boltzmann Physical Kinetics*. Elsevier, Amsterdam, The Netherlands**.**

Alexeev, B.V. 2008 Generalized Quantum Hydrodynamics and Principles of Non-Local Physics, *J. Nanoelectron. Optoelectron*. **3**, 143 - 158.

Alexeev, B.V. 2008 Application of generalized quantum hydrodynamics in the theory of quantum soliton's evolution. *J. Nanoelectron. Optoelectron*. **3**, 316 - 328.

Alexeev, B.V. 2009a Generalized Theory of Landau Damping *J. Nanoelectron. Optoelectron*. **4**, 186 - 199.

Alexeev, B.V. 2009b Generalized Theory of Landau Damping in Collisional Media *J. Nanoelectron. Optoelectron*. **4**, 379 - 393.

Alexeev, B.V. & Ovchinnikova, I. V. 2010a The Generalized Relativistic Kinetic and Hydrodynamic Theory—Part 1. *J. Nanoelectron. Optoelectron*. **5**, 1 - 14.

Alexeev, B.V. & Ovchinnikova, I. V. 2010b The Generalized Relativistic Kinetic and Hydrodynamic Theory—Part 2. *J. Nanoelectron. Optoelectron.***5**, 15 - 27.

Bell, J.S. 1964 On the Einstein Podolsky Rosen paradox, *Physics*, **1**, 195.

Bogolyubov, N.N. 1946 *Problemy Dinamicheskoi Teorii v Statisticheskoi Fizike* (Dynamic theory problems in statistical physics) (Moscow Leningrad Gostekhizdat 1946) [Translated into English *The Dynamical Theory in Statistical Physics* (Delhi Hindustan Publ. Corp. 1965)]

Boltzmann, L. 1872 Weitere Studien über das Wärmegleichgewicht unter Gasmolekulen. *Sitz.Ber.Kaiserl. Akad. Wiss*. **66(2)** s.275.

Boltzmann, L. 1912 *Vorlesungen über Gastheorie.* (Leipzig: *Verlag von Johann Barth*. Zweiter unveränderten Abdruck. 2 Teile.

Born, M. & Green H.S. 1946 A General Kinetic Theory of Liquids 1. The molecular distribution function. *Proc. Roy. Soc*. 188 (1012) p.10.

Chapman S. & Cowling T.G. 1952 *The mathematical theory of non-uniform gases*. Cambridge University Press.

Chernin, A.D. 2008 Dark energy and universal antigravitation. *Physics – Uspekhi*, 51 (3), 253 – 282.

Gliner, E.B. 2002 Inflationary universe and the vacuum-like state of physical medium. *Physics – Uspekhi*, **172**, N2, 213-220.

Green, H.S. 1952 *The molecular theory of fluids*. Amsterdam.

Hirschfelder I.O., Curtiss Ch.F., Bird R.B. 1954 *Molecular Theory of Gases and Liquids*. John Wiley and sons, inc. New York. Chapman and Hall, lim., London.

Kirkwood, J.G. 1947 The Statistical Mechanical Theory of Transports Processes II. Transport in gases. *J. Chem. Phys*. 15 (1) p.72.

Landau, L.D. 1946 On the vibrations of the electronic plasma *J. Phys. USSR* **10** 25-34.

Racusin, J.L. *et al*. 2008 Broadband observations of the naked-eye gamma-ray burst GRB080319B. *Nature* **455** (7210) 183–188.

Rubin, V., Ford W. K., Jr. 1970 Rotation of the Andromeda Nebula from a spectroscopic survey of emission regions. *Astrophysical Journal* **159**: 379.

Rubin, V., Thonnard N., Ford W.K., Jr, 1980 Rotational properties of 21 Sc galaxies with a large range of luminosities and radii from NGC 4605 (R=4kpc) to UGC 2885 (R=122kpc). *Astrophysical Journal* **238**, 471.